\documentclass[nofootinbib,aps,twocolumn,showpacs]{revtex4-2}

\usepackage{physics}

\usepackage{graphicx}
\usepackage{amsfonts}
\usepackage{amssymb}
\usepackage{amsbsy}
\usepackage{amsmath}
\usepackage{mathrsfs}
\usepackage{latexsym}
\usepackage{natbib}
\usepackage{bm}
\usepackage{subfigure} 
\usepackage{color}
\usepackage{wasysym}
\usepackage{bigints}
\usepackage{bbold}
\usepackage{esvect}
\usepackage{float}
\usepackage{booktabs}

\begin{document}

	\title{Fate of entanglement between two Unruh-DeWitt detectors due to their motion and background temperature}

	\author{Pratyusha Chowdhury}
	\email{cpratyusha03@gmail.com}

	\author{Bibhas Ranjan Majhi}
	\email{bibhas.majhi@iitg.ac.in}

	\affiliation{Department of Physics, Indian Institute of Technology Guwahati, Guwahati 781039, Assam, India}
	
	\pacs{04.62.+v, 04.60.Pp}

	\date{\today}

\begin{abstract}
		We investigate the fate of initial entanglement between two accelerated detectors with respect to an observer attached to one of the detectors. Both $(1+1)$ and $(1+3)$ spacetime dimensions are being considered here, with the detectors interacting with real massless scalar fields through monopole terms. The investigation is being performed for both non-thermal as well as thermal fields. In general, irrespective of the detectors moving in the same Rindler wedge or opposite wedges, increase of the field temperature reduces the initial entanglement. In all situations, degradation of entanglement is high for high acceleration $a_A$ of our observer. Interestingly, the degradation depends on the measure of initial entanglement. For $(1+1)$ dimensions, the degradation saturates for small values of $a_A$, whereas the same fluctuates in $(1+3)$ dimensions with the decrease of $a_A$.  For motions in opposite Rindler wedges, a noticeable feature we observe in $(1+1)$ dimensions is that, depending on the strength of initial entanglement, there is a possibility of entanglement harvesting in the system for certain values of the observers' acceleration. However the same is absent in $(1+3)$ dimensions. The whole analysis is operationally different from earlier similar investigations. The thermal equilibrium is satisfied throughout the calculations here, by considering the Wightman functions with respect to the Rindler modes evaluated in the vacuum of Unruh modes, contrary to the use of Minkowski modes.
\end{abstract}

	\maketitle

	\section{Introduction}
		\label{Introduction}
Recently, quantum entanglement in a relativistic setting has been investigated quite extensively. It is believed to give new insights on various fields such as AdS/CFT, black hole information, etc. 
Using the approach of algebraic quantum field theory in \cite{summers1985vacuum,summers1988maximal,summers1987bell}, it has been shown that the vacuum state of quantum field maximally violates the Bell's inequality \cite{bell1964einstein}, implying that the vacuum contains non-local correlations, and hence entanglement. 
Now, in the pioneering work of Reznik \cite{reznik2003entanglement,reznik2005violating}, a system of initially non-entangled point-like Unruh-DeWitt detectors (UD) \cite{hawking2010general} in causally-disconnected regions, interacting with the intermediate massless scalar quantum field, is shown to become entangled at a later time. Thus, the entanglement existing between the causally-disconnected regions in the vacuum can be extracted with the help of Unruh-Dewitt detectors. This phenomenon is known as \textit{entanglement harvesting} from the vacuum. Entanglement harvesting is a relativistic effect, that occurs due to vacuum fluctuations as seen by different observers. Entanglement harvesting has a great significance in the context of utilization of the extracted entanglement \cite{hotta2008quantum,hotta2009quantum,frey2014strong}, hence it is an active area of research. There has been plenty of work in the area of entanglement harvesting for different spacetimes and setups of different kinds of observers \cite{gallock2021harvesting,cong2019entanglement,zhang2020entanglement,henderson2019entangling,ardenghi2018entanglement,koga2018quantum}. A very recent work also deals with entanglement harvesting from the vacuum in presence of a  thermal background field \cite{Barman:2021bbw}.

 As mentioned that entanglement harvesting is an observer-dependent quantity, it raises the question -- if a system is initially entangled with respect to a certain observer, how does this entanglement vary with respect to other relativistic observers? 
Several works along this line show how the entanglement between the field modes vary for different kind of observers. In \cite{fuentes2005alice}, it is shown that a state of two field modes, maximally entangled in an inertial frame, is less entangled with respect to an accelerating observer. Similar results are also found in \cite{martin2010unveiling}. Extension of such studies to Fermion fields have also been made \cite{alsing2006entanglement,pan2008degradation,wang2011multipartite} and its comparison to bosonic fields have been further studied \cite{richter2015degradation}. 
On the other hand, if two observers (UD detectors) are entangled initially with respect to one frame, how will the entanglement vary with other observers due to interaction with the intermediate quantum field? This question has been addressed in several works. In \cite{lin2008disentanglement}, the entanglement dynamics between two UD detectors modeled by simple harmonic oscillators, one at rest (Alice) and another in constant acceleration (Rob) interacting with the quantum field has been studied. Here the acceleration of one detector brings about a horizon, thus introducing a nontrivial global spacetime structure, causing disentanglement between the two detectors in a finite time. This is referred to as the sudden death of entanglement.
 In \cite{dai2015killing}
	a system of two qubits entangled with each other, coupled to the field modes have been considered, and the dynamics of entanglement for the accelerated qubits, either both in the right Rindler wedge or both in the left Rindler wedge. 
	 It is seen that the logarithmic negativity and concurrence decrease with
	the increase of the acceleration-frequency ratio of each qubit and, especially, suddenly dies at a finite value of the acceleration-frequency ratio. 
Some other works that deal with entanglement dynamics in various scenarios: the dynamics of entanglement in de Sitter spacetime \cite{huang2017dynamics,tian2014dynamics},  for a non-uniformly accelerated detector and a detector at rest \cite{ostapchuk2012entanglement}, the role  of anti- Unruh effect on the dynamics of entanglement enhancement \cite{li2018would}. Further, Ref. \cite{hu2012relativistic} gives an overview of Unruh-DeWitt detectors and other models of detector-field interaction in a relativistic quantum field theory setting as a tool for extracting entanglement.
In a very recent work \cite{zhou2021entanglement}, following the framework of open quantum systems, the entanglement dynamics for a quantum system composed of two uniformly accelerated Unruh-DeWitt detectors interacting with a bath of massive scalar fields in the Minkowski vacuum has been studied. It is seen that the entanglement evolution rate for the quantum system coupled with massive fields is always slower compared to that of the one coupled with massless fields. Surprisingly, this time delay can be counteracted with large acceleration, but for a static system in a thermal bath it can't be affected by temperature.
 Degradation here is caused by dissipation and decoherence.

 Now as mentioned earlier, for two detectors on opposite Rindler wedge, we observe entanglement harvesting. Now, given the possibility of entanglement harvesting, it seems interesting to study how the initially existing entanglement between two detectors on opposite Rindler wedge (anti-parallelly accelerating) will evolve:
 will there be entanglement harvesting or degradation of the initial entanglement or both depending on the set of parameters? To the best of our knowledge, the previous works investigate the process of entanglement degradation for both the detectors on the same wedge.
 Also, practically as the universe has a finite temperature, studying the dynamics of entanglement where the field has a background temperature should give us results closer to the real environment. With this motivation, we analyze the variation of entanglement dynamics with the temperature of the background field? As seen in \cite{Barman:2021bbw}, the presence of thermal fields gives rise to several interesting features in the entanglement harvesting phenomenon, so it might be interesting to look at how the presence of a thermal field will affect the dynamics of entanglement as well. To the best of our knowledge, no such study exists in present literature.

Thus, in this work we investigate the dynamics of entanglement between two initially entangled UD detectors for two cases, accelerating parallelly and anti-parallelly, interacting with the massless thermal scalar field in (1+1)-dimension and (1+3)-dimension with the help of the entanglement measure called Negativity. We obtain the time-evolved density matrix of the system of UD detectors, following the approach given in \cite{koga2018quantum}, and compute the eigenvalues of the partially transposed matrix.
To obtain the eigenvalues we compute the Greens functions following \cite{Barman:2021oum} and compute certain integrals as in \cite{Barman:2021bbw}. 
We finally analyze Negativity by numerical analysis of upto $\mathcal{O}(C^2)$, where $C$ is the coupling parameter between the detector and the field. 
\begin{itemize}
	\item We obtain that for both non-thermal and thermal field in (1+1)-dimension and (1+3)-dimension, for the parallel motion of the detectors, there is always entanglement degradation, though for (1+3)-dimension, there is some undulation in the negativity with change in its acceleration. 
 \item We obtain that for both non-thermal and thermal field in (1+1)-dimension, for the anti-parallel motion of the detectors, there is entanglement degradation or entanglement harvesting depending on the range of parameters. Strikingly, for (1+3)-dimension, there is some undulation in the negativity, with change in its acceleration, but there is always entanglement degradation. No entanglement harvesting is observed. The rate of entanglement degradation varies on the set of parameters. It is to be noted that this entanglement harvesting is possible only when the initial detectors are non-maximally entangled. In the case of initially maximally entangled anti-parallelly accelerating detectors, there is always entanglement degradation.
\end{itemize}

In the present analysis is technically different from the earlier investigations \cite{lin2008disentanglement} -- \cite{zhou2021entanglement}. There the Wightman functions were considered using the expansion of the field in terms of Minkowski modes. It has been noticed in several places (e.g. see \cite{Chowdhury:2019set,Barman:2021oum}) that such quantity is not time translational invariant when one finds the corresponding thermal Wightman function in Rindler spacetime. This inherently introduces the non-equilibrium situation in the system. It implies that the Minkowski modes sometimes do not comply with the equilibrium situation of the system. To avoid such a scenario here we adopt the Wightman functions corresponding to Rindler modes with vacuum is taken to be that for Unruh modes, which is incidentally Minkowski here. In this case the required functions are time translational invariant with respect to Rindler proper time \cite{Barman:2021oum,Barman:2021bbw,Kane:2021rhg}. Therefore all our observations of this study are strictly valid for later modes.

In section \ref{model}, we introduce our model setup of a pair of two-level Unruh-DeWitt detectors. The detectors are initially entangled and interact with the intermediate real massless scalar field. We construct the time evolved density matrix and also obtain the eigenvalues of the partially-transposed matrix. We derive the Negativity expression that we shall subsequently analyze to comment on the dynamics of entanglement. We also introduce a set of integrals that needs to be further solved for obtaining the value of Negativity. In section \ref{greens-func intro}, we summarize the Greens functions used to calculate the integrals.
In sections \ref{parallel} and \ref{anti-parallel} , the results after calculating the integrals for parallelly and anti-parallelly accelerating detectors respectively have been presented. 
Finally, in section \ref{numerical analysis}, we analyze the dynamics of entanglement for different cases and conclude with the discussions on the results.

	\section{Model and basic setup}\label{model}
	
		We consider a pair of  two-level Unruh-DeWitt detectors, with discrete energy levels $E^j_n$, where $j \in (A,B)$ (unless specified otherwise) and $n \in (0,1)$, carried by A (Alice) and B (Bob), along trajectories given by $x^a(\tau^j)$. Initially the detectors are entangled and the composite state we denote as $\alpha \ket{E_0^A E_0^B}+ \gamma \ket{E_1^A E_1^B}$. Here coefficients $\alpha$ and $\gamma$, taken to be real, satisfy $\alpha^2+\gamma^2=1$, and for initial state to be entangled we must have $\alpha \neq 1$ (or $\gamma\neq 0$) and $\alpha\neq 0$ (or $\gamma\neq 1$). Consider that they are interacting with background field and for simplicity the field is taken to be {\it massless real scalar}. If the field initially is in vacuum state $\ket{0}$ (Minkowski vacuum), then the initial state of the field plus the two detectors system is given by 
		\begin{equation}
			\ket{in} = \alpha \ket{0; E_0^A E_0^B}+ \gamma \ket{0; E_1^A E_1^B}~,
			\label{V26}
		\end{equation}
		where $\ket{0; E_n^A E_m^B}\equiv \ket{0}\otimes \ket{E_n^A} \otimes \ket{E_m^B}$.
		The detectors individually interact with the field through the monopole coupling of the form:
		\begin{eqnarray}
			S_{int} &=& \int C_A \chi_A(\tau_A)m_A(\tau_A)\phi_A(x_A)d\tau_A 
			\nonumber
			\\
			&& + \int C_B \chi_B(\tau_B)m_B(\tau_B)\phi_B(x_B)d\tau_B~, 
			\label{V24}
		\end{eqnarray}
		where $C_j$ is the coupling constant, $m_j(\tau_j)$ is the monopole operator representing the detector and $\chi_j(\tau_j)$ is the switching function which determines the duration of interaction between the field and the detectors. The same has been chosen in different occasions (e.g. see \cite{koga2018quantum}). Here $\tau_j$ is the $j^{th}$ detector's proper time. 
In the interaction picture we have
\begin{equation}
					m_j(\tau)=e^{iH_{0}\tau_j}m_j(0)e^{-iH_0\tau_j}~,
					\label{B2}
				\end{equation}
with $H_0$ is the unperturbed Hamiltonian for the $j^{th}$ detector.	For our later purpose we choose
\begin{equation}
m_j(0) = \ket{E^j_0}\bra{E^j_1}	+ \ket{E^j_1}\bra{E^j_0}~.
\label{R3}
\end{equation}
The later time state is then determined as
		\begin{equation}
			\ket{out} = T e^{iS_{int}} \ket{in}~,
			\label{out}
		\end{equation}
where ``$T$'' refers to the time order product.

		Now since we are interested to detectors only, the reduced density matrix for our analysis is determined from Eq. \eqref{out} with taking trace over the field degrees of freedom. The initial (i.e. at $(t=0)$) and the later time (i.e. at $t$) density matrices are then given by
		\begin{eqnarray}
			&&\rho_{AB} (t=0)= \mathrm{Tr_{\phi}} \left[ \ket{in} \bra{in} \right]~;
			\nonumber
			\\
			&&\rho_{AB} (t) = \mathrm{Tr_{\phi}} \left[ \ket{out} \bra{out} \right]~.
			\label{RAB}
		\end{eqnarray}
These, in matrix notation (in the basis $\{\ket{E_1^AE_1^B}, \ket{E_1^AE_0^B}, \ket{E_0^AE_1^B}, \ket{E_0^AE_0^B}\}$), are as follows
		\begin{equation}
			\rho(t=0) = \begin{pmatrix}
				|\gamma|^2 & 0 & 0 & \alpha^* \gamma \\
				0 & 0 & 0 & 0 \\
				0 & 0 & 0& 0 \\
				\alpha \gamma^* & 0 & 0 & |\alpha|^2
			\end{pmatrix}~,
			\label{B4}
		\end{equation}
		and 
		\begin{equation}
			\rho_{AB} (t) = \begin{pmatrix}
				a_1 & 0 & 0 & a_2 \\
				0 & b_1 & b_2 & 0 \\
				0 & c_1 & c_2 & 0 \\
				d_1 & 0 & 0 & d_2
			\end{pmatrix}~.
			\label{gen-mat}
		\end{equation}

		\begin{widetext}
			The matrix elements in the latter one are calculated till second order in the perturbation series. They are given by
			\begin{eqnarray} 
				&&a_1 =	\rho_{1111}= \gamma \gamma^* - C_A^2\gamma \gamma^* P_A^{\prime \prime} -  C_B^2\gamma \gamma^* P_B^{\prime \prime} -  C_A C_B\alpha^*\gamma \zeta^*_{AB}
				- C_A C_B\alpha\gamma^* \zeta_{AB}~,
				\nonumber 
				\\
				&&a_2=\rho_{1100}=\gamma \alpha^* -  C_A C_B\alpha \alpha^* \zeta_{AB}- C_A C_B\gamma \gamma^* Y^*_{AB} -  C_A^2\alpha ^* \gamma M_A 
				-   C_B^2\alpha^*\gamma { M_B}~,
				\nonumber 
				\\
				&&b_1=\rho_{1010}= C_A^2\alpha \alpha^*  P_A + C_B^2\gamma \gamma^*  P_B^{\prime \prime} + C_A C_B \alpha^*\gamma  P^{\prime}_{AB}
				+ C_A C_B\alpha\gamma^*  P^{\prime *}_{AB}~, 
				\nonumber 
				\\
				&&b_2=\rho_{1001}=C_A^2\alpha \gamma^* \bar P_A^\prime + C_A C_B\alpha^* \alpha P_{AB}+ C_A C_B\gamma \gamma^* X^*_{AB}+ C_B^2\gamma \alpha^*\bar P_B~, 
				\nonumber 
				\\
				&&c_1=\rho_{0110}=C_A^2\alpha^*\gamma \bar P_A + C_B^2\alpha \gamma^* \bar P_B^\prime + C_A C_B\gamma \gamma^* X_{AB} + C_A C_B\alpha \alpha^* P_{AB}^*~, 
				\nonumber 
				\\
				&&c_2=\rho_{0101}= C_A^2\gamma^*\gamma P_A^{\prime \prime} + C_B^2\alpha \alpha^* P_B +C_A C_B\gamma \alpha^*\bar P^{\prime}_{AB}
				+C_A C_B\alpha \gamma^*\bar P^{\prime *}_{AB}~,
				\nonumber 
				\\
				&&d_1=\rho_{0011}=\alpha \gamma^* -C_A^2\gamma^*\alpha M_A^*- C_B^2\alpha \gamma^* M_B^*
				-C_A C_B\gamma \gamma^* Y_{AB}- C_A C_B\alpha \alpha^*\zeta^*_{AB}~,
				\nonumber 
				\\
				&&d_2=\rho_{0000}=\alpha \alpha^*-C_A^2\alpha \alpha^*P_A- C_B^2\alpha \alpha^*P_B- C_A C_B\alpha^*\gamma Y_{AB}
				-C_A C_B\alpha \gamma^* Y_{AB}^*~,
				\label{eqn_dyna_elements}
			\end{eqnarray}
			where
			\begin{eqnarray}
				&&P_j = \left|\bra{E_0^j}m_j(0)\ket{E_1^j}\right|^2 \int \int d\tau_j d\tau^\prime_j \chi(\tau_j)\chi(\tau^\prime_j)e^{-i\Delta \tau_j \Delta E_j}G_w(x^\prime_j,x_j)~,
				\nonumber 
				\\
				&&P_j^{\prime \prime} = \left|\bra{E_0^j}m_j(0)\ket{E_1^j}\right|^2 \int \int d\tau_j d\tau^\prime_j \chi(\tau_j)\chi(\tau^\prime_j)e^{-i\Delta \tau_j \Delta E_j}G_w(x_j,x^\prime_j)~, 
				\nonumber 
				\\
				&&\bar P_j = \bra{E_0^j}m_j(0)\ket{E_1^j}\bra{E_0^j}m_j(0)\ket{E_1^j}\int \int d\tau_j d\tau^\prime_j \chi(\tau_j)\chi(\tau^\prime_j)e^{-i \bar \tau_j \Delta E_j}G_w(x^\prime_j,x_j)~, 
				\nonumber 
				\\
				&&\bar P_j^\prime = \bra{E_1^j}m_j(0)\ket{E_0^j}\bra{E_1^j}m_j(0)\ket{E_0^j}\int \int d\tau_j d\tau^\prime_j \chi(\tau_j)\chi(\tau^\prime_j)e^{i \bar \tau_j \Delta E_j}G_w(x^\prime_j,x_j)~,
				\nonumber 
				\\
				&&M_j = \left|\bra{E_0^j}m_j(0)\ket{E_1^j}\right|^2 \int \int d\tau_j d\tau^\prime_j \chi(\tau_j)\chi(\tau^\prime_j)e^{i\tau_j \Delta E_j} e^{-i\tau_j^\prime \Delta E_j}\Theta(\tau-\tau^\prime_j)\left\{G_w(x^\prime_j,x_j)+G_w(x_j,x^\prime_j)\right\}~,
				\nonumber 
				\\
				&&P^{\prime}_{AB} = \bra{E_0^A}m_A\ket{E_1^A}\bra{E_0^B}m_B\ket{E_1^B}\int \int d\tau_A d\tau^\prime_B \chi(\tau_A) \chi(\tau^\prime_B) e^{-i \tau_A \Delta E_A}  e^{-i \tau^\prime_B \Delta E_B}G_w(x_A,x^\prime_B)~, 
				\nonumber 
				\\
				&&\bar P^{\prime}_{AB} = \bra{E_0^A}m_A(0)\ket{E_1^A}\bra{E_0^B}m_B(0)\ket{E_1^B}\int \int d\tau_A d\tau^\prime_B \chi(\tau_A)\chi(\tau^\prime_B)e^{-i \tau_A \Delta E_A} e^{-i \tau^\prime_B \Delta E_B}G_w(x^\prime_B,x_A)~, 
				\nonumber 
				\\
				&&X_{AB} = \bra{E_1^B}m_B(0)\ket{E_0^B}\bra{E_0^A}m_A(0)\ket{E_1^A}\int \int d\tau_A d\tau^\prime_B \chi(\tau_A)\chi(\tau^\prime_B)e^{-i \tau_A \Delta E_A} e^{i \tau^\prime_B \Delta E_B}G_w(x^\prime_B,x_A)~,
				\nonumber 
				\\
				&&P^*_{AB} = \bra{E_1^B}m_B(0)\ket{E_0^B}\bra{E_0^A}m_A(0)\ket{E_1^A}\int \int d\tau_A d\tau^\prime_B \chi(\tau_A)\chi(\tau^\prime_B) e^{-i \tau_A \Delta E_A}e^{i \tau^\prime_B \Delta E_B}G_w(x_A,x^\prime_B)~,
				\nonumber 
				\\
				&&Y_{AB} = \bra{E_0^B}m_B(0)\ket{E_1^B}\bra{E_0^A}m_A(0)\ket{E_1^A}\int \int d\tau_A d\tau^\prime_B \chi(\tau_A)\chi(\tau^\prime_B)e^{-i \tau_A \Delta E_A} e^{-i \tau^\prime_B \Delta E_B}\left\{iG_F(x_A,x^\prime_B)\right\}~,
				\nonumber 
				\\
				&&\zeta^*_{AB} =\bra{E_0^B}m_B(0)\ket{E_1^B}\bra{E_0^A}m_A(0)\ket{E_1^A}\int \int d\tau_A d\tau^\prime_B \chi(\tau_A)\chi(\tau^\prime_B)e^{-i \tau_A \Delta E_A} e^{-i \tau^\prime_B \Delta E_B}\left\{- iG_F^*(x_A,x^\prime_B)\right\}~. 
				\label{eqn_dyna_integrals}
			\end{eqnarray}
In the above we denoted $\Delta \tau = \tau^\prime-\tau$, $\bar{\tau} = \tau^\prime+\tau$ and $\Delta E_{j} = E_1^{j} - E_0^{j}$. For our choice of $m_j(0)$, given by (\ref{R3}), we have $\bra{E^j_n}m_j(0)\ket{E_n^j}=0 $, where $n\in 0,1$ and $j\in A,B$. We refer to Appendix \ref{App1} for a detailed calculation of the above results.
		\end{widetext}

		From now on, we consider $\alpha = \alpha^*$ and $\gamma = \gamma^*$  as they have been taken as real. Further for simplicity we choose $C_A = C_B = C $.
In order to investigate the fate of the initial entanglement with the progress of time, we adopt the prescription outlined by Peres \cite{peres1996separability,horodecki1996necessary}. According to the Peres partial transpose (PPT) criteria, for a $2 \cross 2$ system, if any one of the eigenvalues of the partially transposed density matrix is negative, it implies that the state is no longer separable, and hence, is an entangled state. Here, in our analysis, we use the measure Negativity, given by the sum of the negative eigenvalues of the partially transposed density matrix. Now, as the negative eigenvalues can be obtained by applying the PPT criteria, we use it for our further analysis. 
		The partial transposes of Eq. \eqref{B4} and Eq. \eqref{gen-mat} are 
		\begin{equation}
			\rho^{T_{A}}(t=0) = \begin{pmatrix}
				\gamma^2  & 0 & 0 & 0 \\
				0 & 0 &  \alpha \gamma & 0 \\
				0 & \alpha \gamma & 0& 0 \\
				0 & 0 & 0 & \alpha^2
			\end{pmatrix}~,
			\label{B5}
		\end{equation}
		and 
		\begin{equation}
			\rho^{T_{A}}_{AB} (t) = \begin{pmatrix}
				a_1 & 0 & 0 & b_2 \\
				0 & b_1 & a_2 & 0 \\
				0 & d_1 & c_2 & 0 \\
				c_1 & 0 & 0 & d_2
			\end{pmatrix}~,
			\label{B6}
		\end{equation}
		respectively.
		The eigenvalues of matrix Eq. \eqref{B5} are $\{\lambda_1'=\alpha \gamma,\lambda_2'= -\alpha \gamma,\lambda_3' =\gamma^2,\lambda_4' = \alpha^2 \}$, while those of  Eq. \eqref{B6} are
		\begin{eqnarray} 
			&&\lambda_1 = \frac{1}{2}\left\{b_1+c_2+\sqrt{(b_1-c_2)^2+4a_2d_1}\right\}~,
			\label{L1}
			\\
			&&\lambda_2 = \frac{1}{2}\left\{b_1+c_2-\sqrt{(b_1-c_2)^2+4a_2d_1}\right\}~,
			\label{L2}
			\\
			&&\lambda_3 = \frac{1}{2}\left\{a_1+d_2+\sqrt{(a_1-d_2)^2+4b_2c_1}\right\}~,
			\label{L3}
			\\
			&&\lambda_4 = \frac{1}{2}\left\{a_1+d_2-\sqrt{(a_1-d_2)^2+4b_2c_1}\right\}~.
			\label{L4}
		\end{eqnarray}
It may be noted that among all the eigenvalues of matrix \eqref{B5}, depending on the signs of $\alpha$ and $\gamma$, either $\lambda_1'$ or $\lambda_2'$ must be negative. It implies that the initial state was taken to be entangled one. One can also check that the above eigenvalues are indeed real (see Appendix \ref{AppJHEP1}).

		In order to study the fate of entanglement due the detectors' motion and field temperature, we will analyze the nature of one of the quantities which quantifies the existing entanglement in the system. 
		In this study \textit{negativity} is taken to be an entity for investigation to quantify entanglement dynamics of our system. This is defined as \cite{vidal2002computable,zyczkowski1998volume}
		\begin{eqnarray}
			\mathcal{N}=\sum_{i} \left|\lambda_i^{-}\right|~,
			\label{Ne}
		\end{eqnarray}
		where $\lambda_i^{-}$  are the negative eigenvalues of the partially transposed density matrix.
		Therefore, in order to investigate the fate of this entanglement in the final state, now we need to investigate the eigenvalues given by Eq. \eqref{L1} - \eqref{L4}.

		\subsection{About later time eigenvalues} \label{about_late_eigen}
			The eigenvalues Eq. \eqref{L1} - \eqref{L4} has been calculated by using perturbation analysis and are valid till second order in $C$. Under this circumstances, it is now customary to analyze them till $\mathcal{O}(C^2)$ in order to validate our conclusion.

			First we look into eigenvalues $\lambda_1$ and $\lambda_2$. Up to order $C^2$ has been computed in  Appendix \ref{AppJHEP1} (see Eq. \eqref{gen-L12-R-derive}). These are given by 
\begin{equation}
\lambda_{1,2} \approx {\bar{A}}_r + A_c + A_c^*~,
\end{equation} 			
where we denote
\begin{eqnarray}
				\bar{A}_{r} &=& \Big[\frac{C^2}{2} \big\{\alpha^2 ( P_A + P_B) + \gamma^2 (P_B^{\prime \prime} + P_A^{\prime \prime})\big\} \pm  (\gamma \alpha)\Big]~; \nonumber \\ 
				\nonumber \\
				A_c &=& \frac{C^2}{2}\Big[\alpha \gamma (P^{\prime}_{AB} + \bar P^{\prime}_{AB}) \nonumber \\ && \mp (\alpha^2 \zeta_{AB}+\gamma^2 Y_{AB} + \alpha \gamma M_A +  \alpha \gamma M_B) \Big]~.
			\end{eqnarray}
In the above $\bar{A}_{r}$	 is a real quantity, while individual term of $A_c$ can be complex.	But since in the eigenvalues we have $A_c + A_c^*$, the real part of $A_c$ will contribute and hence, as expected, these eigenvalues are real. Then we write the eigenvalues as		
\begin{align}
  \lambda_{1,2} &\approx {\bar{A}}_r + A_c + A_c^*   
   \nonumber \\
   & \approx \mathfrak{R}\Big[\frac{C^2}{2} \{\sum_{j}(\alpha^2  P_j + \gamma^2 P_j^{\prime \prime})  + 2\alpha \gamma  ({ {P^{\prime}_{AB}}} + {{\bar P^{\prime}_{AB}}})\}
   \nonumber \\
                 &\pm [(\gamma \alpha) - C^2\{\alpha^2 \zeta_{AB}  + \gamma^2  Y_{AB} + \alpha \gamma   z \}]\Big]
 \label{gen-L12}
\end{align}
where $z= M_A + M_B $. In the above $\mathfrak{R}$ denotes the real part.
Now since only the real parts of the individual quantities are relevant for determining the eigenvalues from now on we will concentrate only on the real parts of the relevant expressions. Keeping this in mind in the rest of the manuscript, for the notational simplification, we will omit the explicit mention of notation $\mathfrak{R}$. But remember that only the real part is relevant in the later discussion.
Therefore, we now have
			\begin{eqnarray}
				\lambda_1 &\approx& \alpha \gamma + C^2 \left\{\alpha \gamma \left(P^{\prime}_{AB} + \bar P^{\prime}_{AB} - z\right) - \alpha^2 \zeta_{AB} - \gamma^2 Y_{AB} \right\}
				\nonumber \\
				&& + \frac{C^2}{2} \left\{\sum_j \left(\alpha^2 P_j + \gamma^2 P_j^{\prime \prime} \right) \right\}~,
				\label{e1}
			\end{eqnarray}
			and
			\begin{eqnarray}
				\lambda_2 &\approx& - \alpha \gamma + C^2 \left\{\alpha \gamma \left(P^{\prime}_{AB} + \bar P^{\prime}_{AB} + z\right) + \alpha^2 \zeta_{AB} + \gamma^2 Y_{AB} \right\}
				\nonumber  \\
				&& + \frac{C^2}{2} \left\{\sum_j \left(\alpha^2 P_j + \gamma^2 P_j^{\prime \prime} \right) \right\}~,
				\label{e2}
			\end{eqnarray}
where only real parts of all the terms are to be included.			

			Similarly the other two eigenvalues $\lambda_3$ and $\lambda_4$, as obtained in \eqref{gen-L34-R-derive} (in Appendix \ref{AppJHEP1}) takes the following forms
			\begin{eqnarray}
					\lambda_3 \approx \gamma^2 + \mathcal{O}(C^2)~,
				\label{e3}
			\end{eqnarray}
			and
			\begin{eqnarray}
				\lambda_4 \approx \alpha^2 + \mathcal{O}(C^2)~,
				\label{e4}
			\end{eqnarray}
			respectively.
			Now note that in Eq. \eqref{e1} - \eqref{e4} the first terms correspond to initial state. These are now modified by $\mathcal{O}(C^2)$ terms and such terms are evaluated by perturbation technique. Therefore in order to validity of the perturbation technique we must have the restriction on the coupling constant $C$ such  that $\mathcal{O}(C^0)$ term must be very very large compared to $\mathcal{O}(C^2)$ term in each eigenvalue.  Therefore for any values of $\alpha$ and $\gamma$ (with restriction $\alpha^2+\gamma^2=1$), $\lambda_3$ and $\lambda_4$ are always positive. Since the values are always positive we are not interested to deal with the explicit expression in the $O(C^0)$ term. For the same reason, depending on the values of $\alpha$ and $\gamma$, either $\lambda_1$ or $\lambda_2$ is negative.

			Here we choose both $\alpha$ and $\gamma$ are positive (or negative). In this situation again $\lambda_1$ in Eq. \eqref{e1} will be positive as the leading order term is always positive. 
			Thus the only negative eigenvalue is $\lambda_2$. So now onwards we shall concentrate on this particular eigenvalue to investigate the negativity Eq. \eqref{Ne}. 

		\subsection{Quantifying entanglement}\label{quantifying entanglement}
			From the above discussion, for our choice of $\alpha$ and $\gamma$, it is observed that the only negative eigenvalue of the matrix Eq. \eqref{B6} is $\lambda_2$. Thus the negativity of the time evolved system is
			\begin{eqnarray}
				\mathcal{N}(t) &=& \Big| - \alpha \gamma 
				+ C^2 \Big\{\alpha \gamma \left(P^{\prime}_{AB} + \bar P^{\prime}_{AB} + z\right)
				\nonumber
				\\ 
				&&+ \alpha^2 \zeta_{AB} + \gamma^2 Y_{AB} \Big\}
				\nonumber  \\
				&& + \frac{C^2}{2} \left\{\sum_j \left(\alpha^2 P_j + \gamma^2 P_j^{\prime \prime} \right) \right\} \Big|~.
				\label{N1}
			\end{eqnarray}
			The same for initial state (i.e. corresponding to Eq. \eqref{B5}) is given by 
			\begin{eqnarray}
				\mathcal{N}(0) =	\Big|-\alpha \gamma\Big|~. 
				\label{N0}
			\end{eqnarray} 
			Now to study the entanglement dynamics of our system we will analyze how the negativity changes with respect to our choices of parameters. 
			The part which is $\mathcal{O}(C^2)$ in Eq. \eqref{N1}, provides the change in negativity of the system in later time. We denote this term as $\delta \mathcal{N} $ which is here given by
			\begin{eqnarray} 
				&&\delta \mathcal{N} = \left\{\alpha \gamma \left(P^{\prime}_{AB} + \bar P^{\prime}_{AB} + z\right) + \alpha^2 \zeta_{AB} + \gamma^2 Y_{AB} \right\}
				\nonumber  \\
				&& + \frac{1}{2} \left\{\sum_j \left(\alpha^2 P_j + \gamma^2 P_j^{\prime \prime} \right) \right\}~.
				\label{negativity}
			\end{eqnarray}
			Thus the total negativity of our system upto $\mathcal{O}(C^2)$ takes the following form:
			\begin{eqnarray}
				\mathcal{N}(t) = \Big|-\alpha \gamma + C^2\delta \mathcal{N}\Big|~.
				\label{deltaN}
			\end{eqnarray}
			The the above expression signifies that the positivity of $\delta \mathcal{N}$ leads to decrease in negativity and hence entanglement decreases. While if $\delta \mathcal{N}$ is negative, then the entanglement will increase.

			Following the above argument we will now concentrate on $\delta \mathcal{N}$ in order to understand the fate of initial entanglement when the detectors are interacting with background fields and they are in motion.  Before proceeding further, let us point out that the exact numerical value of $\delta \mathcal{N}$ is not important for the present purpose (at least for qualitative investigation of the fate of entanglement). We need to only check the change in $\delta \mathcal{N}$ with respect to our choice of parameters. If it increases (decreases) in the positive side, then according to Eq. \eqref{deltaN}, we will have decrease (increase) in entanglement. On the other hand increase (decrease) of $\delta \mathcal{N}$ in the negative side implies increase (decrease) in entanglement. Here we want to investigate the effect of accelerations of the detectors and the temperature of the scalar fields. Note that $\delta \mathcal{N}$ consists to different terms which contain two point correlation functions like positive frequency Wightman function and Feynman propagator. Therefore below we will first give the expressions of them which will be needed for our main purpose. 

	\section{Wightman functions: summary of previous results}\label{greens-func intro}
		Within this model we want to investigate two situations. In one case both the detectors are in the right Rindler wedge (RRW) and in other case the observers' detector is in RRW while second detector is in the left Rindler wedge (LRW). The corresponding positive frequency Wightman functions are evaluated in \cite{Barman:2021oum,Barman:2021bbw}. They are evaluated by decomposing the scalar fields in Rindler modes and choosing the vacuum as that of Unruh modes (which is incidentally the Minkowski vacuum; see \cite{Book1} for details). Such choice has several advantages which have been mentioned in several places, like in \cite{Barman:2021oum,Barman:2021bbw}. One noticeable feature of such Wightman functions is  that they are inherently time translational invariant and therefore mimic the thermal equilibrium situation.  Therefore we will consider the same here as well.

		\subsection{Coordinate systems}
			The coordinates of a uniformly accelerated object is confined to specific regions in Minkowski spacetime. 
			These specific regions form the Rindler wedges. 
			Let $(x,t,y,z)$ be the coordinates assigned by inertial observer, then the line element for the  Minkowski spacetime will be
			\begin{eqnarray}
				&& ds_M^2 = -dt^2 + dx^2 + dy^2 + dz^2~.   
			\end{eqnarray}
			Let $(\zeta,\eta,y,z)$ be the coordinates assigned by an observer who is uniformly accelerating along $x$. 
			The transformation between these two coordinate systems $(x,t)$  and $(\zeta,\eta)$  is given by 
			\begin{eqnarray}
				&& x = \frac{1}{a}e^{a\zeta}\cosh(a\eta), \nonumber \\
				&& t = \frac{1}{a}e^{a\zeta}\sinh(a\eta)~.
				\label{transformation-rindler}
			\end{eqnarray}
			The other two coordinates $(y,z)$ remains unchanged by Rindler transformation. For $(1+1)$ dimensions, these transverse coordinates can be ignored.
			From Eq. \eqref{transformation-rindler} we see, that for all values of $(\zeta,\eta)$, all values of $(x,t)$ are not accessible, i.e for $\zeta\in (-\infty,\infty)$, and $\eta\in (-\infty,\infty)$, the domain of (x,t) reduces to, $x\geq 0$ and $\left|x\right| \geq \left|t\right|$. 
			Thus for the Rindler observer following the trajectory Eq. \eqref{transformation-rindler}, only a quadrant of minkowski spacetime can be accessible. The observer following Eq. \eqref{transformation-rindler} can access the region of Minkowski spacetime which is known as the RRW.
			Similarly a transformation corresponding to the region $x\leq 0$ and  $\left|x\right| \geq \left|t\right|$ , i.e. LRW can be defined as 
			\begin{eqnarray}
				&& x = -\frac{1}{a}e^{a\zeta}\cosh(a\eta)~, \nonumber \\
				&& t = \frac{1}{a}e^{a\zeta}\sinh(a\eta)~. \label{transformation-rindler-L}
			\end{eqnarray}
			Both Eq. \eqref{transformation-rindler} and Eq. \eqref{transformation-rindler-L} gives: 
			\begin{eqnarray}
				&& x^2-t^2=\frac{1}{a^2}e^{2a\zeta}~,
				\label{trajectory}
			\end{eqnarray}
			and the line element corresponding to the right and left Rindler wedges is
			\begin{eqnarray} 
				&& ds_R^2 = e^{2a\zeta} (dy^2 + dz^2 + d\zeta^2 -d\eta^2)~.
				\label{rindler-metric}
			\end{eqnarray}

			In RRW  proper time is given by  $\tau=e^{a\zeta}\eta$
			and proper acceleration is given by $b=a e^{a\zeta}$. 
			The coordinate relations in terms of proper acceleration and proper time are
			\begin{eqnarray}
				&& x = \frac{1}{b}\cosh(b\tau)~, \nonumber \\
				&& t = \frac{1}{b}\sinh(b\tau)~. 
			\end{eqnarray}
			Whereas in LWR   proper time is given by  $\tau=-e^{a\zeta}\eta$
			and proper acceleration is given by $b=a e^{a\zeta}$. Here the relations between the coordinates in terms of proper acceleration and proper time are 
			\begin{eqnarray}
				&& x = -\frac{1}{b}\cosh(b\tau)~, \nonumber \\
				&& t = \frac{1}{b}\sinh(b\tau)~.
			\end{eqnarray}	
		
		\subsection{Thermal Wightman functions}	
			\label{greens-exp}
			All the required correlation function were calculated in \cite{Barman:2021oum,Barman:2021bbw}. We below summarize the final forms. 
			\begin{widetext}
				In $(1+1)$ dimensions, when both detectors are uniformly accelerating in the RRW, the Wightman function becomes 
				\begin{eqnarray} 
					G_R^{+\beta}(x_B^\prime,x_A)&=&\int_0^\infty \frac{dw}{4 \pi w}\frac{1}{\sqrt{\sinh \left( \frac{\pi w}{a_A}\right) \sinh \left(\frac{\pi w}{a_B}\right)}}
					\Big[\frac{e^{\frac{\pi}{2}w\left(\frac{1}{a_A}+\frac{1}{a_B}\right)} e^{ik\Delta \zeta-iw\Delta\eta}+e^{-\frac{\pi}{2}w\left(\frac{1}{a_A}+\frac{1}{a_B}\right)}e^{ik\Delta \zeta+ iw\Delta\eta}}{1-e^{-\beta w}} \nonumber \\
					&&+\frac{e^{\frac{\pi}{2}w\left(\frac{1}{a_A}+\frac{1}{a_B}\right)}e^{-ik\Delta \zeta + iw\Delta\eta}+e^{-\frac{\pi}{2}w\left(\frac{1}{a_A}+\frac{1}{a_B}\right)}e^{-ik\Delta \zeta -iw\Delta\eta}}{e^{\beta w}-1}\Big]~,
					\label{V21}
				\end{eqnarray} 
				where $a_A$ and $a_B$ are acceleration parameters of the two detectors.
				The same in $(1+3)$ dimensions turns out to be
				\begin{eqnarray}
					G^{+\beta}_{{3D_R}}(x_B^\prime,x_A)&=&\int_{0}^{\infty} dw \frac{2}{\sqrt{a_A a_B}}\Big[\frac{e^{-iw\Delta \eta }e^{\frac{\pi w}{2}\left(\frac{1}{a_A}+\frac{1}{a_B}\right)}+e^{iw\Delta \eta }e^{\frac{-\pi w}{2}\left(\frac{1}{a_A}+\frac{1}{a_B}\right)}}{1-e^{-\beta w}}
					\nonumber
					\\
					&&+	\frac{e^{iw\Delta \eta }e^{\frac{\pi w}{2}\left(\frac{1}{a_A}+\frac{1}{a_B}\right)}+e^{-iw\Delta \eta }e^{\frac{-\pi w}{2}\left(\frac{1}{a_A}+\frac{1}{a_B}\right)}}{1-e^{-\beta w}}\Big] \nonumber \\ 
					&&\times\int_{-\infty}^{\infty} \frac{d^2k_p}{{(2 \pi)}^{4}}\mathcal{K}\left[\frac{iw}{a_A},\frac{\left|k_p\right|e^{a_A\zeta_A}}{a_A}\right] \mathcal{K}\left[\frac{iw}{a_B},\frac{\left|k_p\right|e^{a_B\zeta_B}}{a_B}\right]~, 
					\label{BRMP1}
				\end{eqnarray}
				where $\mathcal{K}\left[\frac{iw}{a_j},\frac{\left|k_p\right|e^{a_j\zeta_j}}{a_j}\right]$ denotes the modified Bessel function of second order.

				When both the detectors are in LRW, then the Wightman functions are
				\begin{eqnarray} 
					G_L^{+\beta}(x_B^\prime,x_A)&=&\int_0^\infty \frac{dw}{4 \pi w}\frac{1}{\sqrt{\sinh \left(\frac{\pi w}{a_A}\right) \sinh \left(\frac{\pi w}{a_B}\right)}}
					\Big[\frac{e^{-\frac{\pi}{2}w\left(\frac{1}{a_A}+\frac{1}{a_B}\right)}e^{ik\Delta \zeta-iw\Delta\eta}+e^{\frac{\pi}{2}w\left(\frac{1}{a_A}+\frac{1}{a_B}\right)}e^{ik\Delta \zeta+ iw\Delta\eta}}{1-e^{-\beta w}} \nonumber \\
					&&+\frac{e^{-\frac{\pi}{2}w\left(\frac{1}{a_A}+\frac{1}{a_B}\right)}e^{-ik\Delta \zeta + iw\Delta\eta}+e^{\frac{\pi}{2}w\left(\frac{1}{a_A}+\frac{1}{a_B}\right)}e^{-ik\Delta \zeta -iw\Delta\eta}}{e^{\beta w}-1}\Big]~,
					\label{V22}
				\end{eqnarray} 
				and
				\begin{eqnarray}
					G^{+\beta}_{{3D_L}}(x_B^\prime,x_A)&=&\int_{0}^{\infty} dw \frac{2}{\sqrt{a_A a_B}}\Big[\frac{e^{iw\Delta \eta }e^{\frac{\pi w}{2}\left(\frac{1}{a_A}+\frac{1}{a_B}\right)}+e^{-iw\Delta \eta }e^{\frac{-\pi w}{2}\left(\frac{1}{a_A}+\frac{1}{a_B}\right)}}{1-e^{-\beta w}}
					\nonumber
					\\
					&& + \frac{e^{-iw\Delta \eta }e^{\frac{\pi w}{2}\left(\frac{1}{a_A}+\frac{1}{a_B}\right)}+e^{iw\Delta \eta }e^{\frac{-\pi w}{2}\left(\frac{1}{a_A}+\frac{1}{a_B}\right)}}{1-e^{-\beta w}}\Big]
					\nonumber \\ 
					&&\times\int_{-\infty}^{\infty} \frac{d^2k_p}{{(2 \pi)}^{4}}\mathcal{K}\left[\frac{iw}{a_A},\frac{\left|k_p\right|e^{a_A\zeta_A}}{a_A}\right] \mathcal{K}\left[\frac{iw}{a_B},\frac{\left|k_p\right|e^{a_B\zeta_B}}{a_B}\right]~, 
					\label{V23}
				\end{eqnarray}
				for $(1+1)$ and $(1+3)$ dimensions, respectively.

				In the case when one detector is in the RRW (namely detector $A$) and other one is in LRW (namely detector $B$), then the correlation functions are given by
				\begin{eqnarray}
					G^{+ \beta}_{LR}(x_B^\prime,x_A)&=&\int_0^\infty \frac{dw}{4 \pi w}\frac{1}{\sqrt{\sinh \left(\frac{\pi w}{a_A}\right) \sinh \left(\frac{\pi w}{a_B}\right)}}
					\Big[\frac{e^{\frac{\pi}{2}w\left(\frac{1}{a_A}-\frac{1}{a_B}\right)}e^{ik\Delta \zeta -iw\Delta \eta}+e^{-\frac{\pi}{2}w\left(\frac{1}{a_A}-\frac{1}{a_B}\right)}e^{ik\Delta \zeta+ iw\Delta \eta}}{1-e^{-\beta w}} \nonumber \\
					&&+\frac{e^{\frac{\pi}{2}w\left(\frac{1}{a_A}-\frac{1}{a_B}\right)}e^{-ik\Delta \zeta+iw\Delta \eta}+e^{-\frac{\pi}{2}w\left(\frac{1}{a_A}-\frac{1}{a_B}\right)}e^{-ik\Delta \zeta-iw\Delta \eta}}{e^{\beta w}-1}\Big]~,
					\label{thermal-opp-ab}
				\end{eqnarray}
				and
				\begin{eqnarray}
					G^{+ \beta}_{3D_{LR}}(x^{\prime}_B,x_A)&=&\int_{0}^{\infty} dw \frac{2}{\sqrt{a_A a_B}}\hspace{.5cm}\int_{-\infty}^{\infty}\frac{d^2k_p}{(2\pi)^4}
					\Big[\frac{e^{-iw\Delta \eta   }e^{\frac{\pi w}{2}\left(\frac{1}{a_A}-\frac{1}{a_B}\right)}+e^{iw\Delta \eta}e^{\frac{-\pi w}{2}\left(\frac{1}{a_A}-\frac{1}{a_B}\right)}}{1-e^{-\beta w}} \nonumber \\ 
					&& + \frac{e^{iw\Delta \eta }e^{\frac{\pi w}{2}\left(\frac{1}{a_A}-\frac{1}{a_B}\right)}+e^{-iw\Delta \eta }e^{\frac{-\pi w}{2}\left(\frac{1}{a_A}-\frac{1}{a_B}\right)}}{1-e^{-\beta w}}\Big]
					\nonumber \\ 
					&&\times\int_{-\infty}^{\infty} \frac{d^2k_p}{{(2 \pi)}^{4}}\mathcal{K}\left[\frac{iw}{a_A},\frac{\left|k_p\right|e^{a_A\zeta_A}}{a_A}\right] \mathcal{K}\left[\frac{iw}{a_B},\frac{\left|k_p\right|e^{a_B\zeta_B}}{a_B}\right]~, 
					\label{thermal-greens-opp-3d}
				\end{eqnarray}
				for $(1+1)$ and $(1+3)$ dimensions, respectively.
				In the above we used the following notations: $\Delta \eta = \eta_{2,i} - \eta_{1,j}$ and $\Delta \zeta = \zeta_{2,i} - \zeta_{1,j} $, where $i,j$ signifies the particular detector we are considering and $1,2$ denote two different spacetime points. The non-thermal ones are obtained by considering $\beta\to\infty$ limit.
				Let us now proceed towards our main investigations.
			\end{widetext}

			To study the entanglement properties of the two detectors system under time evolution through interaction with background scalar fields, as mentioned earlier, here we closely investigate $\delta \mathcal{N}$, given by Eq. \eqref{negativity}. It may be noted that in order to know about $\delta \mathcal{N}$ one needs to compute the integrals in Eq. \eqref{eqn_dyna_integrals}. Here we choose the switching function as $\chi(\tau) = 1$ which means the interaction is on for an infinite proper time {\footnote{Same choice of switching function has also been considered in original detector's response analysis (see e.g. Section $3.3$ of \cite{Book1}) and subsequently in various literature. This helps to perform all the calculations analytically.}}.  Therefore, the limit of integrations on the respective proper times is from $-\infty$ to $+\infty$. Under these circumstances, we will now investigate $\delta \mathcal{N}$ for two situations.

	\section{parallelly accelerating detectors}\label{parallel}
		First, we look at the case when both the detectors are in the same Rindler wedge (take them on the RRW). From Eq. \eqref{trajectory}, one can say that even for detectors with the same $\zeta$ value, the corresponding trajectories will be different if they have different values of $a$. For our further calculation purpose we will hence use $\zeta_A = 0$ and  $\zeta_B = 0$; i.e. detectors are at the vanishing spatial distance in their own frames and  thus  $\Delta \zeta = 0$. Therefore, the parameter $a$ is identified as the proper acceleration while $\eta$ turns out to be the proper time.

		Then for both detectors on the RRW, we use  $\Delta \eta = \Delta \tau= \tau_B-\tau_A$ in the expressions for greens function  and compute the integrals in Eq. \eqref{eqn_dyna_integrals} which are required to obtain expression for Eq. \eqref{negativity}. Our discussion will be both in $(1+1)$ and $(1+3)$ dimensions. Also, at the end consequences of non-thermal limit (i.e. $\beta\to\infty$) will be addressed. 
Before proceeding further let us mention that in the change of Negativity only the real parts of the different quantities which appeared in (\ref{negativity}) will contribute. Therefore only the real contributions of these quantities (which are complex) will be mentioned below. 
		\begin{widetext}
			\subsection{$(1+1)$-dimensions}	
				We calculate the integrals in Eq. \eqref{eqn_dyna_integrals} for thermal  Wightman function in Appendix \ref{app-grn-same} and obtain the following results:
				\begin{eqnarray}
					&&P^{\beta}_{j_R}=\left|\bra{E_0^j}m_j\ket{E_1^j}\right|^2 \frac{\pi}{\Delta E_j}\delta(0)\left[\frac{1}{1-e^{-\beta \Delta E_j}}\times\frac{1}{e^{\frac{2\pi \Delta E_j}{a_j}}-1}+\frac{1}{e^{\beta \Delta E_j}-1}\times\frac{1}{1-e^{-\frac{2\pi \Delta E_j}{a_j}}}\right]~, \nonumber \\
										&&P^{\prime \prime \beta}_{j_R}=\left|\bra{E_0^j}m_j\ket{E_1^j}\right|^2 \frac{\pi}{\Delta E_j}\delta(0)\left[\frac{1}{1-e^{-\beta \Delta E_j}}\times\frac{1}{1-e^-{\frac{2\pi \Delta E_j}{a_j}}}+\frac{1}{e^{\beta \Delta E_j}-1}\times\frac{1}{e^{\frac{2\pi \Delta E_j}{a_j}}-1}\right]~, \nonumber \\
					&&	z^{\beta}=\sum_{j}\left|\bra{E_0^j}m_j\ket{E_1^j}\right|^2 \frac{\pi}{2\Delta E_j}\delta(0)\left[\frac{1+e^{\frac{2\pi \Delta E_j}{a_j}}}{e^{\frac{2\pi\Delta E_j}{a_j}}-1}\times \frac{e^{\beta \Delta E_j}+1}{e^{\beta \Delta E_j}-1}\right]~,
					\nonumber \\
					&&P^{ \prime \beta}_{AB_{same}}=0~, \quad \bar P^{\prime \beta}_{AB_{same}}=0~, \quad Y^{\beta}_{AB_{same}}=0~, \quad \zeta^{\beta}_{AB_{same}}=0~,
					\label{V25}
				\end{eqnarray}
				where we denoted $\bar{E}=\frac{\Delta E_A +\Delta E_B}{2}$ and $\Delta E=\frac{\Delta E_B-\Delta E_A}{2}$. 
					The non-thermal results can be obtained from the above values by considering the limit $\beta \rightarrow \infty $.

			\subsection{$(1+3)$-dimensions}
				In $(1+3)$ dimensions, for the thermal field the values of the integration for our system come out to be as (Appendix \ref{AppB2})
				\begin{eqnarray}
					&&P^\beta_{j_R}=\left|\bra{E_0^j}m_j\ket{E_1^j}\right|^2 \delta(0)\frac{\Upsilon(\Delta E_j,a_j,a_j)}{\pi{a_j}  }\left[\frac{e^{-\frac{\pi \Delta E_j}{a_j}}}{1-e^{-\beta \Delta E_j}}+\frac{e^{\frac{\pi \Delta E_j}{a_j}}}{e^{\beta \Delta E_j}-1}\right]~,\nonumber \\
					&&P^{\prime \prime \beta}_{j_R}=\left|\bra{E_0^j}m_j\ket{E_1^j}\right|^2 \delta(0)\frac{\Upsilon(\Delta E_j,a_j,a_j)}{\pi{a_j}  }\left[\frac{e^{\frac{\pi \Delta E_j}{a_j}}}{1-e^{-\beta \Delta E_j}}+\frac{e^{\frac{-\pi \Delta E_j}{a_j}}}{e^{\beta \Delta E_j}-1}\right]~,\nonumber \\
					&&z^{\beta}=\sum_j\left|\bra{E_0^j}m_j\ket{E_1^j}\right|^2 \delta(0)\frac{\Upsilon(\Delta E_j,a_j,a_j)}{2\pi{a_j}  }\left[\frac{e^{-\frac{\pi \Delta E_j}{a_j}}}{1-e^{-\beta \Delta E_j}}+\frac{e^{\frac{\pi \Delta E_j}{a_j}}}{e^{\beta \Delta E_j}-1}+\frac{e^{\frac{\pi \Delta E_j}{a_j}}}{1-e^{-\beta \Delta E_j}}+\frac{e^{\frac{-\pi \Delta E_j}{a_j}}}{e^{\beta \Delta E_j}-1}\right]~,\nonumber \\
										&&\bar P^{\prime \beta}_{AB_{same}}=0~, \quad  P^{\prime \beta}_{AB_{same}}=0~, \quad Y^{\beta}_{AB_{same}}=0~, \quad \zeta^{\beta}_{AB_{same}}=0~,
					\label{int-3d-same-therm}
				\end{eqnarray}
				where the following notation has been used
				\begin{equation}\label{upsilon}
					\Upsilon(x,a_A,a_B)=\int_0^{\infty}k_p dk_p \mathcal{K}\left[\frac{ix}{a_A},\frac{k_p}{a_A}\right] \mathcal{K}\left[\frac{ix}{a_B},\frac{k_p}{a_B}\right]~,
				\end{equation}
				and 
				\begin{eqnarray}
					&&\Upsilon(\Delta E_j,a_j,a_j)=	\frac{\pi a_j\Delta E_j}{2 \sinh \left(\frac{\pi \Delta E_j}{a_j}\right)}~.
				\end{eqnarray}
				Similar to the previous section, the non-thermal expressions for the integrals are obtained by taking $\beta \rightarrow \infty$ in Eq. \eqref{int-3d-same-therm}.

		\end{widetext}	
		
	\section{Anti-parallelly accelerating detectors} \label{anti-parallel}
		Consider two detectors A and B in RRW and LRW, respectively. Like earlier, we take the detectors are at spatial origin with respect to their own reference frame; i.e. $\zeta_A =0 $ , $\zeta_B =0 $ and hence $\Delta \zeta =0 $. In this situation, we have $\eta_A=\tau_A$ while $\eta_B = -\tau_B$ and so $\Delta \eta = \eta_B-\eta_A= - ( \tau^{\prime}_B+\tau_A )=-\bar \tau $ (say). Here also, as mentioned in the last section, only the real contributions of various quantities will be provided.

		\begin{widetext}
			\subsection{$(1+1)$-dimensions}
				Here we obtain the value of the integrals using the green's functions in Section \ref{greens-exp}. The expressions in $(1+1)$ dimensions have been explicitly calculated in Appendix \ref{app-opp}. These are found to be as follows:
				\begin{eqnarray}
					&&P^\beta_{j_{J}}=\left|\bra{E_0^j}m_j\ket{E_1^j}\right|^2  \frac{\pi}{\Delta E_j}\delta(0)
					\left[\frac{1}{1-e^{-\beta \Delta E_j}}\times\frac{1}{e^{\frac{2\pi \Delta E_j}{a_j}}-1}+\frac{1}{e^{\beta \Delta E_j}-1}\times\frac{1}{1-e^{-\frac{2\pi \Delta E_j}{a_j}}}\right]~,
					\nonumber \\
					&&	P^{\prime \prime \beta}_{j_{J}}=\left|\bra{E_0^j}m_j\ket{E_1^j}\right|^2  \frac{\pi}{\Delta E_j}\delta(0)\left[\frac{1}{1-e^{-\beta \Delta E_j}}\times\frac{1}{1-e^-{\frac{2\pi \Delta E_j}{a_j}}}+\frac{1}{e^{\beta \Delta E_j}-1}\times\frac{1}{e^{\frac{2\pi \Delta E_j}{a_j}}-1}\right]~, \nonumber \\
					&&	z^{\beta}=\sum_{j}\left|\bra{E_0^j}m_j\ket{E_1^j}\right|^2 \frac{\pi}{2\Delta E_j}\delta(0)\left[\frac{1+e^{\frac{2\pi \Delta E_j}{a_j}}}{e^{\frac{2\pi\Delta E_j}{a_j}}-1}\times \frac{e^{\beta \Delta E_j}+1}{e^{\beta \Delta E_j}-1}\right]~,
					\nonumber \\ 
					&&P^{ \prime \beta}_{AB_{opp}}=\langle m_{AB}^\prime \rangle \delta({\Delta E_A - \Delta E_B})\frac{\pi}{\bar E }\frac{1}{\sqrt{\sinh \left(\frac{\pi \bar E}{a_A}\right) \sinh \left(\frac{\pi \bar E}{a_B}\right)}}\left[\frac{e^{\frac{\pi}{2} \bar E(\frac{1}{a_A}-\frac{1}{a_B})}}{e^{\beta \bar E}-1}+\frac{e^{\frac{-\pi}{2} \bar E(\frac{1}{a_A}-\frac{1}{a_B})}}{1-e^{-\beta \bar E}}\right]~, \nonumber \\
					&&\bar P^{ \prime \beta}_{AB_{opp}}=\langle m_{AB}^\prime \rangle \delta({\Delta E_A - \Delta E_B})\frac{\pi}{\bar E }\frac{1}{\sqrt{\sinh\left(\frac{\pi \bar E}{a_A}\right)\sinh\left(\frac{\pi \bar E}{a_B}\right)}}\left[\frac{e^{\frac{-\pi}{2} \bar E(\frac{1}{a_A}-\frac{1}{a_B})}}{e^{\beta \bar E}-1}+\frac{e^{\frac{\pi}{2} \bar E(\frac{1}{a_A}-\frac{1}{a_B})}}{1-e^{-\beta \bar E}}\right]~, \nonumber \\
					&&Y^{\beta}_{AB_{opp}}=\langle m_{AB}^\prime \rangle\delta({\Delta E_A - \Delta E_B})\frac{\pi}{\bar E }\frac{1}{\sqrt{\sinh\left(\frac{\pi \bar E}{a_A}\right)\sinh\left(\frac{\pi \bar E}{a_B}\right)}} \Big[\frac{e^{-\frac{\pi}{2} \bar E(\frac{1}{a_A}-\frac{1}{a_B})}}{1-e^{-\beta \bar E}}+\frac{e^{\frac{\pi}{2} \bar E(\frac{1}{a_A}-\frac{1}{a_B})}}{e^{\beta \bar E}-1}
					\nonumber
					\\
					&&-\sinh{\left\{\frac{\pi}{2}\bar{E}\left(\frac{1}{a_A}-\frac{1}{a_B}\right)\right\}}\Big]~, \nonumber\\
					&&\zeta^{\beta}_{AB_{opp}}=\langle m_{AB}^\prime \rangle \delta({\Delta E_A - \Delta E_B})\frac{\pi}{\bar E }\frac{1}{\sqrt{\sinh\left(\frac{\pi \bar E}{a_A}\right)\sinh\left(\frac{\pi \bar E}{a_B}\right)}}\Big[\frac{e^{\frac{\pi}{2} \bar E(\frac{1}{a_A}-\frac{1}{a_B})}}{1-e^{-\beta \bar E}}+\frac{e^{\frac{-\pi}{2} \bar E(\frac{1}{a_A}-\frac{1}{a_B})}}{e^{\beta \bar E}-1}
					\nonumber
					\\
					&&-\sinh{\left\{\frac{\pi}{2}\bar{E}\left(\frac{1}{a_A}-\frac{1}{a_B}\right)\right\}}\Big]~,
				\label{therm-int-opp}
				\end{eqnarray}
				where $J=R$ ($L$) when $j=A$ ($B$) and  $\langle m_{AB}^\prime \rangle = \bra{E_0^A}m_A\ket{E_1^A}\bra{E_0^B}m_B\ket{E_1^B}$.
				It should be noticed that $z^{\beta}$, $P^{\beta}_{j_{J}}$, $P^{\prime \prime \beta}_{j_{J}}$ values are same irrespective of if the two detectors are parallelly or antiparallelly accelerating.  
				The same for non-thermal  fields are found in the limit $\beta \rightarrow \infty$.

			\subsection{$(1+3)$-dimensions}
				In the same way the integrals can be evaluated in $(1+3)$ dimensions as well. The results are as follows:
				\begin{eqnarray}
					&&P^\beta_{j_J}=\left|\bra{E_0^j}m_j\ket{E_1^j}\right|^2 \delta(0)\frac{\Upsilon(\Delta E_j,a_j,a_j)}{\pi{a_j}  }\left[\frac{e^{-\frac{\pi \Delta E_j}{a_j}}}{1-e^{-\beta \Delta E_j}}+\frac{e^{\frac{\pi \Delta E_j}{a_j}}}{e^{\beta \Delta E_j}-1}\right]~,\nonumber \\
										&&P^{\beta \prime \prime}_{j_J}=\left|\bra{E_0^j}m_j\ket{E_1^j}\right|^2 \delta(0)\frac{\Upsilon(\Delta E_j,a_j,a_j)}{\pi{a_j}  }\left[\frac{e^{\frac{\pi \Delta E_j}{a_j}}}{1-e^{-\beta \Delta E_j}}+\frac{e^{\frac{-\pi \Delta E_j}{a_j}}}{e^{\beta \Delta E_j}-1}\right]~,\nonumber \\
					&&z^{\beta}=\sum_j\left|\bra{E_0^j}m_j\ket{E_1^j}\right|^2 \delta(0)\frac{\Upsilon(\Delta E_j,a_j,a_j)}{2\pi{a_j}  }\left[\frac{e^{-\frac{\pi \Delta E_j}{a_j}}}{1-e^{-\beta \Delta E_j}}+\frac{e^{\frac{\pi \Delta E_j}{a_j}}}{e^{\beta \Delta E_j}-1}+\frac{e^{\frac{\pi \Delta E_j}{a_j}}}{1-e^{-\beta \Delta E_j}}+\frac{e^{\frac{-\pi \Delta E_j}{a_j}}}{e^{\beta \Delta E_j}-1}\right]~,\nonumber \\
					&&P^{\prime \beta}_{AB_{opp}}=\langle m_{AB}^\prime \rangle\delta({\Delta E_A - \Delta E_B})\frac{\Upsilon(\bar E,a_B,a_A)}{\pi\sqrt{a_A a_B}  }\left[\frac{e^{\frac{\pi}{2} \bar E(\frac{1}{a_A}-\frac{1}{a_B})}}{e^{\beta \bar E}-1}+\frac{e^{\frac{-\pi}{2} \bar E(\frac{1}{a_A}-\frac{1}{a_B})}}{1-e^{-\beta \bar E}}\right]~,\nonumber \\
					&&\bar P^{\prime \beta}_{AB_{opp}}=\langle m_{AB}^\prime \rangle\delta({\Delta E_A - \Delta E_B})\frac{\Upsilon(\bar E,a_B,a_A)}{\pi\sqrt{a_A a_B}  }\left[\frac{e^{\frac{-\pi}{2} \bar E(\frac{1}{a_A}-\frac{1}{a_B})}}{e^{\beta \bar E}-1}+\frac{e^{\frac{\pi}{2} \bar E(\frac{1}{a_A}-\frac{1}{a_B})}}{1-e^{-\beta \bar E}}\right]~,\nonumber \\
					&&Y^{\beta}_{AB_{opp}}=\langle m_{AB}^\prime \rangle\delta({\Delta E_A - \Delta E_B})\frac{\Upsilon(\bar E,a_B,a_A)}{\pi\sqrt{a_A a_B}  }\left[\frac{e^{-\frac{\pi}{2} \bar E(\frac{1}{a_A}-\frac{1}{a_B})}}{1-e^{-\beta \bar E}}+\frac{e^{\frac{\pi}{2} \bar E(\frac{1}{a_A}-\frac{1}{a_B})}}{e^{\beta \bar E}-1}-\sinh{\left\{\frac{\pi\bar E}{2}(\frac{1}{a_A}-\frac{1}{a_B})\right\}}\right]~,\nonumber \\
					&&\zeta^{\beta}_{AB_{opp}}=\langle m_{AB}^\prime \rangle\delta({\Delta E_A - \Delta E_B})\frac{\Upsilon(\bar E,a_B,a_A)}{\pi\sqrt{a_A a_B}  }\left[\frac{e^{\frac{\pi}{2} \bar E(\frac{1}{a_A}-\frac{1}{a_B})}}{1-e^{-\beta \bar E}}+\frac{e^{\frac{-\pi}{2} \bar E(\frac{1}{a_A}-\frac{1}{a_B})}}{e^{\beta \bar E}-1}-\sinh{\left\{\frac{\pi\bar E}{2}(\frac{1}{a_A}-\frac{1}{a_B})\right\}}\right]~,
				\end{eqnarray}
				where $\Upsilon(x,a_A,a_B)$ is defined in Eq. \eqref{upsilon}.

				The expression for the non thermal case is obtained by taking $\beta \rightarrow \infty $.
		\end{widetext}

Few comments are to be noted here. In our case the detectors are uniformly accelerating and in this frame (Rindler frame) the scalar field has been decomposed with respect to the Rindler modes (either in left Rindler or in right Rindler). In order to calculate the positive frequency Wightman function we choose the vacuum of Unruh modes which is the Minkowski vacuum. Therefore the Wightman functions are inherently thermal in nature with respect to the Rindler observer (check the expressions (\ref{V21}), (\ref{V22}) for $(1+1)$ dimensions and (\ref{BRMP1}) and (\ref{V23}) for $(1+3)$ dimensions with $\beta\to\infty$ and $a_A=a_B$). This is quite compatible with the well known fact that the positive frequency Wightman function with respect to an accelerated frame in Minkowski vacuum mimics the thermal Wightman function with temperature is identified as $(a_j/2\pi)$.  
Therefore, in this case, a two-level detector carried by this accelerated frame will click as it suffers transition from ground state to excited state and thereby predicts particle content in the Minkowski vacuum. This is the consequence of celebrated Unruh effect. So in this situation the transition probability is non-vanishing and it is quite similar to the Bose distribution when one calculates this using first order perturbation theory with a monopole interaction between detector and scalar field (similar to Eq. (\ref{V24})) for switching function is set to unity. It may be noted that the single detector transition probability is same as our quantities like $P_j$ and $P_{jJ}$. Hence these have to be non-vanishing and must be identical to the known results for the present situation. 
For instance in (1+1) dimensions we have (\ref{V25})
and the same in $(1+3)$ dimension is given by (\ref{int-3d-same-therm}). 
These all reduces to the known expression
\begin{equation}
P_j \sim \delta(0)\frac{1}{e^{\frac{2\pi \Delta E_j}{a_j}} -1}~,
\end{equation}
for non-thermal field (i.e. $\beta\to \infty$). The above ones will vanish if the detector is static or moving with constant velocity (i.e. $a_j =0$). This can be checked by considering both the limits: $\beta\to \infty$ and $a_j\to 0$.
These implies that our obtained expressions are quite consistent with the known forms of detector's transition probability.  

We mentioned above that the single detector transition probability is given by quantities like $P_j(\Delta E_j)$ (for instance see first expression of Eq. (\ref{V25})). We will see, these expressions are quite consistent with Kubo-Martin-Schwinger (KMS) relation. For discussion purpose below, we will concentrate on first result of Eq. (\ref{V25})) for $(1+1)$ dimensions. For other cases the same exactly follows. This expression with dropping index $j$ can be written in the following form as well
\begin{eqnarray}
P(\Delta E) =  \frac{X}{\Delta E}~\Big[n_\beta + n_a(2n_\beta+1)\Big]~,
\label{M1}
\end{eqnarray}
where the following notations have been used:
\begin{eqnarray}
&&X=\left|\bra{E_0}m\ket{E_1}\right|^2 \pi \delta(0)~,
\nonumber
\\
&&
n_\beta = \frac{1}{e^{\beta \Delta E}-1}~;\,\,\,\, n_a = \frac{1}{e^{\frac{2\pi \Delta E}{a}}-1}~.
\label{M2}
\end{eqnarray}
This form of expression for detector's transition amplitude was obtained earlier in \cite{Kolekar:2013hra} under the similar situation (also see \cite{Kolekar:2013xua,Kolekar:2013aka} in the context of the calculation of number of particles and \cite{Bruschi:2013tza} in a little different context). Such an expression in literature \cite{Kolekar:2013hra,Kolekar:2013xua,Kolekar:2013aka} has been explained in terms of spontaneous and stimulated transitions of detector. In absence of temperature of field (i.e. when $n_\beta=0$), the term in (\ref{M1}) with $+1$ only survives and represents the spontaneous excitation of detector due to its acceleration (the usual Unruh effect). Therefore the coefficient of $+1$ is called as the coefficient of spontaneous excitation due to acceleration:
\begin{eqnarray}
A_{\uparrow}^{(a)} = \frac{X}{\Delta E}n_a~.
\label{M3}
\end{eqnarray}
While the cross term signifies the stimulated excitation due to presence of the ambient real thermal bath, represented by inverse temperature $\beta$. Therefore the coefficient $n_a$ can be identified as coefficient of stimulated excitation:
\begin{eqnarray}
B_{\uparrow}^{(a)} =\frac{X}{\Delta E} n_a~.
\label{M4}
\end{eqnarray}
The corresponding de-excitation coefficients are found by replacing $\Delta E \rightarrow -\Delta E$ in the expressions given by \eqref{M2}, \eqref{M3} and \eqref{M4}:
\begin{eqnarray}
A_{\downarrow}^{(a)} = \frac{X}{\Delta E}(n_a+1) = B_{\downarrow}^{(a)}~.
\label{M5}
\end{eqnarray}
Similarly expressing (\ref{M1}) as
\begin{eqnarray}
P(\Delta E) =  \frac{X}{\Delta E}~\Big[n_a + n_\beta(2n_a+1)\Big]~,
\label{M6}
\end{eqnarray}
and using the above argument the coefficients of excitation and de-excitation with respect to $\beta$ are identified. These are
\begin{eqnarray}
&&A_{\uparrow}^{(\beta)} = \frac{X}{\Delta E}n_\beta = B_{\uparrow}^{(\beta)}~;
\nonumber
\\
&&A_{\downarrow}^{(\beta)} = \frac{X}{\Delta E}(n_\beta+1) = B_{\downarrow}^{(\beta)}~.
\label{M7}
\end{eqnarray}
The above approach is inspired by an original investigation done in \cite{Wald:1976ka,Bekenstein:1977mv} in presence of black hole spacetime. Particularly \cite{Bekenstein:1977mv} showed that these coefficients can be interpreted as Einstein's $A$, $B$ coefficients in the context of atomic physics. One can check that these satisfies Einstein's relation (excitation to de-excitation ratio):
\begin{eqnarray}
\frac{B_{\uparrow}^{(a)}}{B_{\downarrow}^{(a)}} = e^{-\frac{2\pi \Delta E}{a}}~,
\label{M8}
\end{eqnarray}
and similar for others. This is reminiscent to the excitation to de-excitation ratio as discussed in \cite{garay2016thermalization} when one considers switching function as unity. We will show now that this is because the Wightman function (\ref{V21}) for single detector is time translational invariant with respect to proper time and satisfies the KMS conditions:
\begin{eqnarray}
&&G\Big(\Delta\tau - \frac{2\pi i}{a}\Big) = G(-\Delta\tau)~; 
\nonumber
\\
&&G(\Delta\tau - i\beta) = G(-\Delta\tau)~.
\label{M9}
\end{eqnarray}
The proof of the KMS relations are given explicitly in Appendix \ref{KMS}. One may note that the transition amplitude with $\chi = 1$ can be seen as the Fourier transformation of Wightman function with respect to $\Delta\tau$ and $\Delta E$ (for instance see first expression of Eq. (\ref{eqn_dyna_integrals})). Therefore the relevant quantity we will investigate is of the following form:
\begin{eqnarray}
P(\Delta E) = \int_{-\infty}^{+\infty} d(\Delta\tau)e^{-i\Delta E \Delta\tau} G(\Delta\tau)~.
\label{M10}
\end{eqnarray} 
Similarly we define
\begin{eqnarray}
P(-\Delta E) 
&=&\int_{-\infty}^{+\infty} d(\Delta\tau)e^{-i\Delta E \Delta\tau} G(\Delta\tau - i\beta')~,
\label{M11}
\end{eqnarray} 
where $\beta' = \frac{2\pi}{a}$ or $\beta$. 
Note that the above result can be thought as the Fourier transformation of complexified-time version of Wightman function. Now, as shown in Appendix \ref{KMS}, the meaning of $G(\Delta\tau-i\beta')$ is not just substitution of $\Delta\tau$ by $\Delta\tau - i\beta'$ in all terms of $G(\Delta\tau)$ particularly when both $\beta$ and $a$ are there (rather one part is related to $\Delta\tau\to\Delta\tau - i\beta'$ while the other part is complex conjugate i.e. replace $\Delta\tau\to\Delta\tau + i\beta'$). The final expression (\ref{M11}) signifies that $P(-\Delta E)$ should not be obtained by replacing $\Delta E\to -\Delta E$ in the integrated value of (\ref{M10}). Therefore it is instructive to calculate $P(\Delta E)$ and $P(-\Delta E)$ directly from (\ref{M10}) and (\ref{M11}), respectively.  Now changing the variable as $\Delta\tau-i\beta^{\prime}\to \Delta\tau$ one finds
\begin{eqnarray}
P(-\Delta E) = e^{\beta'\Delta E}P(\Delta E)~.
\label{M12}
\end{eqnarray}
This can be checked by explicitly evaluating the integrations (\ref{M10}) and (\ref{M11}) for our given Wightman function as well.
Therefore we conclude that once the Wightman function satisfies KMS relations, then the excitation to de-excitation ratio satisfies above type of relation. The same can be shown for $(1+3)$ dimensions as well.

	\section{Numerical analysis}\label{numerical analysis}
	
	Before going to the main aim of this section, let us first make few comments. In the calculations of the above analysis, there exist few divergences,  which we explain below. It is known that the Wightman function contains two types of divergences: a Hadamard short-distance property is observed for both $(1+1)$-D and $(1+3)$-D cases \cite{Radzikowski:1996pa}, and an infrared (IR) divergence is additionally captured in the $(1+1)$-dimensional Wightman function \cite{Book1}. The short-distance divergence is distinctly visible in the position-space representation of the Wightman function and is tackled by introducing an ultraviolet (UV) regulator. Whereas, the IR ambiguity in two dimensions appears as an additive logarithmic divergence in the position-space representation. The same is reflected in our work through the divergence of the integrand at the mode frequency $w = 0$ when one performs the $w$ integration (for instance, see presence of $w$ in the denominators of Eqs. (\ref{V21}), (\ref{V22}) and (\ref{thermal-opp-ab})). It must be mentioned that the IR divergence is very specific to two-dimensional correlation functions \cite{Book1}. In some situations, the IR divergence creates problem in the  detector's response function, particularly for the choice of monopole type of interaction as the detector's density matrix may depend on the choice of the IR cut-off (e.g. see \cite{Pozas-Kerstjens:2015gta,Juarez-Aubry:2014jba,Martin-Martinez:2014qda}). Normally, this cut-off is chosen based on the characteristic scale of the system or is  removed by hand (for instance see the treatment in \cite{Book1}). Since we have chosen the same interaction, a similar problem may arise in calculating different quantities in the negativity. In this regard, it may be pointed out that for a unity switching function, the IR cut-off can be safely removed from each term (e.g. see a discussion on $P_j$ in the Appendix C of our manuscript using the position-space representation of Wightman function). Since our choice is also the same, it is expected that the negativity will not depend on the IR cut-off parameter. This is evident in the process of our calculation where the $w$ integration is safely done. We do not adopt the ``exact'' position-space representation of the Wightman function as the $w$ integrations are not explicitly done at the Wightman function level. Rather, they are evaluated while calculating the different terms in negativity. Therefore, we see that in the present situation, the negativity is IR-divergence safe. A similar scenario has been observed earlier in \cite{Cong:2018vqx} where the negativity is, by definition, IR-safe. This helps to avoid the involvement of any extra prescription in order to tackle IR divergence in $(1+1)$ dimension.

	In doing so, although we have managed the problem of IR divergence for the choice $\chi_j(\tau_j)=1$, we encounter another divergent factor in each term of $\delta\mathcal{N}$. This appears as a Dirac-delta function $\delta(0)$ due to the interaction between the detectors and the scalar field for an infinite time (as we chose $\chi_j(\tau_j)=1$).  In all the cases, as the $\delta(0)$  arises from a relation of the type $\delta(0) =\lim_{a\to 0} \frac{1}{2\pi} \int_{-\infty}^{+\infty} du e^{iax} \equiv \frac{1}{2\pi} \int_{-\infty}^{+\infty} du$, it has nothing to do with the mentioned UV or IR divergences. And, since the same switching function has been chosen for $(1+1)$- and $(1+3)$-dimensions, $\delta(0)$ pops up for both the cases in an identical manner. It is to be noted that such a divergence here is not a new issue; the original single Unruh-DeWitt detector's transition probability also suffers the same problem due to this specific choice of switching function. In such a situation, drawing an analogy with the well known Fermi's golden rule, the transition probability per unit time (known as detector's response function) is considered to be the relevant physical quantity (see a detailed discussion in section $3.3$ of \cite{Book1}). This quantity is uniform throughout the span of interaction (also reflected by the fact that Wightman function is time translational invariant) and hence is useful for the main purpose. Using the same spirit,  here we investigate the rate of change of $\delta\mathcal{N}$, defined in the following way. Our $\delta\mathcal{N}$ has the form $\delta \mathcal{N} = \textrm{finite quantity} \times \delta(0)$. Now, it might seem that the terms like $P^{\prime}_{AB}$; $\bar P^{\prime}_{AB}$; $\zeta_{AB}$ and $Y_{AB}$ in $\delta\mathcal{N}$ do not contain the $\delta(0)$ factor, but notice in
	\eqref{therm-int-opp}, they contain a factor of $\delta(\Delta E_A - \Delta E_B)$. It should be noted, these terms will contribute only if $\Delta E_A = \Delta E_B$, hence, that is what we have considered for our analysis. A similar argument can be seen in \cite{koga2018quantum,Barman:2021bbw}. Thus $\delta(\Delta E_A - \Delta E_B)$ will essentially reduce to $\delta(0)$: 
	\begin{eqnarray}
  \delta(0) &=& \lim_{\Delta  E^{\prime} \to 0} \delta(\Delta E^{\prime}) = \lim_{\Delta  E^{\prime} \to 0}\frac{1}{2\pi} \int_{-\infty}^{+\infty} due^{i\Delta E^{\prime}u} 
\nonumber
\\
&=& \frac{1}{2\pi}\int_{-\infty}^{+\infty}du, 
\end{eqnarray}
where $\Delta  E^{\prime} = \Delta E_A - \Delta E_B$.
Hence we have,
	\begin{eqnarray}
		\delta \mathcal{N} &=& \underbrace{\textrm{finite quantity} \times \delta(0)}_{\textrm{from terms}~ P_j^\prime, P_j^{\prime \prime}, M_j} 
\nonumber
\\
&+& \underbrace{\textrm{finite quantity} \times \delta(0)}_{\textrm{from terms}\hspace{.1cm} P_{AB}^\prime, \bar P_{AB}^{\prime \prime}, Y_{AB},\zeta_{AB}\hspace{.1cm}\textrm{with}\hspace{.1cm}\Delta E^{\prime} \to 0}~.
	  \end{eqnarray}
	We now define the rate of change of $\delta\mathcal{N}$ as
\begin{equation}
\delta \mathcal{\dot{N}}= (\textrm{finite quantity}) \times \lim_{T\to\infty} \frac{1}{2\pi T}\int _{-T/2}^{+T/2} du~,
\end{equation}
where we have used the fact $\delta(0) = \lim_{T\to\infty}\frac{1}{2\pi} \int _{-T/2}^{+T/2} du$. The above then reduces to 
$\delta \mathcal{\dot{N}}=(\textrm{finite quantity})/2\pi $.
Using the above idea in this paper we define the relevant quantity to investigate as
\begin{equation}
\delta\textrm{N} = 2\pi \delta\mathcal{\dot{N}}\Delta E_j~,
\end{equation}
 for $(1+1)$ D and 
 \begin{equation}
\delta\textrm{N} = \frac{2\pi \delta\mathcal{\dot{N}}}{\Delta E_j}~,
\end{equation}
in $(1+3)$ D. These are regarded as scaled versions of the rate of change of $\delta\mathcal{N}$ and must be finite. But one must be careful in handeling $\delta\mathcal{N}$ as it grows linearly with time. To make pertubative calculation viable we must have the condition -- initial entaglement per unit interaction time is greater than $\delta\dot{\mathcal{N}}$ and then for a span of time one has $|{\mathcal{N}}(0)|>>|C^2\delta\mathcal{N}|$. 
Therefore, following the discussion at the end of Section II B, we conclude that increase (decrease) in $\delta\mathrm{N}$ implies decrease (increase) in entanglement. In this work, we numerically investigate these quantities. The same idea is being used in defining detector's response function for the Unruh-DeWitt case \cite{Book1} and also has been adopted in \cite{Koga:2019fqh} to investigate the entanglement harvesting phenomenon. Of course such an issue can be avoided by invoking different switching function ( e.g. Gaussian type \cite{Tjoa:2020eqh}), but then the analytic expression of the integrations will be difficult to obtain. Therefore for simplicity of the calculation and in order to achieve an analytic expression for $\delta\mathcal{N}$, like in various earlier investigations, here we adopted $\chi =1$.

Therefore to study the fate of entanglement, we plot $\delta \text{N}$  with respect to different parameters. 
Note that our analysis is considered up to the second order of perturbation theory ${O(C^2)}$, thus the parameters for which $\delta \mathcal{N}$ values become large has to be excluded. Otherwise, there will be a violation of the perturbation method.
	We define three dimensionless parameters as $\mathrm{A} = (\Delta E_A/a_A)$, $\mathrm{B} = (\Delta E_B/a_B)$ and $\sigma = \beta \Delta E $. 
		\subsection{Non-thermal field}

			\subsubsection{$(1+1)$-dimensions}

\begin{figure}[h!]
					\centering
					\includegraphics[width=0.40\textwidth]{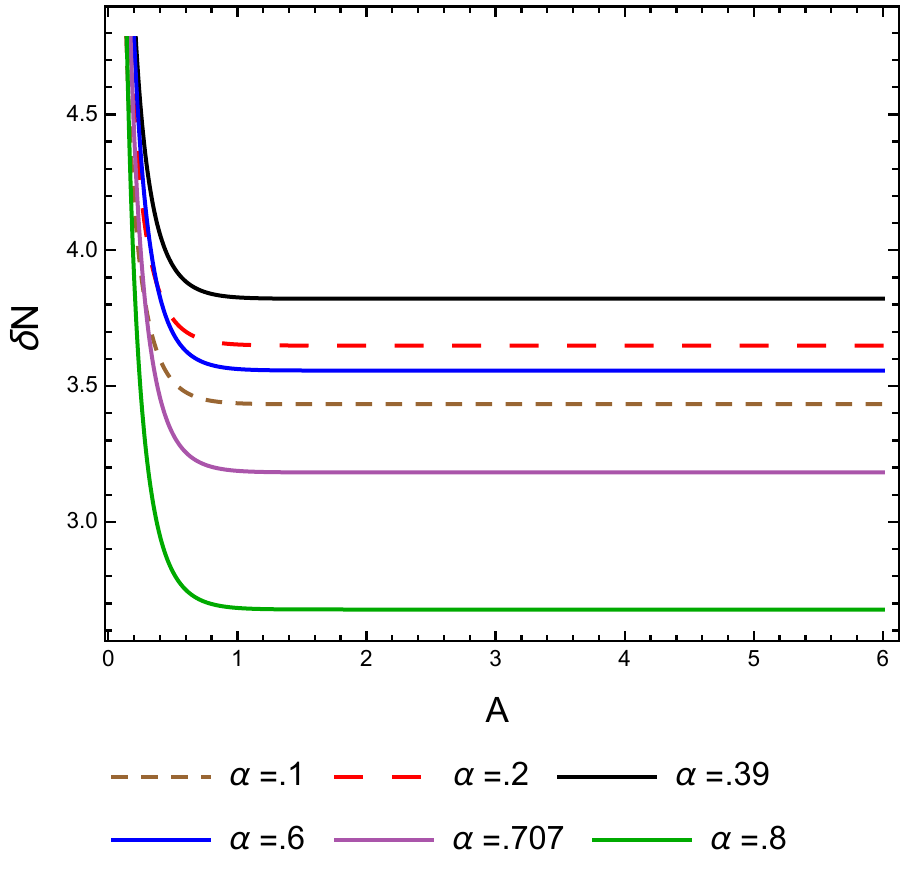}
					\caption{Parallelly accelerating detectors in $(1+1)$ dimensions: $\delta \text{N} $ VS $\mathrm{A}= \frac{\Delta E_A}{a_A}$ plot for different fixed values of the coefficient $\alpha $ with $\frac{\Delta E_B}{a_B} = 1$.}
					\label{fig1-new}
				\end{figure}

\begin{figure}[h!]
					\centering
					\includegraphics[width=0.45\textwidth]{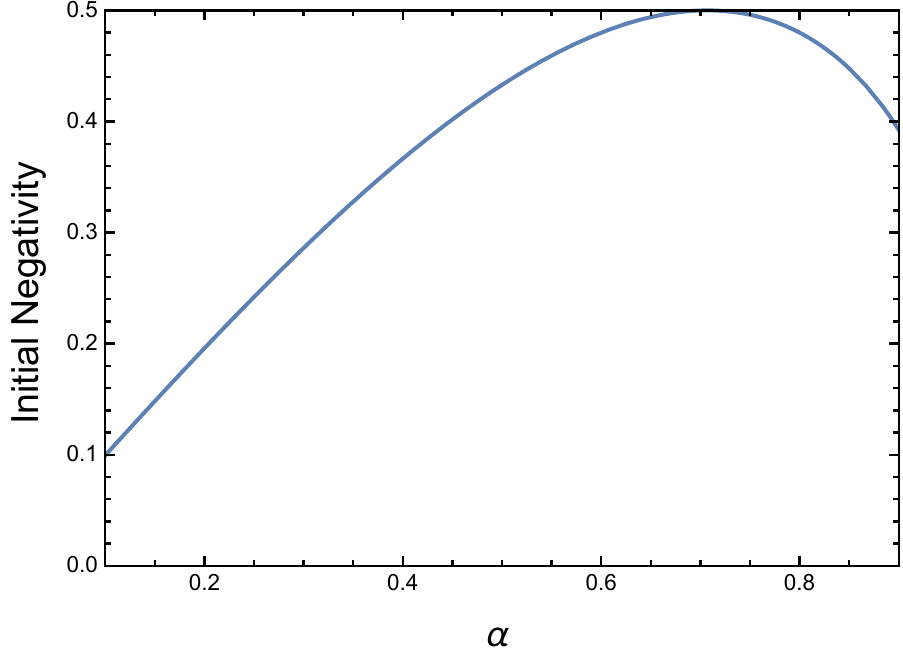}
					\caption{$\mathcal{N} = |-\alpha\gamma|$ VS $\alpha$ plot under the constraint $|\alpha|^2+|\gamma|^2=1$.}
					\label{Bibhas1}
				\end{figure}

\begin{figure}[h!]
					\centering
					\includegraphics[width=0.63\textwidth]{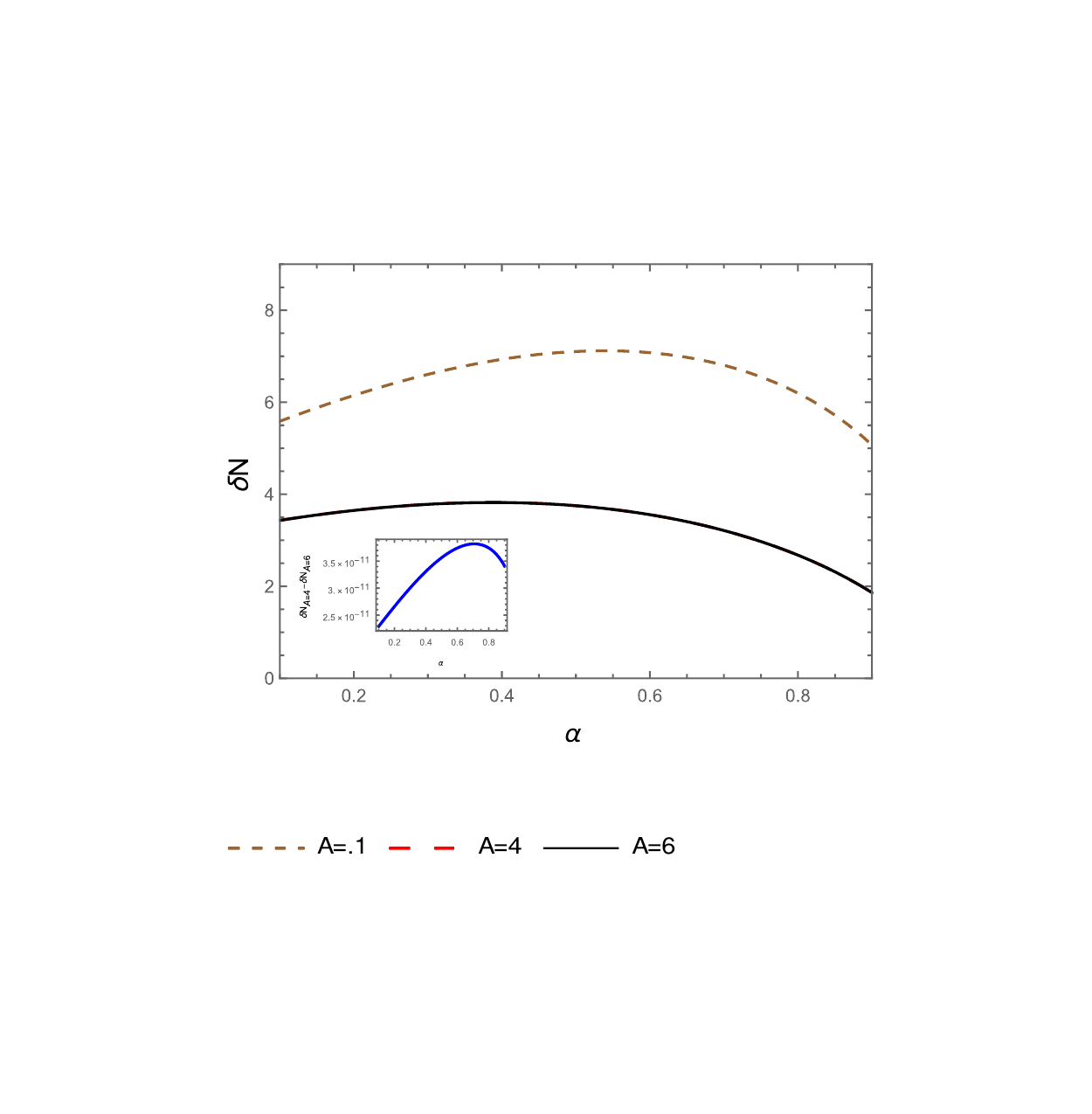}
					\caption{Parallelly accelerating detectors in $(1+1)$ dimensions: $\delta \text{N} $ VS $\alpha$ plot for different fixed values of  $\mathrm{A}$ with $\frac{\Delta E}{a_B} = 1$. Here the two different cases $\mathrm{A}=4$ and $\mathrm{A}=6$ are not distinguishable as the difference is very less.}
\label{fig2-new}
				\end{figure}

{\bf Parallel motion:}	
In Fig.  \ref{fig1-new} we plot $\delta \textrm{N}$ as a function of  $\mathrm{A}$  for different choices of $\alpha$ when two detectors are accelerating parallelly. Here  $\mathrm{A}$ is proportional to the inverse of the proper acceleration of our observer.
Note that $\delta\textrm{N}$ is always positive, which implies that due to the acceleration of the detectors	the initial entanglement must degrade. This degradation is higher when $\mathrm{A}$ is low (i.e. $a_A$ is high). Whereas the entanglement degradation becomes less as observers' acceleration decreases and ultimately this saturates after a particular low value of $a_A$.
Furthermore, it may be noted that the degradation also depends on the measure of initial entanglement. The initial entanglement is quantified by the values of $\alpha$ and $\gamma$ (See Fig. \ref{Bibhas1}). 
Fig. \ref{fig1-new} shows that as $\alpha$ increases the degradation of entanglement increases and this happens till a certain value of $\alpha$. After this particular value the situation reverses i.e. $\delta\textrm{N}$, with a positive value, decreases as $\alpha$ increases. For example, in Fig. \ref{fig1-new} we see that as  $\alpha$ increases from $0.1$ to $0.39$, the degradation increases. Further as it increases from $0.39$ the degradation decreases. This is visualized by plotting $\delta \textrm{N}$ as a function of $\alpha$ for a fixed $\mathrm{A}$ in Fig. \ref{fig2-new}, 
which supports the aforesaid observation.  Thus, the parallelly accelerating entangled detectors always suffer the degradation of entanglement due to the vacuum fluctuations in the Minkowski vacuum when the observer is accelerating.

\begin{figure}[h!]
					\centering
					\includegraphics[width=.45\textwidth]{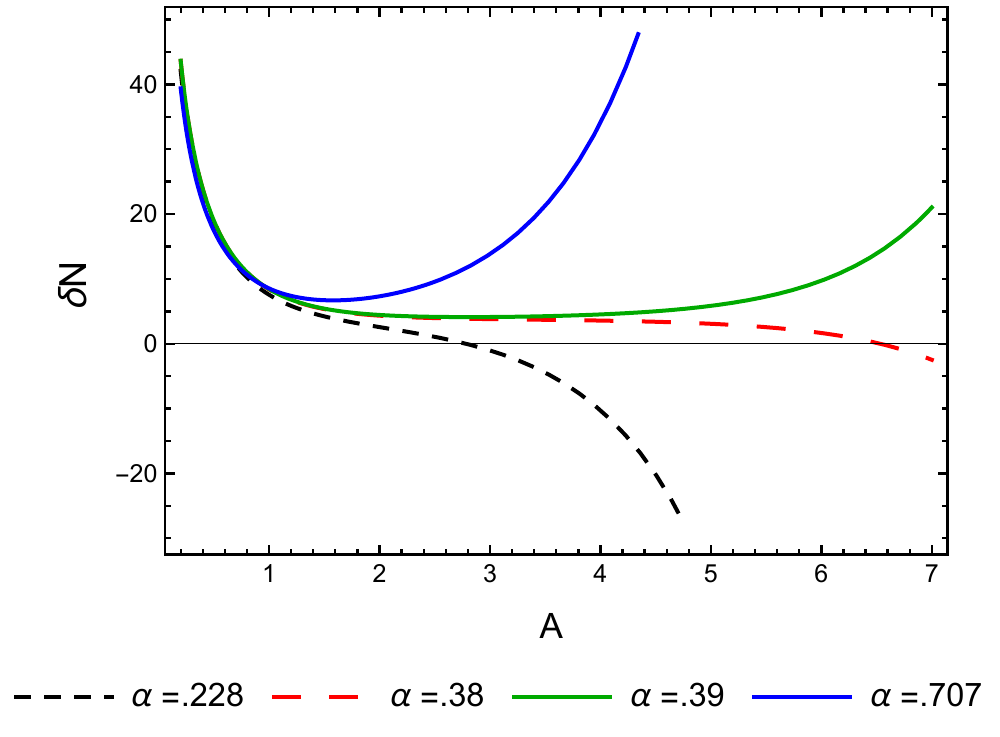}
					\caption{Anti-parallelly accelerating detectors in $(1+1)$ dimensions: $\delta \text{N} $ VS $\mathrm{A}= \frac{\Delta E_A}{a_A}$ plot for different fixed values of the coefficient $\alpha $ with $\frac{\Delta E_B}{a_B} = 1$.}
					\label{fig3-new}
				\end{figure}

{\bf Anti-parallel motion: }
From Fig. \ref{fig3-new} we see that when the detectors are accelerating anti-parallelly, depending on the values of $\mathrm{A}$ and $\alpha$, the system either suffers entanglement degradation or enjoys entanglement harvesting. 
When $\mathrm{A}$ is low, for all $\alpha$,degradation is high. Above a fixed certain value of $\alpha$ (e.g. which is here $\alpha \geq 0.39$), the detectors always suffer degradation in initial entanglement for the full range of $\mathrm{A}$. In this range of $\alpha$,  for low $\mathrm{A}$ degradation is more and then the degradation decreases as $\mathrm{A}$ increases. Interestingly further increase of $\mathrm{A}$ leads to increase in the degradation; i.e. the degradation becomes minimum at a certain value of $\mathrm{A}$ for any  $\alpha \geq 0.39$. From Fig. \ref{fig3-new}, it is to be noted that $\delta \text{N} $ is less (e.g. $\alpha = 0.707$ ) when $\mathrm{A}$ and $\mathrm{B}$ have comparable values. 

\begin{figure}[h!]
					\centering
					\includegraphics[width=.45\textwidth]{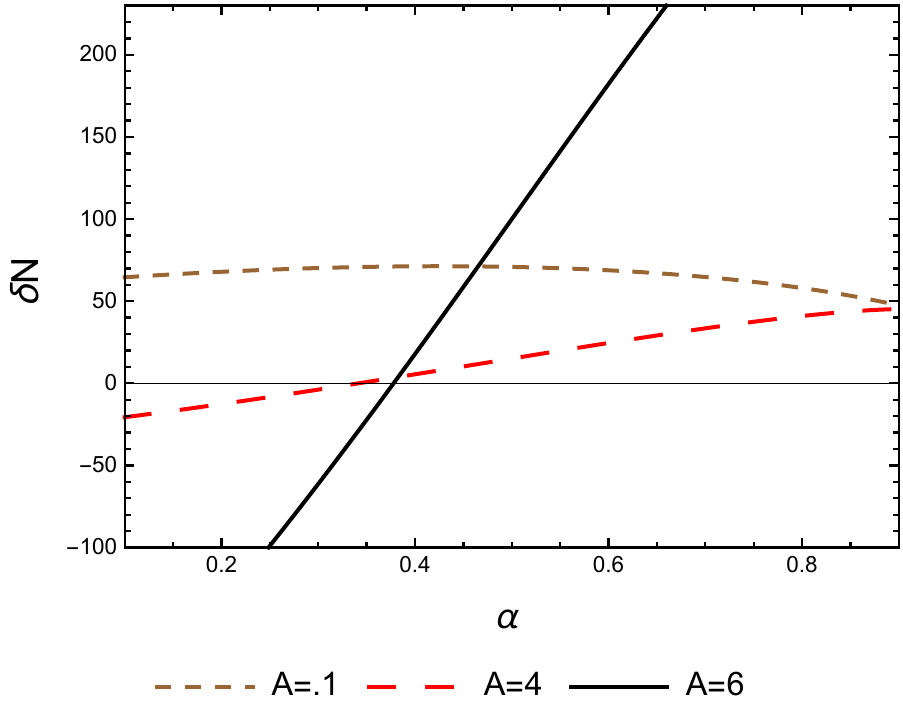}
					\caption{Anti-parallelly accelerating detectors in $(1+1)$ dimensions: $\delta \text{N} $ VS $\alpha$ plot for different fixed values of  $\mathrm{A}$ with $\frac{\Delta E}{a_B} = 1$.}
					\label{fig5-new}
				\end{figure}

On the other hand, for any $\alpha$ which is below the aforesaid critical value (i.e. when $\alpha \leq 0.38 $ here), the entanglement degradation decreases as $\mathrm{A}$ increases from a very low value. But after a certain value of $\mathrm{A}$ (where $\delta\textrm{N}=0$), we have now entanglement harvesting as $\delta\textrm{N}$ becomes negative. This harvesting increases as observers' acceleration decreases more. The value of $\mathrm{A}$ after which we can have entanglement harvesting increases as $\alpha$ increase as long as the chosen $\alpha$ is within the above mentioned range.
The same feature is depicted in Fig. \ref{fig5-new} as well 
where we plot $\delta\textrm{N}$ as a function of $\alpha$. We observe that $\delta\textrm{N}$ can be negative within a range of low values of $\alpha$ for certain fixed high values of $\mathrm{A}$. 
Thus when the detectors are not maximally entangled (i.e. for $\alpha\neq 1/\sqrt{2}\simeq 0.707$) we can have entanglement harvesting within a specific ranges of system parameters.

\begin{figure}[h!]
					\centering
					\includegraphics[width=.61\textwidth]{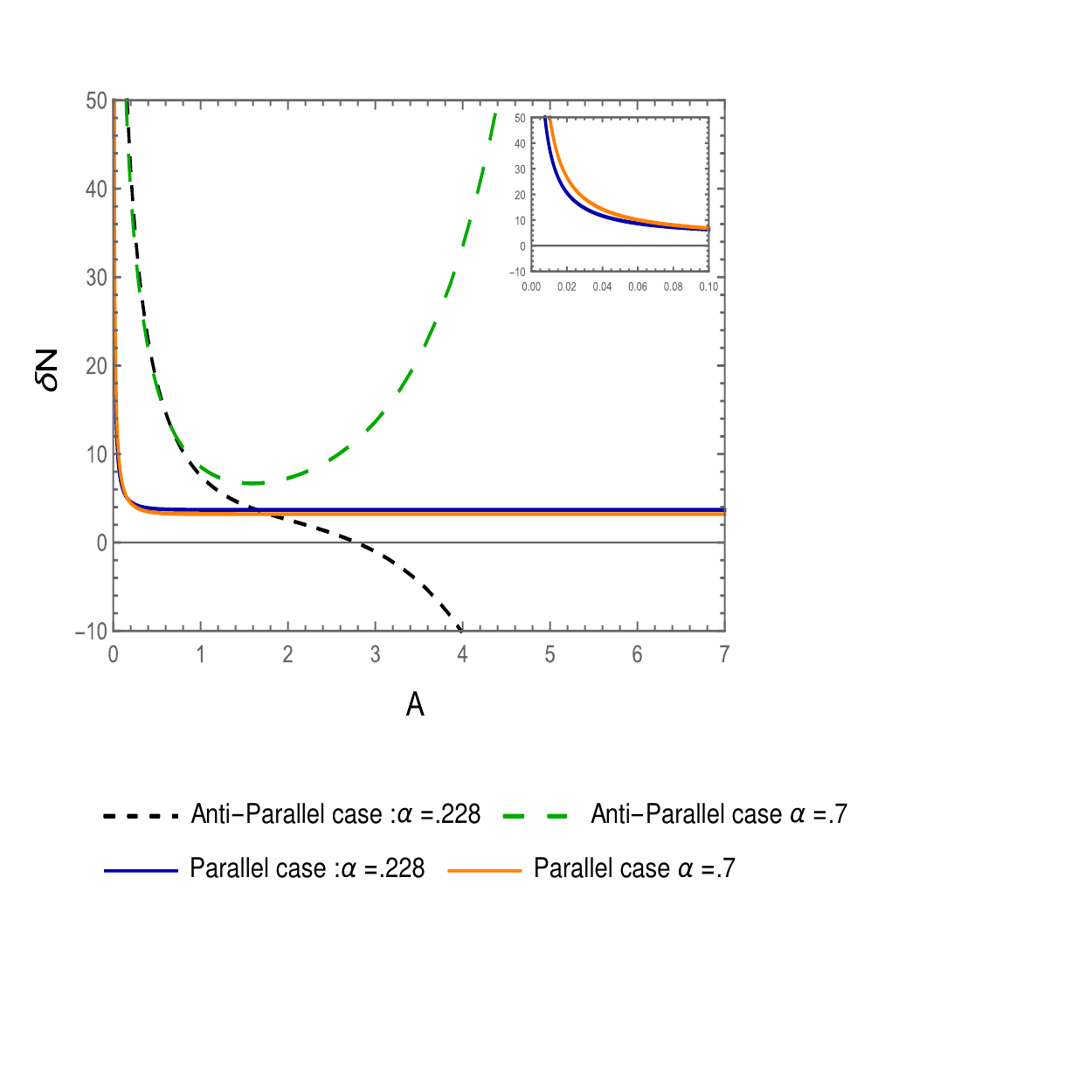}
					\caption{Parallelly and anti-parallelly accelerating detectors in $(1+1)$ dimensions:  $\delta \text{N} $ VS $\mathrm{A}$ plot for different fixed values of the coefficient $\alpha $ with $\frac{\Delta E}{a_B} = 1$.}
					\label{fig7-new}
				\end{figure}

{\bf Comparative study: }	
  In the above discussion we found that for both parallel and anti-parallel cases, depending on the values of the parameters, there exists entanglement degradation. Therefore, now we aim to investigate in which situation the degradation is more for the same values of the parameters.
 Fig. \ref{fig7-new}, shows that 
for very low $\mathrm{A}$, degradation is more for anti-parallel case compared to the parallel situation for all fixed $\alpha$. On the other hand, the rate of its change is more in the case of parallel motion. For higher range of $A$ degradation keeps dominating in anti-parallel case as well, but that happens above a certain choice of $\alpha$ (like $\alpha\geq 0.39$ here). Whereas for low values of $\alpha$ (like $\alpha < 0.39$ here), the situation is different, cause remember that in this range of $\alpha$, after a particular value of $\mathrm{A}$ entanglement harvesting is possible in the anti-parallel motion. Now, if we keep our attention within the range of $\mathrm{A}$ where only entanglement degradation is happening for both parallel as well as anti-parallel motions, then we observe that after a particular value of $\mathrm{A}$ the degradation in parallel case dominates over anti-parallel motion (e.g  $\mathrm{A}$ $\approx$ 2.1 to 3 ). Here the change in degradation is almost negligible for parallel case whereas same for anti-parallel motion is not so. 
			
\begin{figure}[h!]
					\centering
					\includegraphics[width=.45\textwidth]{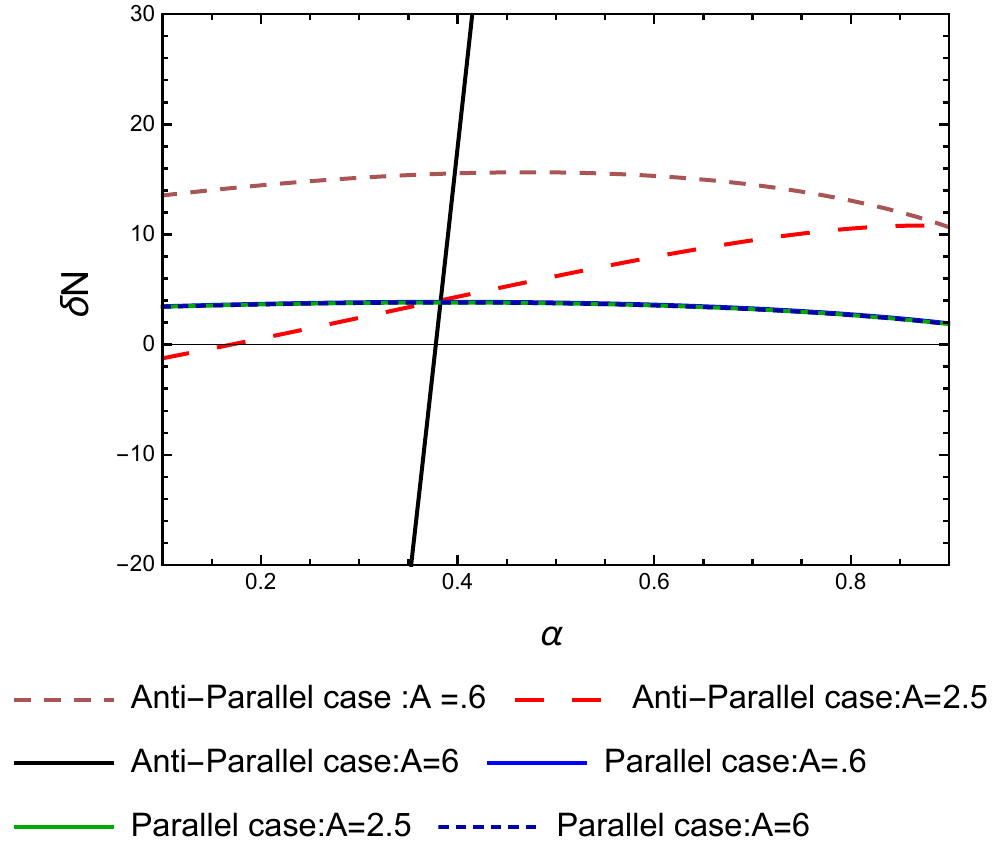}
					\caption{Parallelly and anti-parallelly accelerating detectors in $(1+1)$ dimensions:  $\delta \text{N} $ VS $\alpha$ plot for different fixed values $\mathrm{A}$ with $\frac{\Delta E}{a_B} = 1$. All the parallel plots are very close to one another like in the previous figures.}
					\label{fig6-new}
				\end{figure}

				In Fig. \ref{fig6-new} we further bolstered the above observations by plotting $\delta\textrm{N}$ as a function of $\alpha$ for both parallel and anti-parallel cases by considering same fixed value of $\mathrm{A}$. 
For a fixed low value of $\mathrm{A}$ (here $\mathrm{A}=.6$ ), $\delta\textrm{N}$ for anti-parallel motion is greater than that for parallel motion. This happens over the full range of $\alpha$. The same is not true for higher fixed values of $\mathrm{A}$. In this case, within a certain range of $\alpha$ starting from the low value of it (within which entanglement decrement in the anti-parallel case can happen), degradation in entanglement is more in parallel motion, but beyond that range of $\alpha$ the situation completely reversed.


			\subsubsection{$(1+3)$-dimensions}
\begin{figure}[h!]
				\centering
				\includegraphics[width=.45\textwidth]{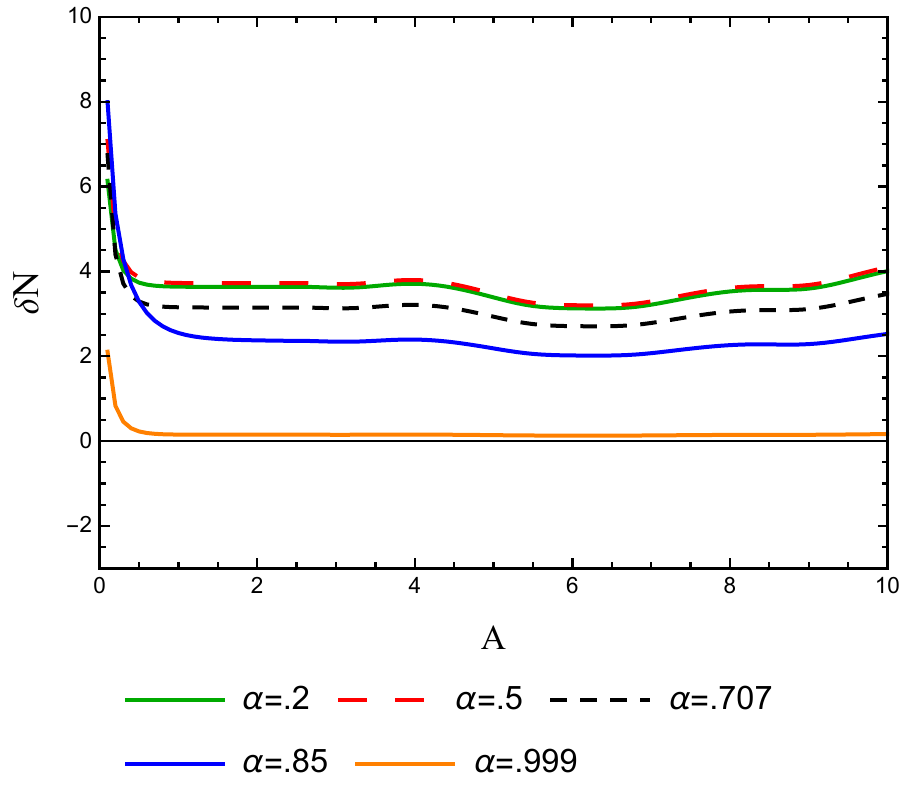}
				\caption{ Parallelly accelerating detectors in $(1+3)$ dimensions :  $\delta \text{N} $ VS $\mathrm{A}= \frac{\Delta E_A}{a_A}$ plot for different fixed values of the coefficient $\alpha $ with $\frac{\Delta E_B}{a_B} = 1$. }
				\label{fig8-new}
			\end{figure}

\begin{figure}[h!]
				\centering
				\includegraphics[width=.45\textwidth]{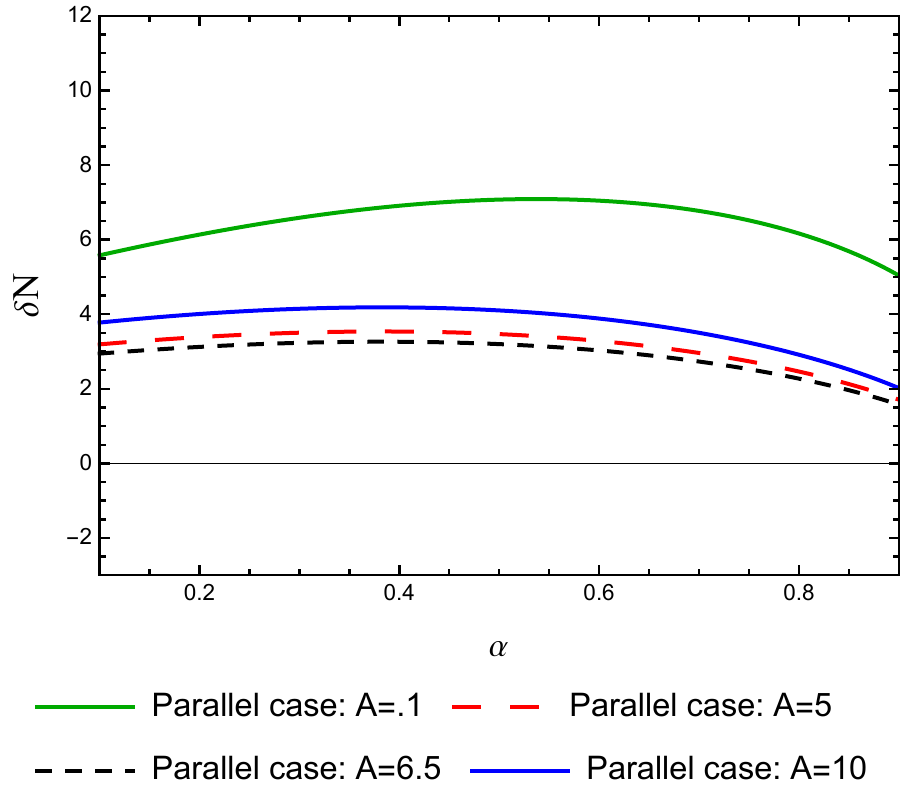}
				\caption{ Parallelly accelerating detectors in $(1+3)$D: $\delta \text{N} $ VS $\alpha$ plot for different fixed values of A with $\frac{\Delta E_B}{a_B} = 1$. }
				\label{fig9-new}
			\end{figure}

	\textbf{Parallel motion:}
			In Fig. \ref{fig8-new}	we plot $\delta\text{N}$ as a function of A for different choices of $\alpha$. 
Note that $\delta\text{N}$ is always positive which implies that due to the acceleration of the detectors the initial entanglement has degraded just like in (1+1) dimensional case. Here also the degradation is higher when $\mathrm{A}$ is very low. As $\mathrm{A}$ increases, unlike (1+1)-dimensional feature, degradation has a variation and does not saturate after a certain $\mathrm{A}$. It is interesting to note that for a large value of fixed $\alpha$  (e.g. see $\alpha = 0.999 $ in Fig. \ref{fig8-new}) this variation almost dies out and saturation starts appearing. Similar to the (1+1) dimensional case, here also the entanglement degradation depends on the initial entanglement. The entanglement degradation first increases as one increases $\alpha$ up to a certain value of it, and then starts to decrease with the further increase in $\alpha$. This is further clarified in Fig. \ref{fig9-new}. 
Also from this figure, the variation in nature of degradation with change in $\mathrm{A}$ is observed. Thus the parallelly accelerating entangled detectors always suffer degradation.

\begin{figure}[h!]
					\centering
					\includegraphics[width=.45\textwidth]{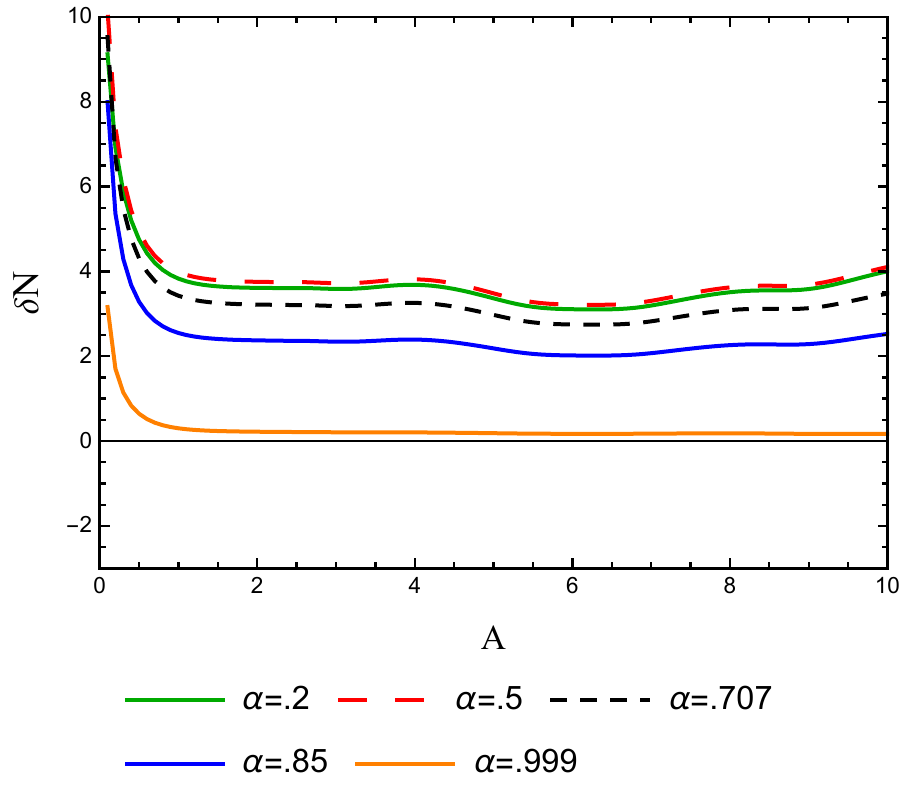}
					\caption{ Anti-parallelly accelerating detectors in $(1+3)$ dimensions: $\delta \text{N} $ VS $\mathrm{A}= \frac{\Delta E_A}{a_A}$ plot for different fixed values of the coefficient $\alpha $ with $\frac{\Delta E_B}{a_B} = 1$. }
					\label{fig10-new}
				\end{figure}

\begin{figure}[h!]
					\centering
					\includegraphics[width=.45\textwidth]{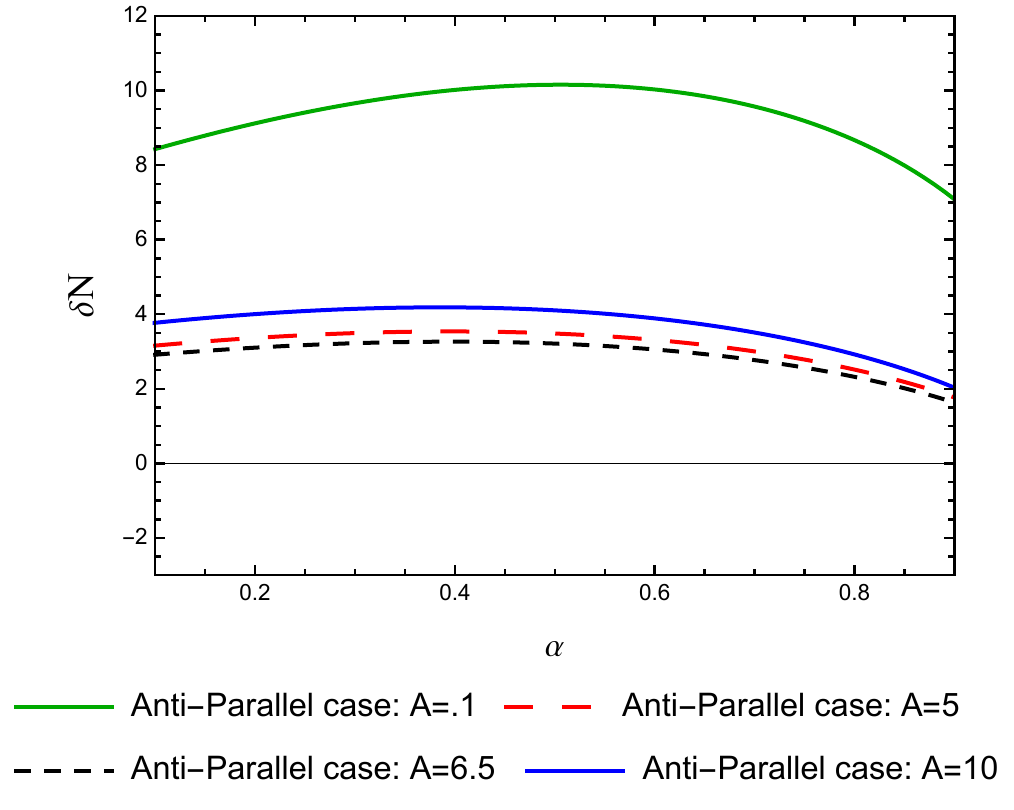}
					\caption{Anti-parallelly accelerating detectors in $(1+3)$ dimensions: $\delta \text{N} $ VS $\alpha$ plot for different fixed values of $\mathrm{A}$ with $\frac{\Delta E_B}{a_B} = 1$. }
					\label{fig11-new}
				\end{figure}

	\textbf{Anti-Parallel motion:}	
	We plot $\delta\textrm{N}$ as a function of $\mathrm{A}$ in Fig. \ref{fig10-new} for different fixed values of $\alpha$. Interestingly, unlike $(1+1)$ dimensional case, the pair of anti-parallelly accelerating detectors in $(1+3)$ dimension suffers only degradation of entanglement. This is because, for all values of $\mathrm{A}$ and $\alpha$   
here $\delta\text{N}$ is always positive. Degradation is high for low $\mathrm{A}$ and as $\mathrm{A}$ increases the degradation varies for a wide range of values of $\alpha$. When $\alpha$ approaches to a very large value, then $\delta\text{N}$ tends to saturate as one increases $\mathrm{A}$  (e.g see the curve for $\alpha = 0.999$).  
Here also we see that the nature of degradation depends on the initial entanglement. This is studied by plotting $\delta\text{N}$ as a function of $\alpha$ in Fig. \ref{fig11-new} for different fixed $\mathrm{A}$. 
For fixed a $\mathrm{A}$ the degradation first increases and then after a certain $\alpha$, it decreases. The nature obtained is similar to that in the parallel case.

\textbf{Comparative study:}
We have seen in this section that for both parallel and anti-parallel cases, there exists entanglement degradation. We will now compare the two scenarios and see for which set of parameters the degradation is more. 
  
\begin{figure}[h!]
					\centering
					\includegraphics[width=.45\textwidth]{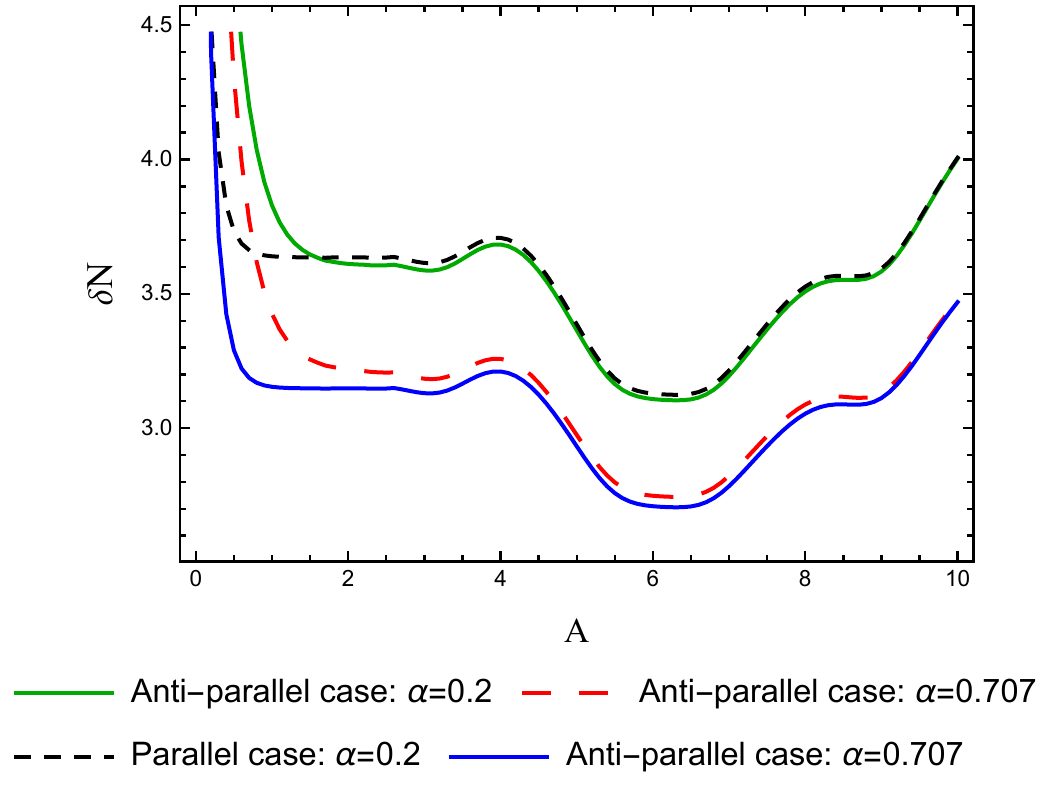}
					\caption{ Anti-parallelly accelerating and parallelly accelerating detectors in $(1+3)$ dimensions : $\delta \text{N} $ VS $\mathrm{A}= \frac{\Delta E_A}{a_A}$ plot for different fixed values of the coefficient $\alpha $ with $\frac{\Delta E_B}{a_B} = 1$.}
					\label{fig12-new}
				\end{figure}

\begin{figure}[h!]
					\centering
					\includegraphics[width=.45\textwidth]{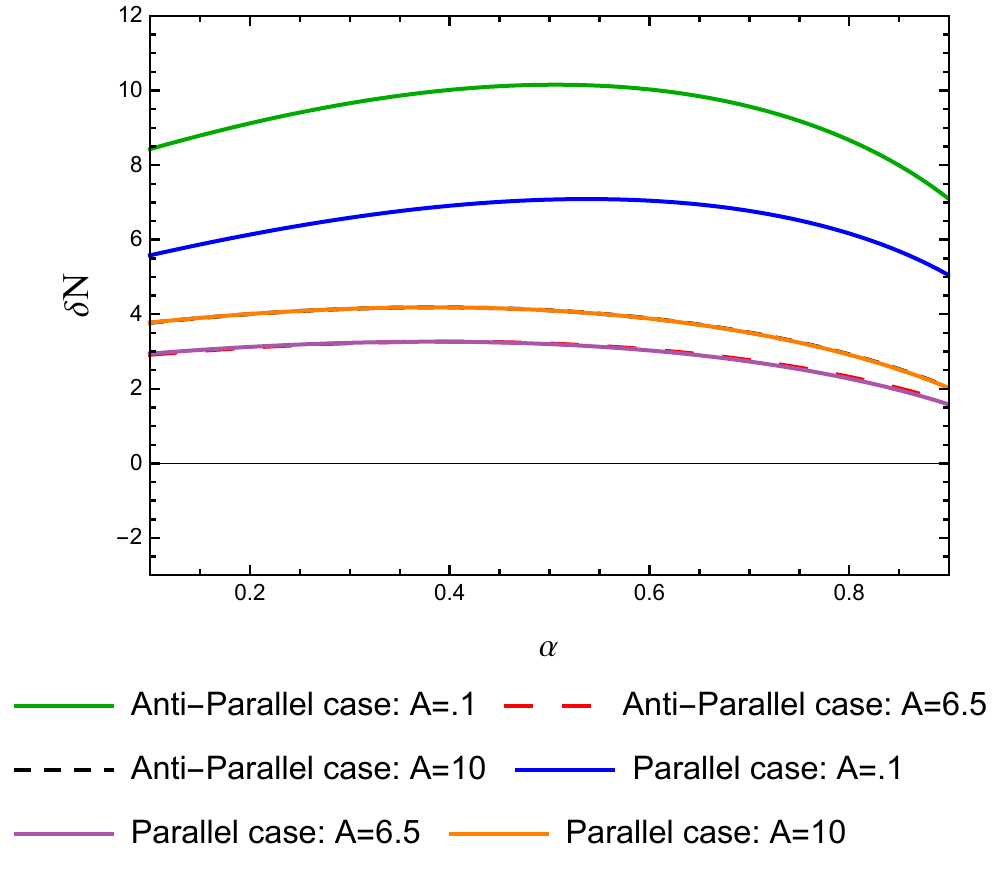}
					\caption{ Parallelly accelerating  and Anti-parallelly accelerating detectors in $(1+3)$ dimensions: $\delta \text{N} $ VS $\alpha$ plot for different fixed values of $\mathrm{A}$ with $\frac{\Delta E_B}{a_B} = 1$. }
					\label{fig13-new}
				\end{figure}

From Fig. \ref{fig12-new} 
it is observed that for very low $\mathrm{A}$ degradation for anti-parallel case is more than the degradation for parallel case. Although as $\mathrm{A}$ increases the degradation for parallel and antiparallel for a fixed value of $ \alpha $ is almost same, for $ \alpha =0.2 $ the degradation for parallel case is a bit more compared to that for anti-parallel case. This also being verified by plotting $\delta\textrm{N}$ as a function of $\alpha$ for both parallel and anti-parallel situations for same fixed values of $\mathrm{A}$ in Fig.  \ref{fig13-new}.
It is observed that for a very low $\mathrm{A}$ degradation for parallel and anti-parallel case has considerable difference with the latter having more degradation over the full range of $\alpha$. While for higher fixed values of $\mathrm{A}$ the degradations for the parallel and antiparallel cases are the same for the entire range of $\alpha$ values.

 The above analysis reveals one striking difference between the $(1+1)$ dimensional and $(1+3)$ dimensional cases. For the former, both entanglement generation, as well as entanglement degradation, is observed for a set of parameters in the anti-parallel case. Whereas for the latter, only entanglement degradation takes place. A common feature between them is that for low $\mathrm{A}$ both suffer more degradation for the anti-parallel case.

\subsection{Thermal field}

\subsubsection{$(1+1)$-dimensions}	

\begin{figure}[h!]
	\centering
	\includegraphics[width=.45\textwidth]{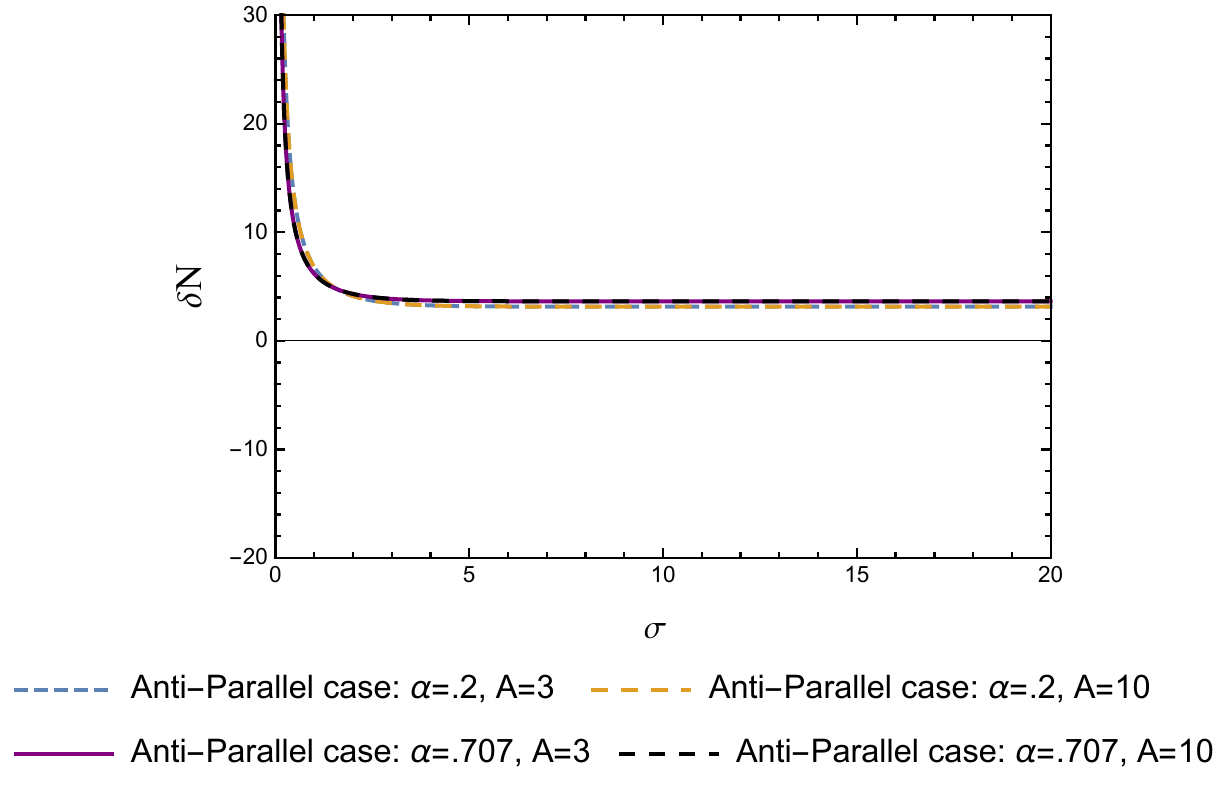}
	\caption{parallelly accelerating detectors in $(1+1)$ dimensions with thermal background: $\delta \text{N} $ VS $ \sigma = \beta \Delta E$ plot for different fixed values of the coefficient $\alpha $ and $\mathrm{A}$ with $\frac{\Delta E_B}{a_B} = 1$.}
	\label{figsigmasame}
\end{figure}

\begin{figure}[h!]
	\centering
	\includegraphics[width=.45\textwidth]{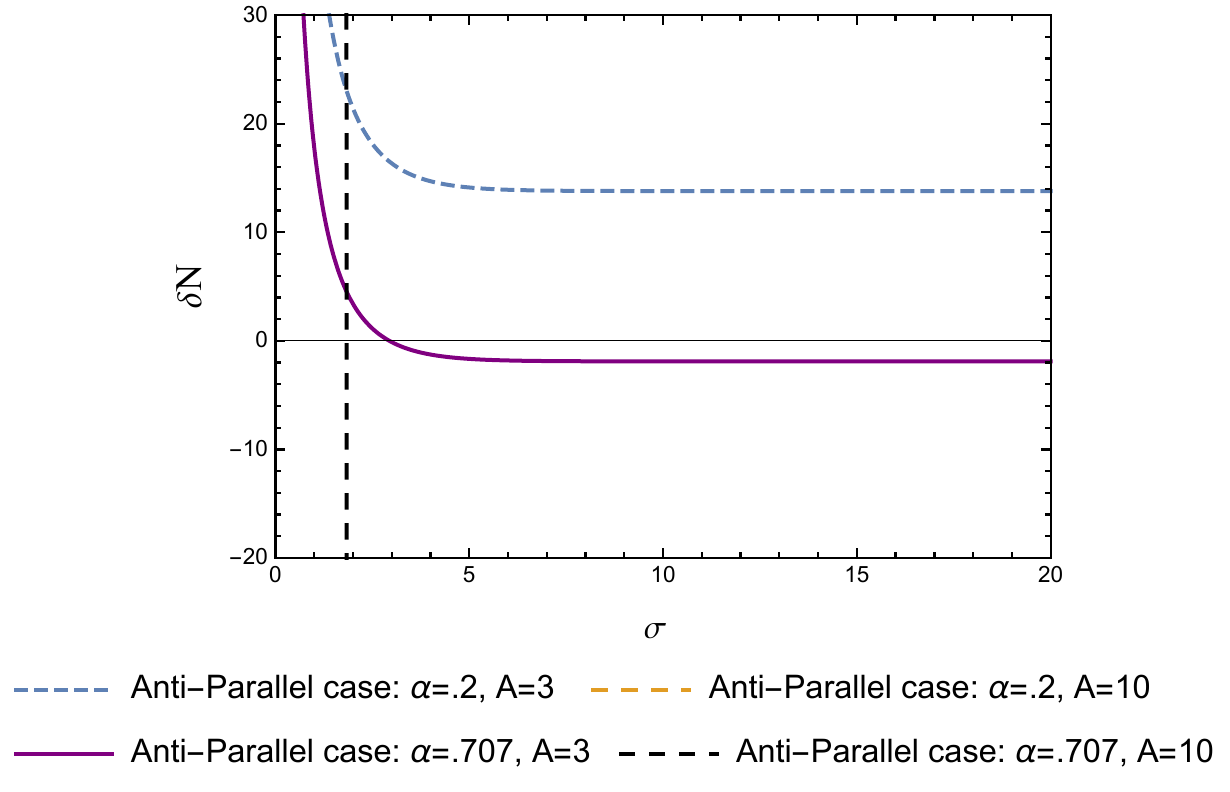}
	\caption{Anti-parallelly accelerating detectors in $(1+1)$ dimensions with thermal background: $\delta \text{N} $ VS $  \sigma = \beta \Delta E$ plot for different fixed values of the coefficient $\alpha $ and $\mathrm{A}$ with $\frac{\Delta E_B}{a_B} = 1$.}
	\label{figsigmaopp}
\end{figure}

{\bf Parallel and Anti-parallel motion:} 
For thermal case in (1+1) dimension, at a fixed temperature (i.e. with fixed $\sigma$), the variation of  $\delta \text{N}$ with $\mathrm{A}$ and $\alpha$ is similar in nature to what have been observed in (1+1) dimensional non-thermal case. Therefore, this investigation will not be repeated here. Hence, we only study how the $\delta \text{N}$ varies with $\sigma$. In Fig. \ref{figsigmasame}, we see variation of $\delta \text{N} $ for parallel motion.
We observe that as $\sigma$ increases (i.e. decrement in background field temperature), the entanglement degradation decreases and finally saturates. Thus, presence of a background temperature decreases entanglement. 
Anti-parallel case has been studied in Fig. \ref{figsigmaopp}.
For certain values of $\mathrm{A}$ and $\alpha$, we notice, as $\sigma $ increases entanglement degradation stops and entanglement harvesting begins.

{\bf Comparative study:}
In the above discussion, we have seen that for different ranges of the parameters there is entanglement degradation or harvesting. Here we compare the parallel and antiparallel cases in the thermal background to see for which set of parameters there is more degradation as a function of the temperature of fields.

\begin{figure}[h!]
	\centering
\includegraphics[width=.45\textwidth]{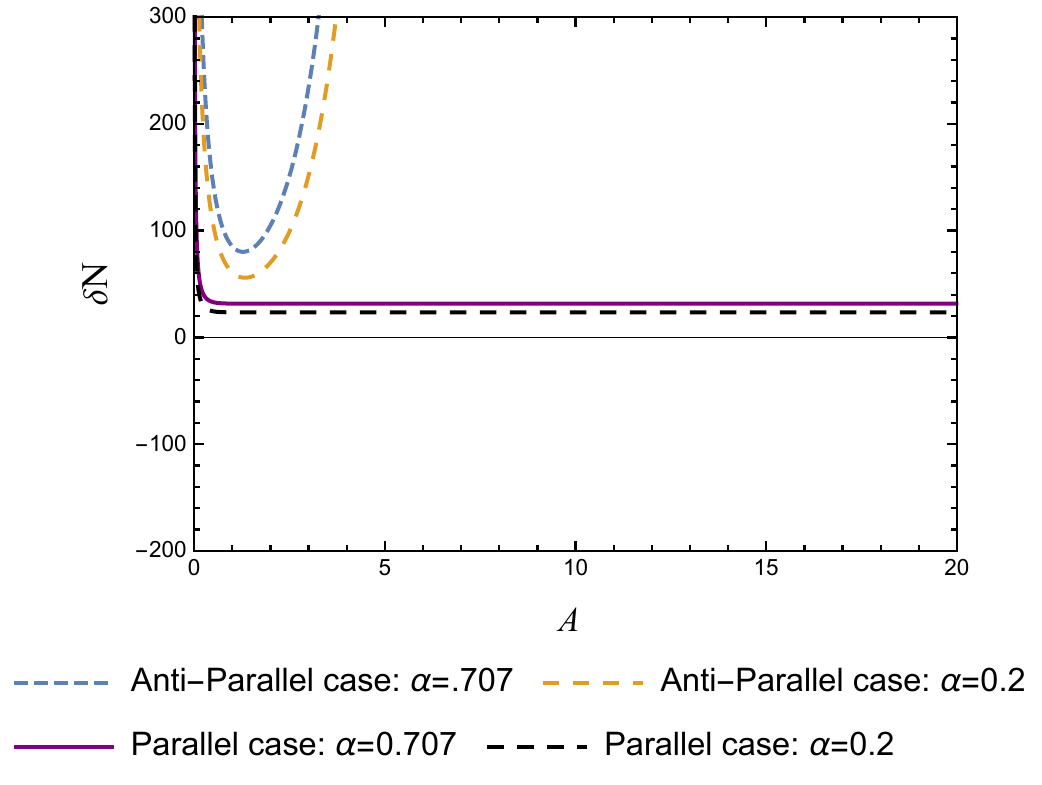}
	\caption{Parallel and Anti-parallel motions in $(1+1)$ dimensions with thermal background:  $\delta\textrm{N}$ Vs $\mathrm{A}$ plot with $\frac{\Delta E}{a_B} = 1$, the $\sigma=0.2$.}
	\label{fig26-new}
\end{figure}

\begin{figure}[h!]
	\centering
\includegraphics[width=.45\textwidth]{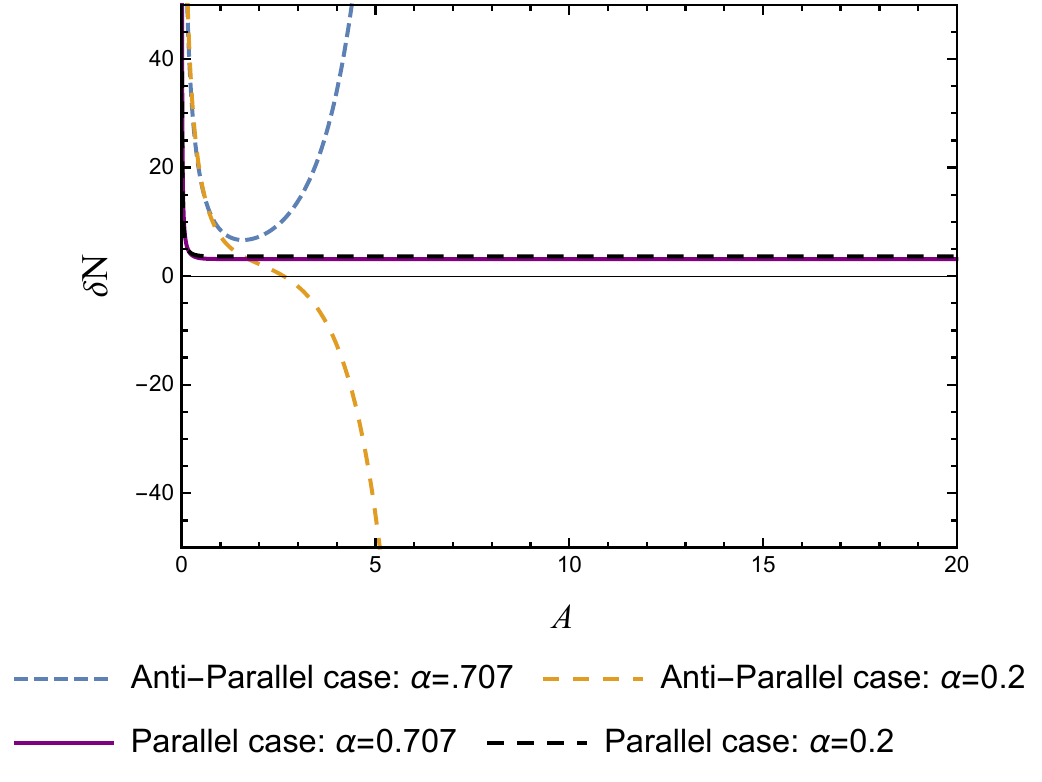}
	\caption{Parallel and Anti-parallel motions in $(1+1)$ dimensions with thermal background:  $\delta\textrm{N}$ Vs $\mathrm{A}$ plot with $\frac{\Delta E}{a_B} = 1$, the $\sigma=200$.}
	\label{fig27-new}
\end{figure}

In Fig. \ref{fig26-new} 
we observe that when  $\sigma=0.2$ (i.e., at high temperature) there is entanglement degradation both for parallel and anti-parallel motions. The degradation for anti-parallel case always dominates the parallel case. For $\sigma=200 $ (i.e. at low temperature) and at high values of $\alpha$ (see Fig. \ref{fig27-new}) 
degradation within anti-parallel case wins over that of parallel motion over that full range of $\mathrm{A}$. Whereas at this low temperature, when $\alpha$ is small, although the degradation for antiparallel case dominates for very low $\mathrm{A}$, but after a particular value of $\mathrm{A}$ within the certain range of it (e.g. here $\mathrm{A}\simeq 2$ to $\mathrm{A}\simeq3$), the degradation for the parallel case dominates. On further increasing in $\mathrm{A}$, entanglement harvesting starts for anti-parallel motion. This nature is exactly similar to that observed for non-thermal case (see Fig. \ref{fig5-new}).

\begin{figure}[h!]
	\centering
\includegraphics[width=.45\textwidth]{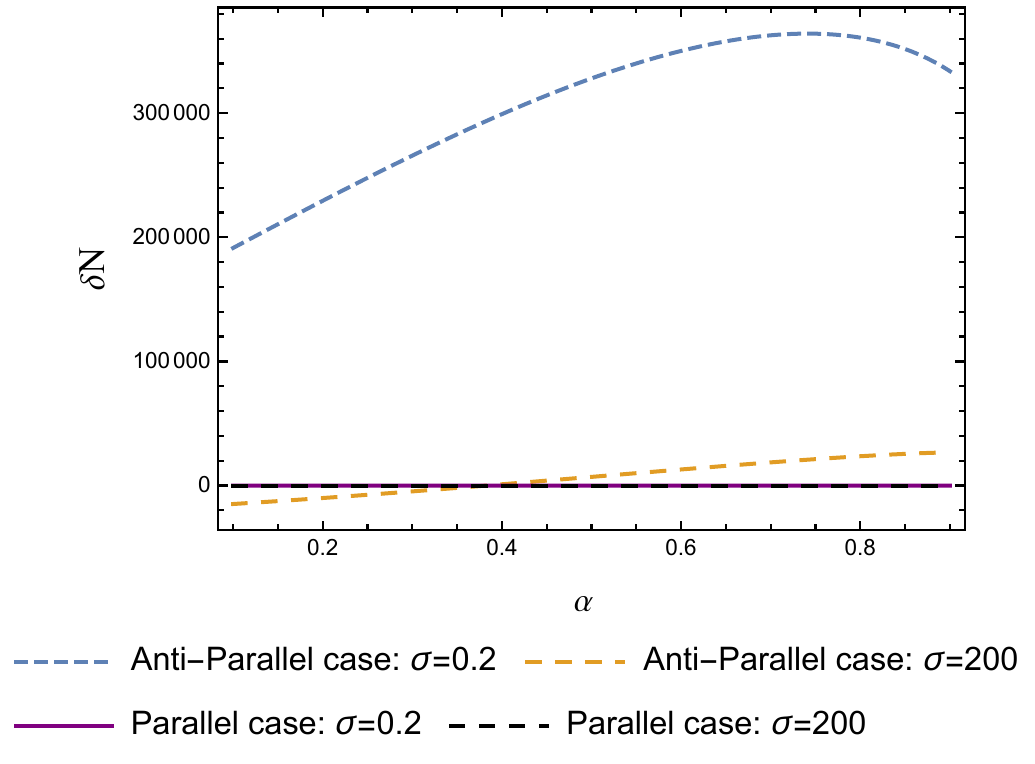}
	\caption{Parallel and Anti-parallel motions in $(1+1)$ dimensions with thermal background:  $\delta\textrm{N}$ Vs $\alpha$ plot with $\frac{\Delta E}{a_B} = 1$, $\mathrm{A}=3$.}
   \label{fig28-new}
\end{figure}

\begin{figure}[h!]
	\centering
\includegraphics[width=.45\textwidth]{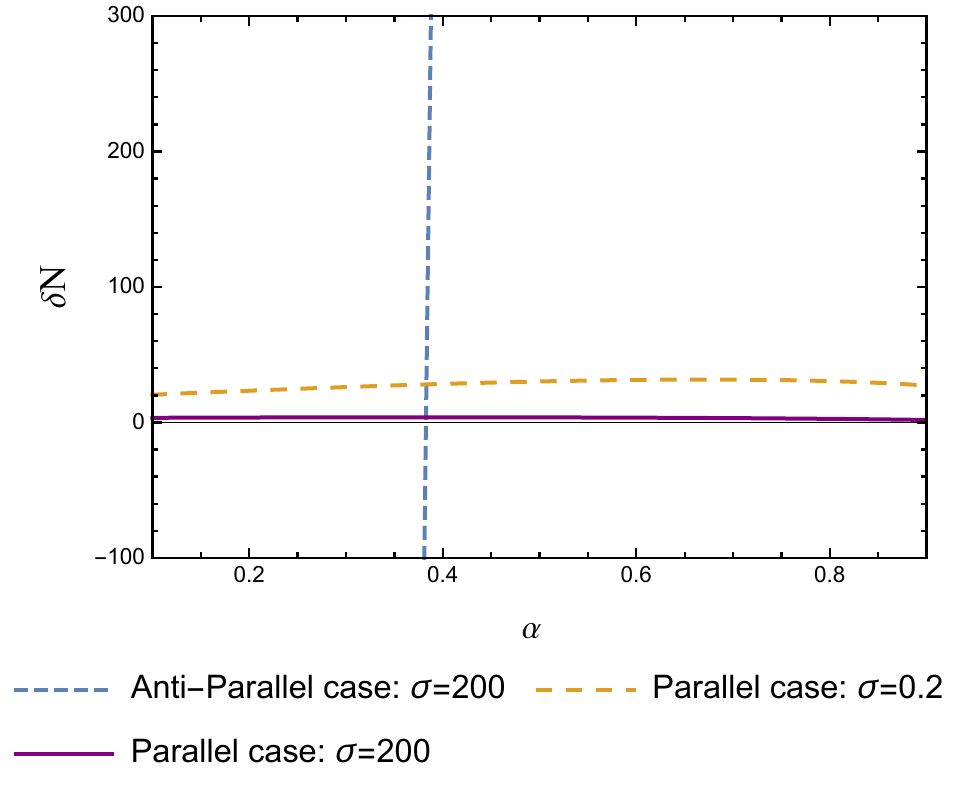}
	\caption{Parallel and Anti-parallel motions in $(1+1)$ dimensions with thermal background:  $\delta\textrm{N}$ Vs $\alpha$ plot with $\frac{\Delta E}{a_B} = 1$,  $\mathrm{A}=10$. }
   \label{fig29-new}
\end{figure}

Next from Fig. \ref{fig28-new} and Fig. \ref{fig29-new} we observe the variation of $\delta \textrm{N}$ with $\alpha$. 
We notice that when $\mathrm{A}=3$ and $\sigma=0.2$, degradation for antiparallel case is more than the degradation for parallel case for all possible values of $\alpha$. But for $\sigma=200$ and low values of $\alpha$, for antiparallel case there is entanglement generation, but as $\alpha $ increases the degradation increases and dominates over the degradation for parallel case. For $\mathrm{A}=10$ also similar behavior is noticed. One noticeable feature is that when $a_A$ is lower, for anti-parallel motion the entanglement harvesting is more prominent. The above observed trend is similar to what is seen in the non-thermal case (see Fig. \ref{fig6-new}).

\begin{figure}[h!]
				\centering
				\includegraphics[width=.45\textwidth]{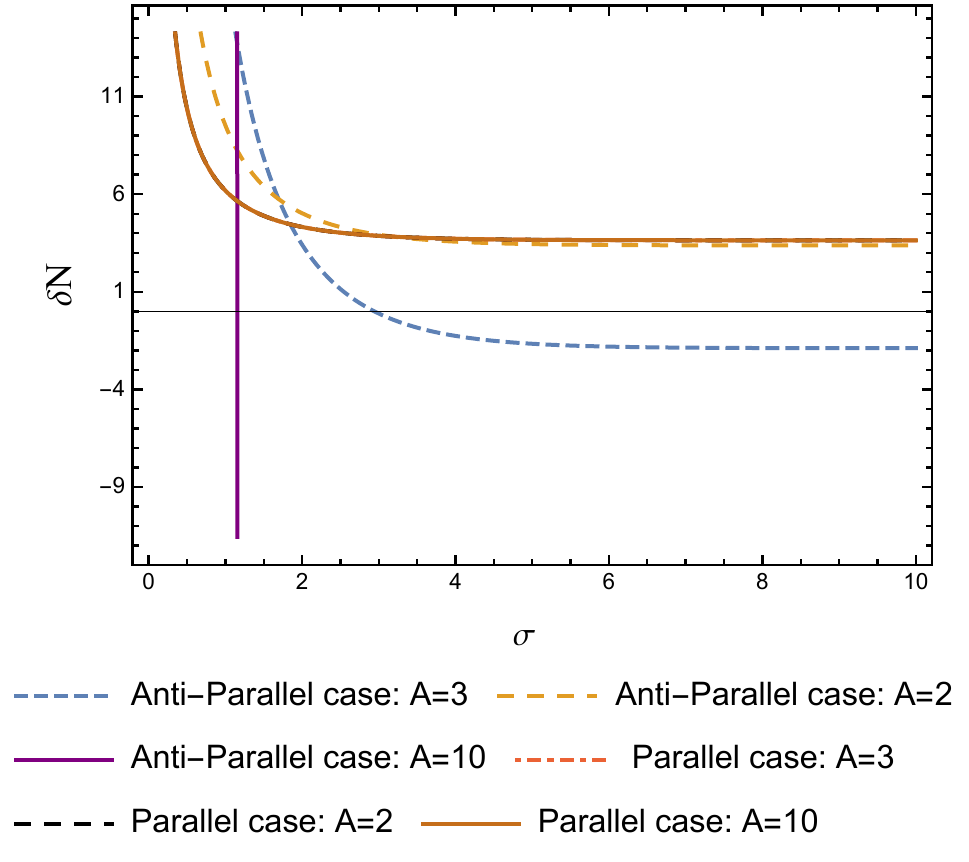}
				\caption{Parallel and Anti-parallel motions in $(1+1)$ dimensions with thermal background:  $\delta\textrm{N}$ Vs $\sigma$ plot with $\frac{\Delta E}{a_B} = 1$, $\alpha=0.2$.}
				\label{fig30-new}
			\end{figure}

\begin{figure}[h!]
				\centering
				\includegraphics[width=.45\textwidth]{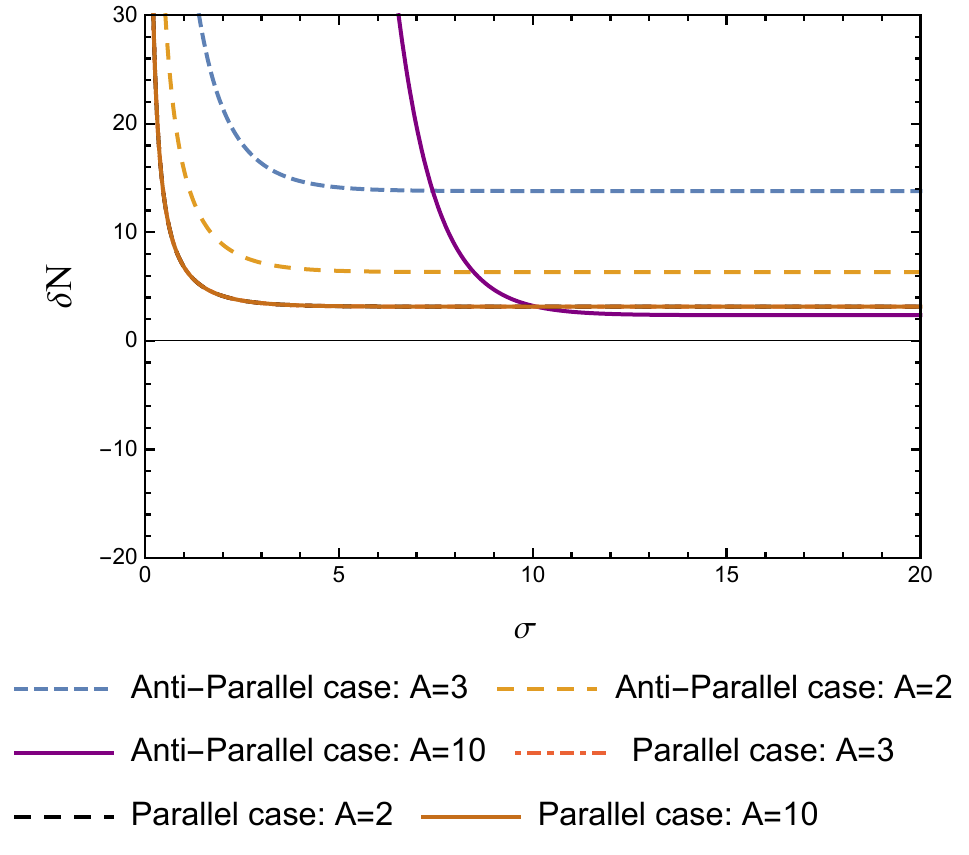}
				\caption{Parallel and Anti-parallel motions in $(1+1)$ dimensions with thermal background:  $\delta\textrm{N}$ Vs $\sigma$ plot with$\frac{\Delta E}{a_B} = 1$, $\alpha=0.707$.}
				\label{fig31-new}
			\end{figure}

\begin{figure}[h!]
				\centering
				\includegraphics[width=.45\textwidth]{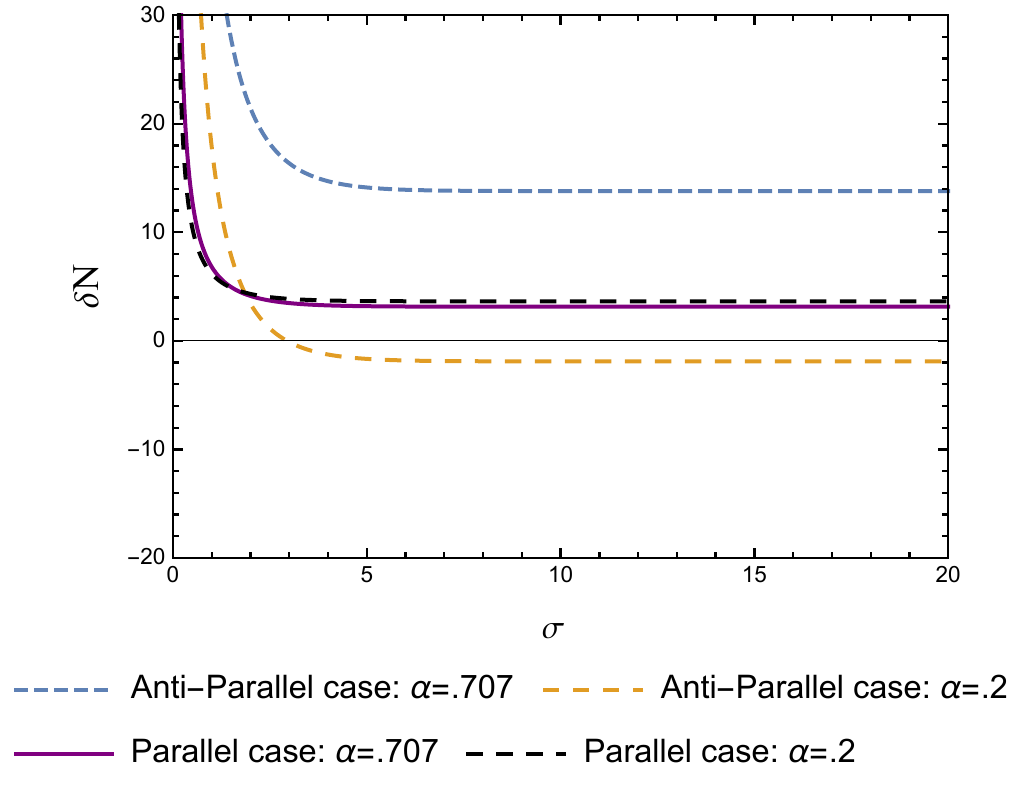}
				\caption{Parallel and Anti-parallel motions in $(1+1)$ dimensions with thermal background:  $\delta\textrm{N}$ Vs $\sigma$ plot with $\frac{\Delta E}{a_B} = 1$, $\mathrm{A}=3$.}
				\label{fig32-new}
			\end{figure}

\begin{figure}[h]
				\centering
				\includegraphics[width=.45\textwidth]{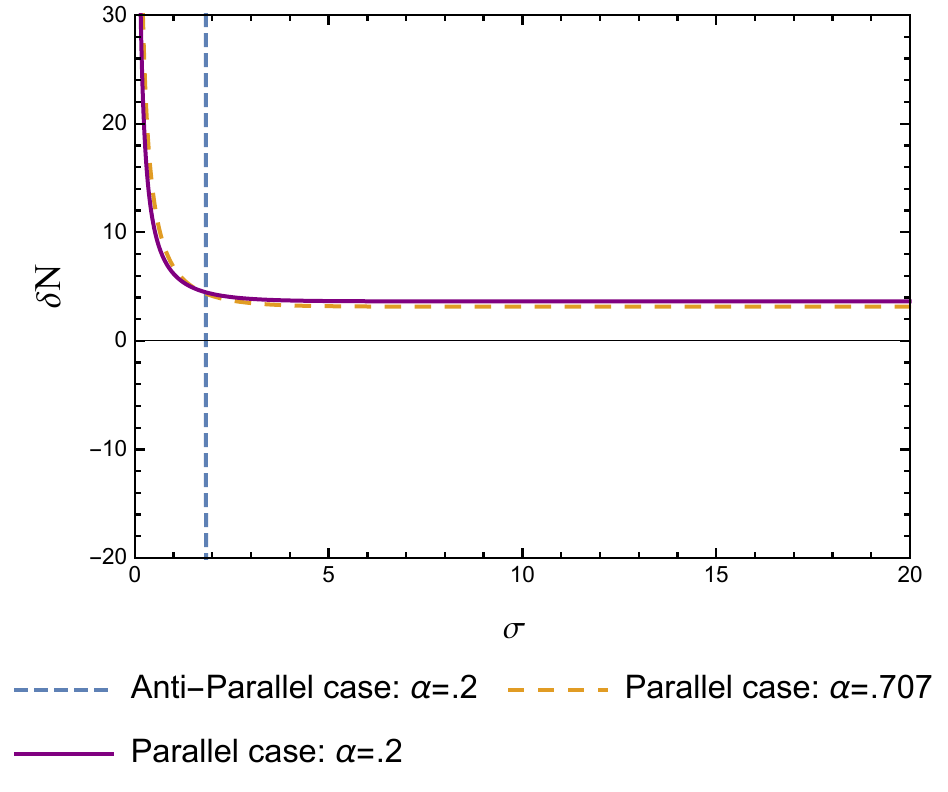}
				\caption{Parallel and Anti-parallel motions in $(1+1)$ dimensions with thermal background:  $\delta\textrm{N}$ Vs $\sigma$ plot with $\frac{\Delta E}{a_B} = 1$, $\mathrm{A}=10$.}
				\label{fig33-new}
			\end{figure}

	Now to bolster our previous observations we further plot $\delta \textrm{N}$ with $\sigma$. 
	For $\alpha=0.2$ and low values of $\sigma$, Fig. \ref{fig30-new} shows that degradation is more for anti-parallel case. 
But after a critical value of $\sigma$ the degradation for parallelly accelerating detectors dominates, and for some values of $\mathrm{A}$ (e.g. $\mathrm{A}$=3 or $\mathrm{A}=10$) it is seen that even entanglement harvesting starts for antiparallelly accelerating case.
Now for $\alpha=0.707$ (see Fig. \ref{fig31-new}), 
again for low $\sigma$, the degradation for antiparallel case is more. But after a critical value of $\sigma$ the degradation for antiparallel case can be less compared to that for parallel case for some higher values of $\mathrm{A}$. This observation is further bolstered by Fig. \ref{fig32-new} and Fig. \ref{fig33-new}. 

			\subsubsection{$(1+3)$-dimensions}

\begin{figure}[h!]
		             \centering
		            \includegraphics[width=.45\textwidth]{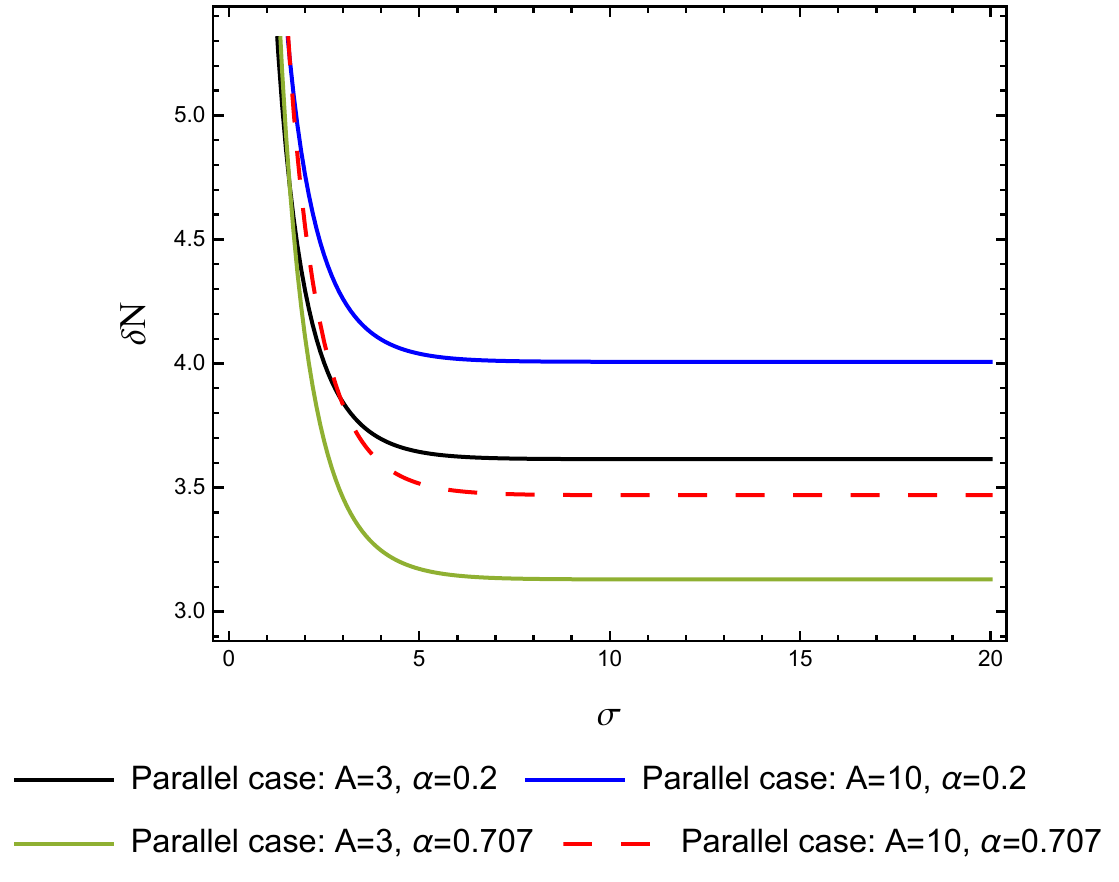}
		            \caption{Parallel motion in $(1+3)$-dimensions with thermal background:  variation of $\delta\textrm{N}$ as a function of $\sigma$ with fixed $\alpha$ and $\mathrm{A}$, with the choice $\frac{\Delta E}{a_B} = 1$}
		           \label{new-sigmasame}
	        \end{figure}

\begin{figure}[h!]
		\centering
		\includegraphics[width=.45\textwidth]{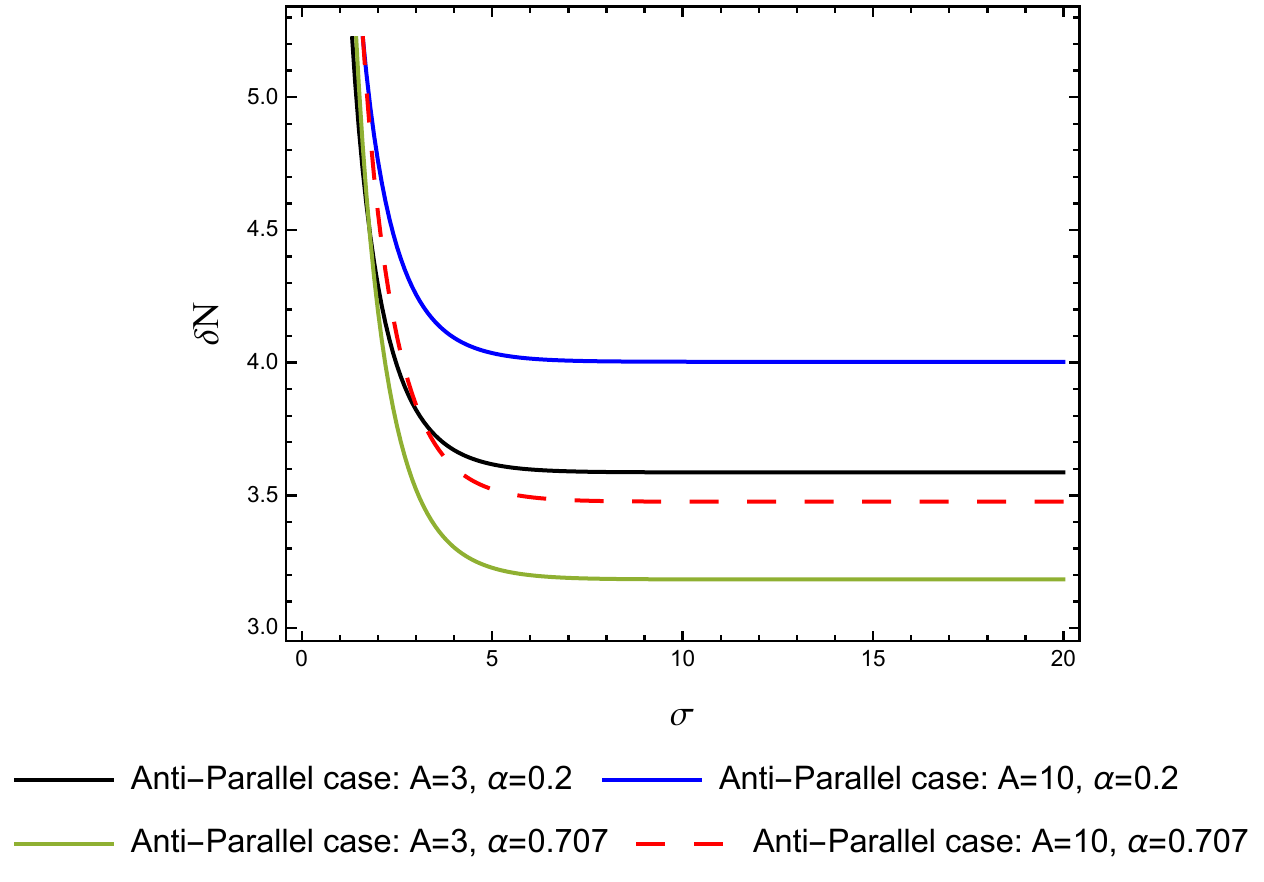}
		\caption{Anti-Parallel motion in $(1+3)$-dimensions with thermal background:  variation of $\delta\textrm{N}$ as a function of $\sigma$ with fixed $\alpha$ and A, with the choice $\frac{\Delta E}{a_B} = 1$}
		\label{new-sigmaopp}
	\end{figure}

\textbf{Parallel and Anti-Parallel motion: }		
	Here also at a fixed $\sigma $, the variation of $\delta\textrm{N}$ with a certain $\mathrm{A}$ and a certain $\alpha$ is similar in nature to that for the $(1+3)$-dimensional non-thermal case.  So in the subsequent analysis, we aren't explicitly including the corresponding detailed analysis. We rather only look at how  $\delta\textrm{N}$ varies with $\sigma$ in Fig. \ref{new-sigmasame} and Fig. \ref{new-sigmaopp}. 
It is observed that the value of $\delta\textrm{N}$ is very close to each other for the parallel and anti-parallel cases. We see that the entanglement degradation decreases as $\sigma$ increases (low temperature). Note that this nature is also observed for the (1+1)-dimensional thermal case.

\begin{figure}[h!]
							\centering
							\includegraphics[width=.45\textwidth]{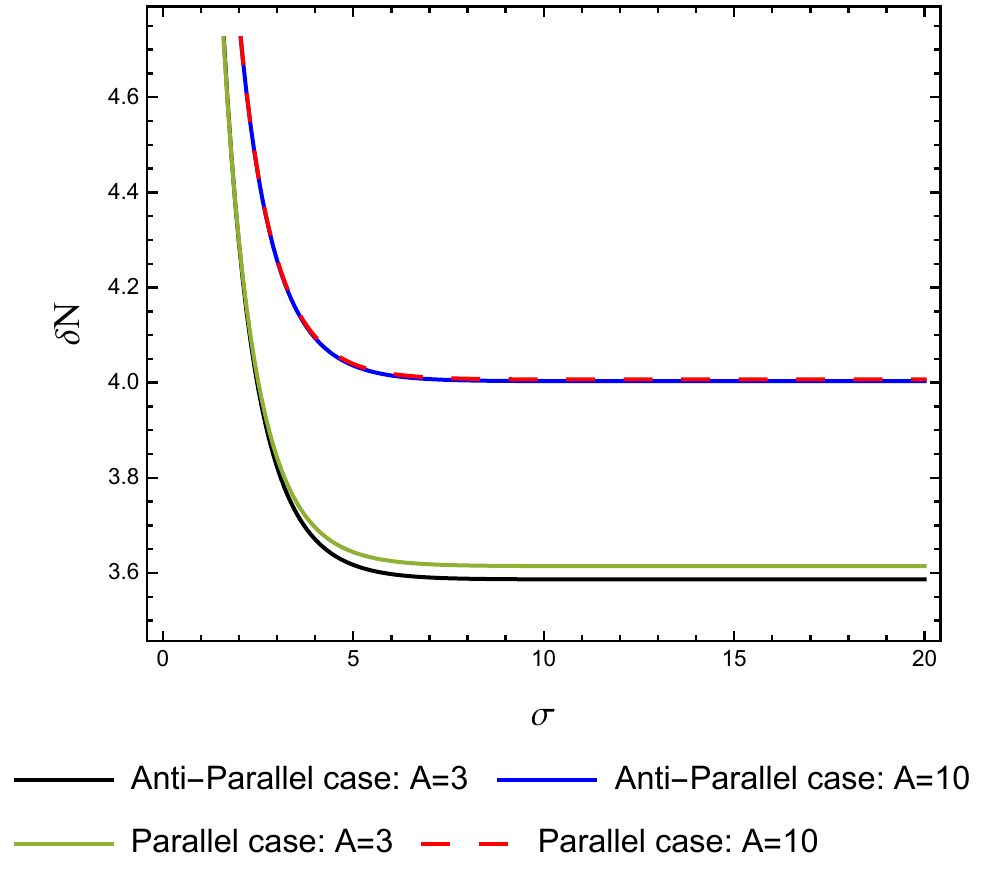}
							\caption{Parallel and Anti-parallel motions in $(1+3)$ dimensions with thermal background:  $\delta\textrm{N}$ Vs $\sigma$ plot with $\frac{\Delta E}{a_B} = 1$, $\alpha=0.2$.}
							\label{fig50-new}
						\end{figure}

\begin{figure}[h!]
							\centering
							\includegraphics[width=.45\textwidth]{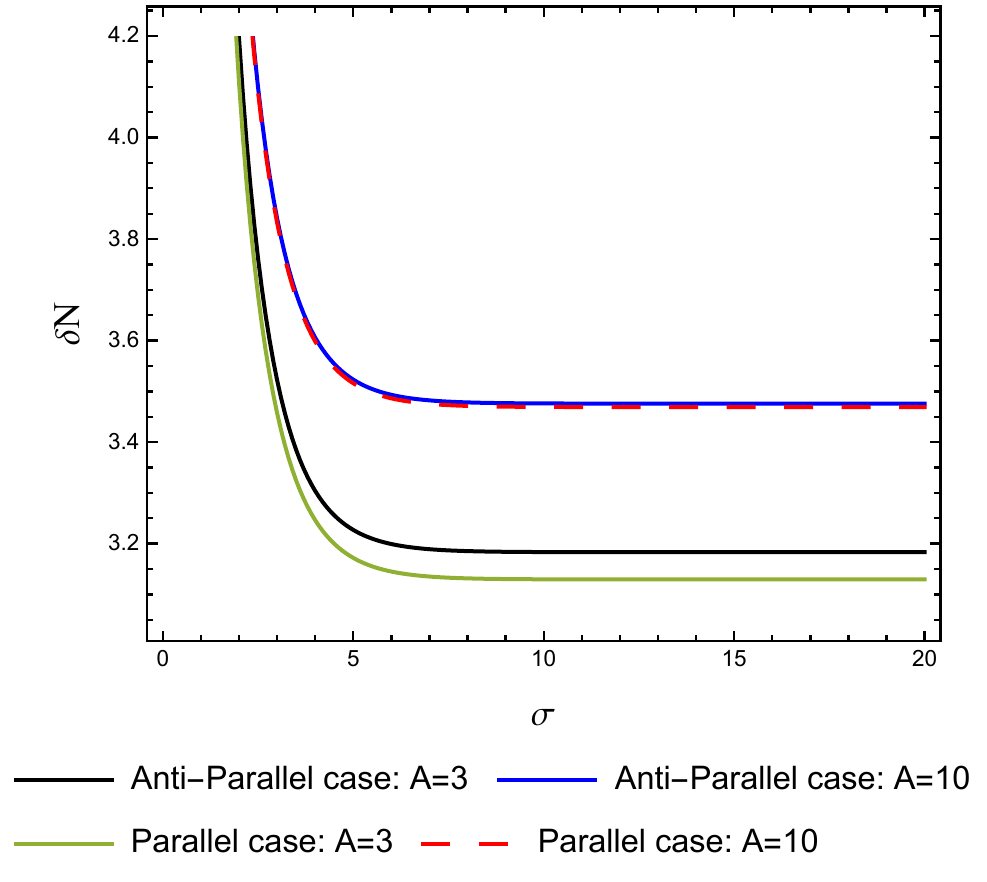}
							\caption{Parallel and Anti-parallel motions in $(1+3)$ dimensions with thermal background:  $\delta\textrm{N}$ Vs $\sigma$ plot with$\frac{\Delta E}{a_B} = 1$, $\alpha=0.707$.}
							\label{fig51-new}
						\end{figure}

\begin{figure}[h!]
							\centering
							\includegraphics[width=.45\textwidth]{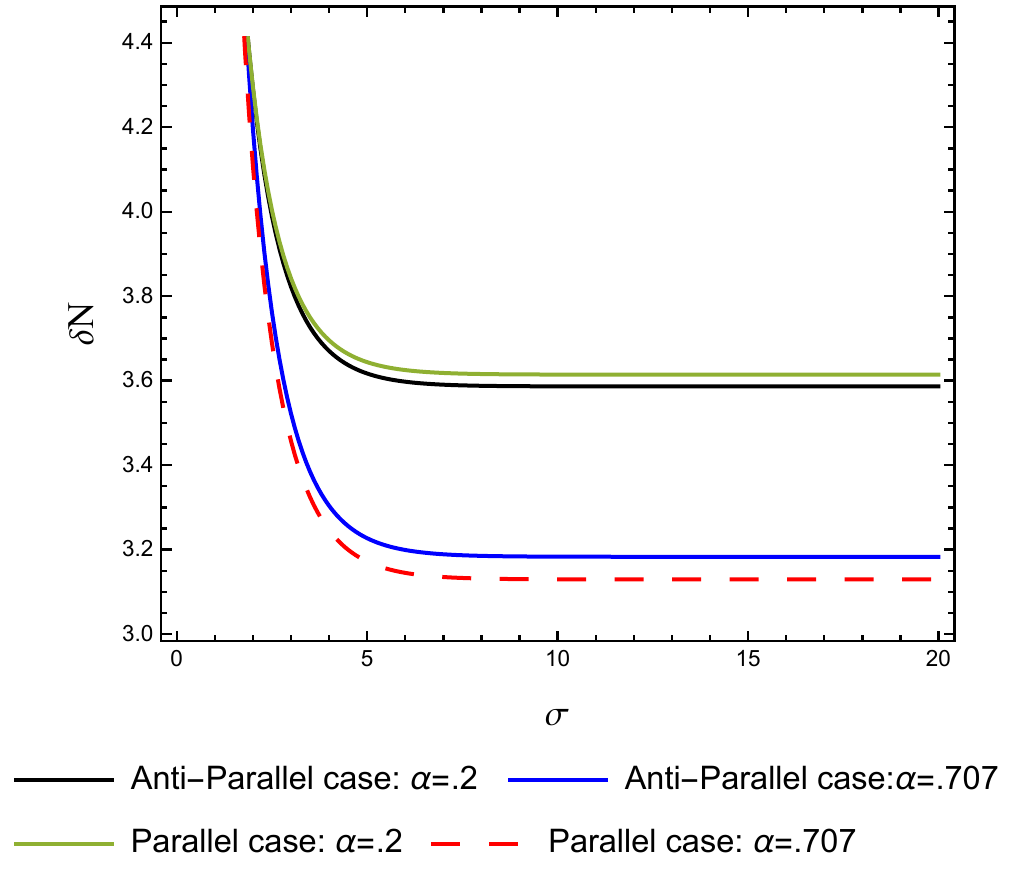}
							\caption{Parallel and Anti-parallel motions in $(1+3)$ dimensions with thermal background:  $\delta\textrm{N}$ Vs $\sigma$ plot with $\frac{\Delta E}{a_B} = 1$, $\mathrm{A}=3$.}
							\label{fig52-new}
						\end{figure}

\begin{figure}[h]
							\centering
							\includegraphics[width=.45\textwidth]{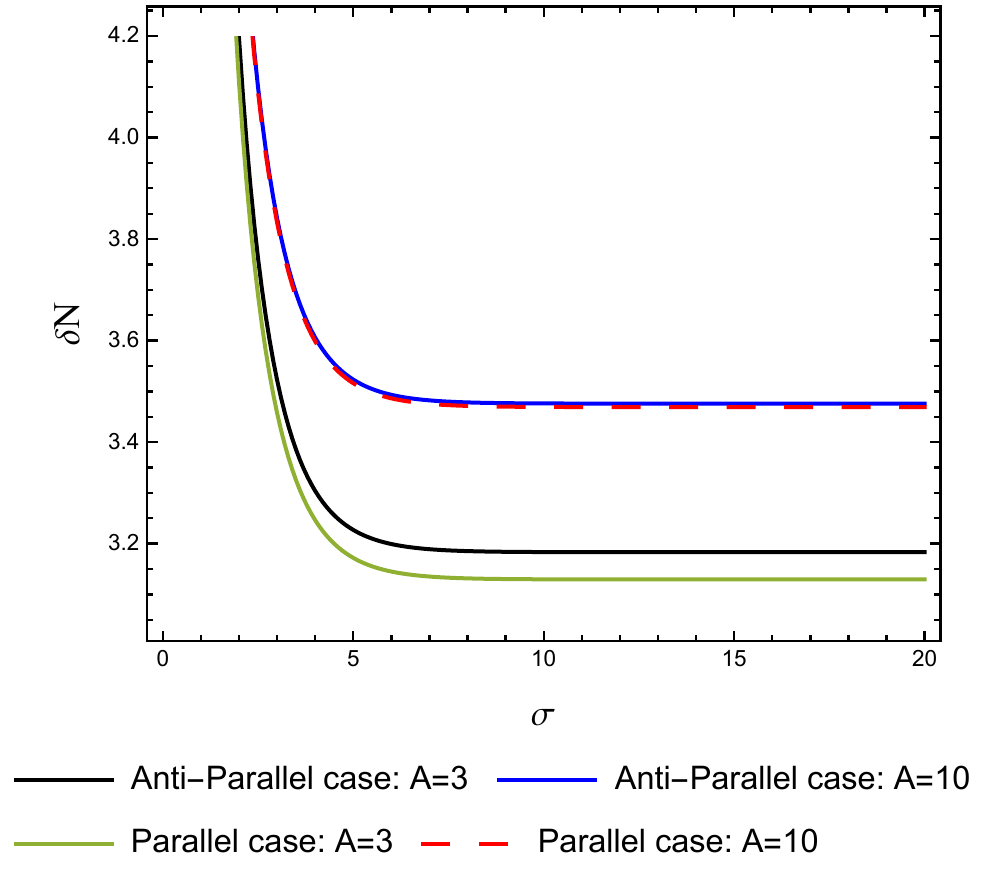}
							\caption{Parallel and Anti-parallel motions in $(1+3)$ dimensions with thermal background:  $\delta\textrm{N}$ Vs $\sigma$ plot with $\frac{\Delta E}{a_B} = 1$, $\mathrm{A}=10$.}
							\label{fig53-new}
						\end{figure}

\textbf{Comparative study:} Comparing the parallel case with anti-parallel for fixed $\sigma$, we again see similar nature of entanglement degradation as in $(1+3)$-dimensional non-thermal case and hence we don't study it explicitly. We move on to comparing nature of degradation with $\sigma $ for parallel and antiparallel case in Fig. \ref{fig50-new}, Fig. \ref{fig51-new}, Fig. \ref{fig52-new} and Fig. \ref{fig53-new}.
Fig. \ref{fig50-new} and  Fig. \ref{fig51-new} show that although the degradation for parallel and anti-parallel case is very close to each other for a particular $\mathrm{A}$, at fixed $\alpha$, for some cases degradation for parallel case is greater than the degradation for anti-parallel case (e.g. see at $\mathrm{A}=3$ and $\alpha=0.2$). This observation is bolstered by Fig. \ref{fig52-new} and  Fig. \ref{fig53-new}. 
So finally, after studying the nature of degradation for (1+3) dimension thermal case, we conclude no entanglement harvesting is observed here.

\section{Conclusion}
 According to the equivalence principle, several local aspects of gravity can be mimicked by an accelerating observer. This serves as one of the main motivations behind studying relativistic entanglement with respect to accelerated frames. So if we can conclude about the fate of entanglement between two observers in an accelerated frame, we can also get some insights regarding entanglement between observers in the context of a black hole.
 The existing works in this direction study mainly the effect of a temperature-independent field on the entanglement dynamics for different kinds of observers and also primarily focus on the entanglement dynamics between observers on the same Rindler wedges. Here we study how the background temperature of the field further contributes to the enhancement of the degradation in entanglement. Also, given the process of entanglement harvesting between UD detectors in opposite Rindler wedges (anti-parallelly accelerating detectors), we investigate how this entanglement harvesting phenomenon interferes with the usual entanglement degradation process. 
 The main results obtained are as follows:
\begin{itemize}
	\item For the parallel motion of the detectors in non-thermal field, in $(1+1)$-dimension and $(1+3)$-dimension, there is always entanglement degradation. In case of $(1+3)$-dimension, there is some undulation in the negativity with change in its acceleration. 
	\item For anti-parallel motion, in the non-thermal field in $(1+1)$-dimension, entanglement harvesting interferes with the entanglement degradation process. Hence there is either entanglement harvesting or entanglement
	 degradation depending on the value of the parameters. 
	 Strikingly, for $(1+3)$-dimension there is no entanglement harvesting.
	\item For the parallel motion of the detectors in thermal field, in $(1+1)$-dimension and $(1+3)$-dimension, we always find entanglement degradation. In case of $(1+3)$-dimension, there is some undulation in the negativity value as in the non-thermal case.
    \item  For anti-parallel motion, in the thermal field in $(1+1)$-dimension, there is entanglement degradation or entanglement harvesting depending on the range of parameters. For $(1+3)$-dimension there is no entanglement harvesting just like the case of non-thermal field.
    \item We see that at higher temperatures, the degradation of entanglement is more.
\end{itemize}

The present study not only differs from earlier ones \cite{lin2008disentanglement} -- \cite{zhou2021entanglement} by the structure of the system but also captures a robust difference in the analysis. The noticeable difference from the earlier ones is here we kept our system always in thermal equilibrium by expanding the field in terms of Rindler modes and Wightman functions are calculated in the vacuum of Unruh modes. This provided time translational invariance in required Green's functions with respect to Rindler time, contrary to those for Minkowski modes. Therefore our present results are unique compared to the existing ones through this constructional difference.

Before closing let us make a comment on the choice of values for $\alpha$. Recall that the eigenvalues of the partially transposed density matrix are given by Eqs. (\ref{e1}), (\ref{e2}), (\ref{e3}) and (\ref{e4}).
Since {\it Negativity} is the sum of all the negative eigenvalues of the partially transposed density matrix,
 for the case $\alpha, \gamma > 0$ (which has been considered in the current analysis),  $\lambda_2$ (Eq. (\ref{e2})) being the only negative eigenvalue, contributes to the negativity. This has been justified and elaborately argued below Eq. (\ref{e4}). Therefore in the subsequent analysis we considered $\lambda_2$ only in order to find the nature of negativity at later time.  
Now, since our initial state (given by Eq. (\ref{V26})) is normalised, for the choice $\alpha=1$,  one needs to choose $\gamma=0$. In that case one has   
$\ket{in} = \ket{0; E_0^A E_0^B}$,
which is no longer an entangled state. The analysis for this initial state cannot be directly obtained by taking the limit $\alpha=1$ and $\gamma=0$ in \eqref{negativity}. 
This is clear if we look at the general expression obtained using \eqref{L1} and \eqref{L2}, 
\begin{eqnarray}
	\lambda_{1,2} = \mathcal{O}(C^2) \pm\sqrt{\alpha^2 \gamma^2 -2\alpha\gamma C^2(\dots)+  C^4(\dots)}~.
\end{eqnarray}
Remember that here the analysis is done upto order $C^2$ in the perturbation series of the density matrix. So the $\lambda$'s are to be determined upto order $C^2$. Therefore
 when $\gamma =0 $, we see that only the $\mathcal{O}(C^4)$ term inside the square-root contributes to eigenvalue at the $C^2$ order. Whereas, if $\gamma \neq 0$, inside the square-root, the $\mathcal {O}(C^0)$ and $\mathcal {O}(C^2)$ terms are nonzero and then the $\mathcal {O}(C^4)$ term is negligible as it contributes in eigenvalue at the order $C^4$ (keep in mind that $\alpha^2\gamma^2$ term within the square-root always greater than other terms as this term signifies the initial entanglement and others are due to perturbation; otherwise the perturbation approach will break down). Hence, we get a different contribution in $\lambda_{1,2}$ in the latter senario. Since, in our analysis we are only interested with the initially entangled states, we can safely ignore the $\mathcal{O}(C^4)$ term and move on to get the approximated values 
 (\ref{e1}) and (\ref{e2}) and further obtain Eq. \eqref{N1}. To consider the situation $\alpha=1$, we need to start full expression for $\mathcal{N}$, defined through negative eigenvalue among (\ref{L1}) -- (\ref{L4}). Hence $\alpha=1$ situation can not be verified from our present result.
 As we mentioned that $\alpha=1$ does not fulfil our purpose and not compatible with the present analysis, in all plots with respect to $\alpha$ as an axis, we have considered the range $\alpha \in [0.1, 0.9]$.

In this connection, we point out that near the black hole horizon, the spacetime metric can be locally approximated as a Rindler metric. Moreover, an accelerated observer in Minkowski spacetime is similar to a static observer in black hole spacetime. Therefore we expect that static detectors in the near horizon regime in black hole space find identical features of entanglement phenomenon, which we have noticed in this investigation. We believe that there may be additional features that will also be observed due to the presence of curvature, which is a future direction of this work. 
	\begin{acknowledgments}
		PC would like to thank Indian Institute of Technology Guwahati, India for providing the opportunity to carry out the work as a part of her master's thesis. PC would also like to thank Subhajit Barman, Sampreet Kalita and Dipankar Barman for their helpful suggestions and comments. 

		The research of one of the authors (BRM) is
supported by a START - UP RESEARCH GRANT (No.
SG/PHY/P/BRM/01) from the Indian Institute of Technology Guwahati, India and by a Core Research Grant
(File no. CRG/2020/000616) from Science and Engineering Research Board (SERB), Department of Science $\&$
Technology (DST), Government of India.
	\end{acknowledgments}

	\begin{widetext}
		\section*{Appendices}
				
		\begin{appendix} 

\section{Calculating the Elements of Density Matrix}
				\label{App1} 
				The matrix elements of $\rho_{AB} (t)$ in (\ref{RAB}) for our choice of basis in general are denoted as $\langle E_{n_A}^A E_{n_B}^B|\rho_{AB}(t)| E_{\bar n_A}^A E_{\bar n_B}^B\rangle $. We will evaluate this till $\mathcal{O}(C^2)$. Therefore expansion of $e^{iS_{int}}$ up to the second order in $C^2$ will be sufficient for our purpose. Also since we are interested in detectors only, the trace over all the field states will be taken. With these then, one finds
				\begin{eqnarray}
					&&R_{n_A,n_B,\bar n_A,\bar n_B}= \textrm{Tr}_{\phi}\langle E_{n_A}^A E_{n_B}^B|\rho_{AB}(t)| E_{\bar n_A}^A E_{\bar n_B}^B\rangle 
					\nonumber
					\\	
					&&=\textrm{Tr}_{\phi}\Big[\bra{ E_{n_A}^A E_{n_B}^B}[1+iS_{int}-\frac{T(S_{int}S_{int}^\prime)}{2}]\ket{0}(\alpha\ket{ E_0^A E_0^B}+\gamma\ket{ E_1^A E_1^B}) 
					\nonumber
					\\
					&&\times \bra{0}(\alpha^\star\bra{ E_0^A E_0^B}+\gamma^\star\bra{ E_1^A E_1^B})[1-iS_{int}-\frac{T(S_{int}S_{int}^\prime)}{2}]\ket{ E_{\bar n_A}^A E_{\bar n_B}^B}\Big] 
					\nonumber 
					\\
					&& =\underbrace{\bra{ E_{n_A}^A E_{n_B}^B}(\alpha\ket{ E_0^A E_0^B}+\gamma\ket{ E_1^A E_1^B})(\alpha^\star\bra{ E_0^A E_0^B}+\gamma^\star\bra{ E_1^A E_1^B})\ket{ E_{\bar n_A}^A E_{\bar n_B}^B}}_{R^{(0)}_{n_A,n_B,\bar n_A,\bar n_B}} \nonumber \\
					&&+\underbrace{\textrm{Tr}_{\phi} \Big[\bra{ E_{n_A}^A E_{n_B}^B}S_{int}(\alpha\ket{ E_0^A E_0^B}+\gamma\ket{ E_1^A E_1^B})\ket{0}\bra{0}(\alpha^\star\bra{ E_0^A E_0^B}+\gamma^\star\bra{ E_1^A E_1^B})S_{int}\ket{ E_{\bar n_A}^A E_{\bar n_B}^B}\Big]}_{R^{(1)}_{n_A,n_B,\bar n_A,\bar n_B}} \nonumber \\
					&&\underbrace{-\textrm{Tr}_{\phi}\Big[\bra{ E_{n_A}^A E_{n_B}^B}\frac{T(S_{int}S_{int}^\prime)}{2}(\alpha\ket{ E_0^A E_0^B}+\gamma\ket{ E_1^A E_1^B})\ket{0}\bra{0}(\alpha^\star\bra{ E_0^A E_0^B}+\gamma^\star\bra{ E_1^A E_1^B})\ket{ E_{\bar n_A}^A E_{\bar n_B}^B}\Big]}_{R^{(2)}_{n_A,n_B,\bar n_A,\bar n_B}} \nonumber \\
					&&\underbrace{-\textrm{Tr}_{\phi}\Big[\bra{ E_{n_A}^A E_{n_B}^B}(\alpha\ket{ E_0^A E_0^B}+\gamma\ket{ E_1^A E_1^B})\ket{0}\bra{0}(\alpha^\star\bra{ E_0^A E_0^B}+\gamma^\star\bra{ E_1^A E_1^B})\frac{T(S_{int}S_{int}^\prime)}{2}\ket{ E_{\bar n_A}^A E_{\bar n_B}^B}\Big]}_{R^{(2)\star}_{\bar n_A,\bar n_B,n_A,n_B}}~.
					\label{eqn:pf}
				\end{eqnarray}
In the above, ``$\textrm{Tr}_{\phi}$'' denotes the trace over all field states and ``$T$'' signifies the time order product.

The first term is straight-forward to calculate and it turns out to be
				\begin{eqnarray}
					R^{(0)}_{n_A,n_B,\bar n_A,\bar n_B} = (\alpha \delta^0_{n_A} \delta^0_{n_B} +\gamma\delta^1_{n_A} \delta^1_{n_B})( \alpha^\star \delta^0_{\bar n_A}\delta^0_{\bar n_B}+\gamma^\star \delta^1_{\bar n_A}\delta^1_{\bar n_B})~. 
					\label{r}
				\end{eqnarray}
The other terms are evaluated as follows. We first concentrate on $R^{(1)}_{n_A,n_B,\bar n_A,\bar n_B}$. This, after using the explicit form of $S_{int}$ from (\ref{R1}), is given by
				\begin{eqnarray}
					&&R^{(1)}_{n_A,n_B,\bar n_A,\bar n_B} =\textrm{Tr}_{\phi}\Big[\bra{ E_{n_A}^A E_{n_B}^B}(\int C_A \chi_A(\tau_A)m_A(\tau_A)\phi_A(x_A)d\tau_A + \int C_B \chi_B(\tau_B)m_B(\tau_B)\phi_B(x_B)d\tau_B)
					\nonumber
					\\
					&&~~~~~~~~~~~~~~~~~~~~~~~~~~~~~~~~(\alpha\ket{ E_0^A E_0^B}+\gamma\ket{ E_1^A E_1^B})\ket{0} 
					\nonumber 
					\\
					&&\times \bra{0}(\alpha^\star\bra{ E_0^A E_0^B}+\gamma^\star\bra{ E_1^A E_1^B})(\int C_A \chi_A(\tau^\prime_A)m_A(\tau^\prime_A)\phi_A(x^\prime_A)d\tau^\prime_A + \int C_B \chi_B(\tau^\prime_B)m_B(\tau^\prime_B)\phi_B(x^\prime_B)d\tau^\prime_B)
					\ket{ E_{\bar n_A}^A E_{\bar n_B}^B}\Big] 
					\nonumber 
					\\
					&&=\textrm{Tr}_{\phi}\Big[\bra{ E_{n_A}^A E_{n_B}^B}\int C_A \chi_A(\tau_A)m_A(\tau_A)\phi_A(x_A)d\tau_A(\alpha\ket{ E_0^A E_0^B}+\gamma\ket{ E_1^A E_1^B})\ket{0} 
					\nonumber
					\\
					&&\times \bra{0}(\alpha^\star\bra{ E_0^A E_0^B}+\gamma^\star\bra{ E_1^A E_1^B})\int C_A \chi_A(\tau^\prime_A)m_A(\tau^\prime_A)\phi_A(x^\prime_A)d\tau^\prime_A \ket{ E_{\bar n_A}^A E_{\bar n_B}^B} 
					\nonumber 
					\\
					&& + \bra{ E_{n_A}^A E_{n_B}^B}\int C_A \chi_A(\tau_A)m_A(\tau_A)\phi_A(x_A)d\tau_A(\alpha\ket{ E_0^A E_0^B}+\gamma\ket{ E_1^A E_1^B})\ket{0}
					\nonumber
					\\
					&& \times \bra{0}(\alpha^\star\bra{ E_0^A E_0^B}+\gamma^\star\bra{ E_1^A E_1^B})\int C_B \chi_B(\tau^\prime_B)m_B(\tau^\prime_B)\phi_B(x^\prime_B)d\tau^\prime_B \ket{ E_{\bar n_A}^A E_{\bar n_B}^B} 
					\nonumber 
					\\
					&& +\bra{ E_{n_A}^A E_{n_B}^B}\int C_B \chi_B(\tau_B)m_B(\tau_B)\phi_B(x_B)d\tau_B(\alpha\ket{ E_0^A E_0^B}+\gamma\ket{ E_1^A E_1^B})\ket{0}
					\nonumber 
					\\
					&& \times \bra{0}(\alpha^\star\bra{ E_0^A E_0^B}+\gamma^\star\bra{ E_1^A E_1^B})\int C_A \chi_A(\tau^\prime_A)m_A(\tau^\prime_A)\phi_A(x^\prime_A)d\tau^\prime_A \ket{ E_{\bar n_A}^A E_{\bar n_B}^B} 
					\nonumber 
					\\
					&& +\bra{ E_{n_A}^A E_{n_B}^B}\int C_B \chi_B(\tau_B)m_B(\tau_B)\phi_B(x_B)d\tau_B(\alpha\ket{ E_0^A E_0^B}+\gamma\ket{ E_1^A E_1^B})\ket{0}
					\nonumber 
					\\
					&& \times \bra{0}(\alpha^\star\bra{ E_0^A E_0^B}+\gamma^\star\bra{ E_1^A E_1^B})\int C_B \chi_B(\tau^\prime_B)m_B(\tau^\prime_B)\phi_B(x^\prime_B)d\tau^\prime_B \ket{ E_{\bar n_A}^A E_{\bar n_B}^B}\Big]~.
					\label{B1}
				\end{eqnarray}
Now using (\ref{R3}), in the interaction picture, in Eq. \eqref{B1} one finds,
				\begin{eqnarray}
					&&R^{(1)}_{n_A,n_B,\bar n_A,\bar n_B}=\alpha \alpha^*\delta^0_{n_B}\delta^0_{\bar n_B}\bra{E_{n_A}^A}m_A\ket{E_0^A}\bra{E^A_{0}}m_A\ket{E_{\bar n_A}^A}
					\nonumber 
					\\
					&&\times \int \int C_A^2 d\tau_A d\tau^\prime_A \chi_A(\tau_A)\chi_A(\tau^\prime_A)e^{i\tau_A(E_{n_A}^A-E_{0}^A)}e^{-i\tau'_A(E_{\bar n_A}^A-E_{0}^A)}\bra{0} \phi(x^\prime_A)\phi(x_A)\ket{0} \nonumber 
					\\
					&& +\alpha \gamma^*\delta^0_{n_B}\delta^1_{\bar n_B}\bra{E_{n_A}^A}m_A\ket{E_0^A}\bra{E_{1}^A}m_A\ket{E_{\bar n_A}^A} \nonumber 
					\\
					&& \times \int \int C_A^2 d\tau_A d\tau^\prime_A \chi_A(\tau_A)\chi_A(\tau^\prime_A)e^{i\tau_A(E_{n_A}^A-E_{0}^A)}e^{-i\tau'_A(E_{\bar n_A}^A-E_{1}^A)}\bra{0} \phi(x^\prime_A)\phi(x_A)\ket{0} \nonumber 
					\\  
					&& +\gamma\alpha^* \delta^1_{n_B}\delta^0_{\bar n_B}\bra{E_{n_A}^A}m_A\ket{E_1^A}\bra{E_{0}^A}m_A\ket{E_{\bar n_A}^A} \nonumber 
					\\
					&& \times \int \int C_A^2 d\tau_A d\tau^\prime_A \chi_A(\tau_A)\chi_A(\tau^\prime_A)e^{i\tau_A(E_{n_A}^A-E_{1}^A)}e^{-i\tau'_A(E_{\bar n_A}^A-E_{0}^A)}\bra{0} \phi(x^\prime_A)\phi(x_A)\ket{0} \nonumber 
					\\ 
					&& +\gamma^* \gamma \delta^1_{n_B}\delta^1_{\bar n_B}\bra{E_{n_A}^A}m_A\ket{E_1^A}\bra{E_{1}^A}m_A\ket{E_{\bar n_A}^A} \nonumber 
					\\
					&& \times \int \int C_A^2 d\tau_A d\tau^\prime_A \chi_A(\tau_A)\chi_A(\tau^\prime_A)e^{i\tau_A(E_{n_A}^A-E_{1}^A)}e^{-i\tau'_A(E_{\bar n_A}^A-E_{1}^A)}\bra{0} \phi(x^\prime_A)\phi(x_A)\ket{0} \nonumber 
					\\
					&& + \alpha \alpha^*\delta^0_{n_B}\delta^0_{\bar n_A}\bra{E_{n_A}^A}m_A\ket{E_0^A}\bra{E^B_{0}}m_B\ket{E_{\bar n_B}^B} \nonumber 
					\\
					&& \times \int \int C_A C_B d\tau_A d\tau^\prime_B \chi_A(\tau_A)\chi_B(\tau^\prime_B)e^{i\tau_A(E_{n_A}^A-E_{0}^A)}e^{-i\tau'_B(E_{\bar n_B}^B-E_{0}^B)}\bra{0} \phi(x^\prime_B)\phi(x_A)\ket{0} \nonumber 
					\\
					&& + \alpha \gamma^*\delta^0_{n_B}\delta^1_{\bar n_A}\bra{E_{n_A}^A}m_A\ket{E_0^A}\bra{E^B_{1}}m_B\ket{E_{\bar n_B}^B} \nonumber 
					\\
					&& \times \int \int C_A C_B d\tau_A d\tau^\prime_B \chi_A(\tau_A)\chi_B(\tau^\prime_B)e^{i\tau_A(E_{n_A}^A-E_{0}^A)}e^{-i\tau'_B(E_{\bar n_B}^B-E_{1}^B)}\bra{0} \phi(x^\prime_B)\phi(x_A)\ket{0} \nonumber 
					\\
					&& + \alpha^* \gamma\delta^1_{n_B}\delta^0_{\bar n_A}\bra{E_{n_A}^A}m_A\ket{E_1^A}\bra{E^B_{0}}m_B\ket{E_{\bar n_B}^B} \nonumber 
					\\
					&& \times \int \int C_A C_B d\tau_A d\tau^\prime_B \chi_A(\tau_A)\chi_B(\tau^\prime_B)e^{i\tau_A(E_{n_A}^A-E_{1}^A)}e^{-i\tau'_B(E_{\bar n_B}^B-E_{0}^B)}\bra{0} \phi(x^\prime_B)\phi(x_A)\ket{0} \nonumber 
					\\
					&& + \gamma^* \gamma\delta^1_{n_B}\delta^1_{\bar n_A}\bra{E_{n_A}^A}m_A\ket{E_1^A}\bra{E^B_{1}}m_B\ket{E_{\bar n_B}^B} \nonumber 
					\\
					&& \times \int \int C_A C_B d\tau_A d\tau^\prime_B \chi_A(\tau_A)\chi_B(\tau^\prime_B)e^{i\tau_A(E_{n_A}^A-E_{1}^A)}e^{-i\tau'_B(E_{\bar n_B}^B-E_{1}^B)}\bra{0} \phi(x^\prime_B)\phi(x_A)\ket{0} \nonumber 
					\\
					&& +\alpha \alpha^*\delta^0_{n_A}\delta^0_{\bar n_B}\bra{E_{n_B}^B}m_B\ket{E_B^0}\bra{E^A_{0}}m_A\ket{E_{\bar n_A}^A} \nonumber 
					\\
					&& \times \int \int C_A C_B  d\tau_B d\tau^\prime_A \chi_B(\tau_B)\chi_A(\tau^\prime_A)e^{i\tau_B(E_{n_B}^B-E_{0}^B)}e^{-i\tau'_A(E_{\bar n_A}^A-E_{0}^A)}\bra{0} \phi(x^\prime_A)\phi(x_B)\ket{0} \nonumber 
					\\
					&& +\alpha \gamma^*\delta^0_{n_A}\delta^1_{\bar n_B}\bra{E_{n_B}^B}m_B\ket{E_B^0}\bra{E^A_{1}}m_A\ket{E_{\bar n_A}^A} \nonumber 
					\\
					&& \times \int \int C_A C_B  d\tau_B d\tau^\prime_A \chi_B(\tau_B)\chi_A(\tau^\prime_A)e^{i\tau_B(E_{n_B}^B-E_{0}^B)}e^{-i\tau'_A(E_{\bar n_A}^A-E_{1}^A)}\bra{0} \phi(x^\prime_A)\phi(x_B)\ket{0} \nonumber 
					\\
					&& +\alpha^* \gamma \delta^1_{n_A} \delta^0_{\bar n_B}\bra{E_{n_B}^B}m_B\ket{E_B^1}\bra{E^A_{0}}m_A\ket{E_{\bar n_A}^A} \nonumber 
					\\
					&& \times \int \int C_A C_B  d\tau_B d\tau^\prime_A \chi_B(\tau_B)\chi_A(\tau^\prime_A)e^{i\tau_B(E_{n_B}^B-E_{1}^B)}e^{-i\tau'_A(E_{\bar n_A}^A-E_{0}^A)}\bra{0} \phi(x^\prime_A)\phi(x_B)\ket{0} \nonumber 
					\\
					&& +\gamma^* \gamma \delta^1_{n_A} \delta^1_{\bar n_B}\bra{E_{n_B}^B}m_B\ket{E_B^1}\bra{E^A_{1}}m_A\ket{E_{\bar n_A}^A} \nonumber 
					\\
					&& \times \int \int C_A C_B  d\tau_B d\tau^\prime_A \chi_B(\tau_B)\chi_A(\tau^\prime_A)e^{i\tau_B(E_{n_B}^B-E_{1}^B)}e^{-i\tau'_A(E_{\bar n_A}^A-E_{1}^A)}\bra \phi(x^\prime_A)\phi(x_B)\ket{0} \nonumber 
					\\
					&& +\alpha \alpha^*\delta^0_{n_A}\delta^0_{\bar n_A}\bra{E_{n_B}^B}m_B\ket{E_B^0}\bra{E^B_{0}}m_B\ket{E_{\bar n_B}^B} \nonumber 
					\\
					&& \times \int \int C_B^2 d\tau_B d\tau^\prime_B \chi_B(\tau_B)\chi_B(\tau^\prime_B)e^{i\tau_B(E_{n_B}^B-E_{0}^B)}e^{-i\tau'_B(E_{\bar n_B}^B-E_{0}^B)}\bra{0} \phi(x^\prime_B)\phi(x_B)\ket{0} \nonumber 
					\\
					&& +\alpha \gamma^*\delta^0_{n_A}\delta^1_{\bar n_A}\bra{E_{n_B}^B}m_B\ket{E_B^0}\bra{E^B_{1}}m_B\ket{E_{\bar n_B}^B} \nonumber 
					\\
					&& \times \int \int C_B^2 d\tau_B d\tau^\prime_B \chi_B(\tau_B)\chi_B(\tau^\prime_B)e^{i\tau_B(E_{n_B}^B-E_{0}^B)}e^{-i\tau'_B(E_{\bar n_B}^B-E_{1}^B)}\bra{0} \phi(x^\prime_B)\phi(x_B)\ket{0} \nonumber 
					\\ 
					&& +\alpha^* \gamma\delta^1_{n_A}\delta^0_{\bar n_A}\bra{E_{n_B}^B}m_B\ket{E_B^1}\bra{E^B_{0}}m_B\ket{E_{\bar n_B}^B} \nonumber 
					\\
					&& \times \int \int C_B^2 d\tau_B d\tau^\prime_B \chi_B(\tau_B)\chi_B(\tau^\prime_B)e^{i\tau_B(E_{n_B}^B-E_{1}^B)}e^{-i\tau'_B (E_{\bar n_B}^B-E_{0}^B)}\bra{0} \phi(x^\prime_B)\phi(x_B)\ket{0} \nonumber 
					\\  
					&& +\gamma^* \gamma\delta^1_{n_A}\delta^1_{\bar n_A}\bra{E_{n_B}^B}m_B\ket{E_B^1}\bra{E^B_{1}}m_B\ket{E_{\bar n_B}^B} \nonumber 
					\\
					&& \times \int \int C_B^2 d\tau_B d\tau^\prime_B \chi_B(\tau_B)\chi_B(\tau^\prime_B)e^{i\tau_B(E_{n_B}^B-E_{1}^B)}e^{-i\tau'_B(E_{\bar n_B}^B-E_{1}^B)}\bra{0} \phi(x^\prime_B)\phi(x_B)\ket{0}~. 
					\label{R1}
				\end{eqnarray}
In the above, we used the notation $m_j(0)=m_j$.
				
				For our choice of $m_j(0)$ (see Eq. (\ref{R3})), we have $\bra{E^j_n}m_j(0)\ket{E_n^j}=0 $, where $n\in 0,1$ and $j\in A,B$. Then, from the above, we find $R^{(1)}_{0000}=0, R^{(1)}_{0001}=0, R^{(1)}_{0010}=0, R^{(1)}_{0011}=0, R^{(1)}_{0100}=0, R^{(1)}_{0111}=0, R^{(1)}_{1000}=0, R^{(1)}_{1011}=0, R^{(1)}_{1100}=0, R^{(1)}_{1111}=0, R^{(1)}_{1101}=0, R^{(1)}_{1110}=0$ and the non-vanishing ones are,
				\begin{eqnarray}
					&&R^1_{0101}=\gamma^*\gamma \left|\bra{E_0^A}m_A\ket{E_1^A}\right|^2 C_A^2\int \int d\tau_A d\tau^\prime_A \chi(\tau_A)\chi(\tau^\prime_A)e^{i\Delta \tau_A \Delta E_A}G_w(x^\prime_A,x_A) 
					\nonumber 
					\\ 
					&& +\gamma \alpha^* C_A C_B\bra{E_0^A}m_A\ket{E_1^A}\bra{E_0^B}m_B\ket{E_1^B}\int \int d\tau_A d\tau^\prime_B \chi(\tau_A)\chi(\tau^\prime_B)e^{-i \tau_A \Delta E_A}  e^{-i \tau^\prime_B \Delta E_B}G_w(x^\prime_B,x_A) 
					\nonumber 
					\\
					&& +\alpha \gamma^* C_A C_B\bra{E_1^A}m_A\ket{E_0^A}\bra{E_1^B}m_B\ket{E_0^B}\int \int d\tau_A d\tau^\prime_B \chi(\tau_A)\chi(\tau^\prime_B)e^{i \tau_A \Delta E_A}  e^{i \tau^\prime_B \Delta E_B}G_w(x_A,x^\prime_B) 
					\nonumber 
					\\
					&& +\alpha \alpha^* C_B^2\left|\bra{E_0^B}m_B\ket{E_1^B}\right|^2 \int \int d\tau_B d\tau^\prime_B \chi(\tau_B)\chi(\tau^\prime_B)e^{-i\Delta \tau_B \Delta E_B}G_w(x^\prime_B,x_B) 
					\nonumber 
					\\ 
					&& \equiv C_A ^2\gamma \gamma^*  P^{ \prime \prime }_A+C_A C_B\gamma \alpha^* \bar P_{AB}^{\prime} +C_A C_B\gamma^* \alpha  \bar P^{\prime *}_{AB}+ C_B^2\alpha \alpha^* P_B~, 
					\\ \label{b0}
					&& R^1_{0110}=\gamma \alpha^* C_A^2\bra{E_0^A}m_A\ket{E_1^A}\bra{E_0^A}m_A\ket{E_1^A}\int \int d\tau_A d\tau^\prime_A \chi(\tau_A)\chi(\tau^\prime_A)e^{-i \tau_A \Delta E_A}  e^{-i \tau^\prime_A \Delta E_A}G_w(x^\prime_A,x_A) \nonumber 
					\\ 
					&& +\gamma \gamma^* C_A C_B\bra{E_0^A}m_A\ket{E_1^A}\bra{E_1^B}m_B\ket{E_0^B}\int \int d\tau_A d\tau^\prime_B \chi(\tau_A)\chi(\tau^\prime_B)e^{-i \tau_A \Delta E_A}  e^{i \tau^\prime_B \Delta E_B}G_w(x^\prime_B,x_A) 
					\nonumber 
					\\
					&& +\alpha \alpha^* C_A C_B\bra{E_0^A}m_A\ket{E_1^A}\bra{E_1^B}m_B\ket{E_0^B}\int \int d\tau_A d\tau^\prime_B \chi(\tau_A)\chi(\tau^\prime_B)e^{-i \tau_A \Delta E_A}  e^{i \tau^\prime_B \Delta E_B}G_w(x_A,x^\prime_B) 
					\nonumber 
					\\
					&& +\alpha \gamma^* C_B^2\left|\bra{E_0^B}m_B\ket{E_1^B}\right|^2 \int \int d\tau_B d\tau^\prime_B \chi(\tau_B)\chi(\tau^\prime_B)e^{i\tau_B \Delta E_B}e^{i\tau^\prime_B \Delta E_B}G_w(x^\prime_B,x_B) 
					\nonumber 
					\\
					&& \equiv C_A^2\gamma \alpha^* \bar{P_A}+C_A C_B\gamma^* \gamma X_{AB} +C_A C_B\alpha \alpha^*P^*_{AB}+C_B^2 \gamma^*  \alpha\bar{P^\prime_B}~,
					\\  \label{b1}
					&& R^1_{1001}=\gamma^* \alpha C_A^2\bra{E_1^A}m_A\ket{E_0^A}\bra{E_1^A}m_A\ket{E_0^A}\int \int d\tau_A d\tau^\prime_A \chi(\tau_A)\chi(\tau^\prime_A)e^{i \tau_A \Delta E_A}  e^{i \tau^\prime_A \Delta E_A}G_w(x^\prime_A,x_A) \nonumber 
					\\
					&& +\alpha \alpha^* C_A C_B\bra{E_1^A}m_A\ket{E_0^A}\bra{E_0^B}m_B\ket{E_1^B}\int \int d\tau_A d\tau^\prime_B \chi(\tau_A)\chi(\tau^\prime_B)e^{i \tau_A \Delta E_A}  e^{-i \tau^\prime_B \Delta E_B}G_w(x^\prime_B,x_A) 
					\nonumber 
					\\ 
					&& +\gamma \gamma^* C_A C_B\bra{E_1^A}m_A\ket{E_0^A}\bra{E_0^B}m_B\ket{E_1^B}\int \int d\tau_A d\tau^\prime_B \chi(\tau_A)\chi(\tau^\prime_B)e^{i \tau_A \Delta E_A}  e^{-i \tau^\prime_B \Delta E_B}G_w(x_A,x^\prime_B) 
					\nonumber 
					\\
					&& +\gamma \alpha^* C_B^2\bra{E_0^B}m_B\ket{E_1^B}\bra{E_0^B}m_B\ket{E_1^B} \int \int d\tau_B d\tau^\prime_B \chi(\tau_B)\chi(\tau^\prime_B)e^{-i\tau_B \Delta E_B}e^{-i\tau^\prime_B \Delta E_B}G_w(x^\prime_B,x_B) 
					\nonumber 
					\\
					&& \equiv C_A^2 \gamma^* \alpha \bar{P^\prime_A}+C_A C_B\gamma^* \gamma X_{AB}^* +C_A C_B\alpha \alpha^*P_{AB}+ C_B^2\gamma  \alpha^*\bar{P_B}~,
					\\  \label{b2}
					&& R^1_{1010}=\alpha^*\alpha \left|\bra{E_0^A}m_A\ket{E_1^A}\right|^2 C_A^2\int \int d\tau_A d\tau^\prime_A \chi(\tau_A)\chi(\tau^\prime_A)e^{-i\Delta \tau_A \Delta E_A}G_w(x^\prime_A,x_A) 
					\nonumber 
					\\ 
					&& +\gamma^* \alpha C_A C_B\bra{E_1^A}m_A\ket{E_0^A}\bra{E_1^B}m_B\ket{E_0^B}\int \int d\tau_A d\tau^\prime_B \chi(\tau_A)\chi(\tau^\prime_B)e^{i \tau_A \Delta E_A}  e^{i \tau^\prime_B \Delta E_B}G_w(x^\prime_B,x_A) 
					\nonumber 
					\\
					&& +\gamma \alpha^* C_A C_B\bra{E_0^A}m_A\ket{E_1^A}\bra{E_0^B}m_B\ket{E_1^B}\int \int d\tau_A d\tau^\prime_B \chi(\tau_A)\chi(\tau^\prime_B)e^{-i \tau_A \Delta E_A}  e^{-i \tau^\prime_B \Delta E_B}G_w(x_A,x^\prime_B) 
					\nonumber 
					\\
					&& +\gamma \gamma^* C_B^2\left|\bra{E_0^B}m_B\ket{E_1^B}\right|^2 \int \int d\tau_B d\tau^\prime_B \chi(\tau_B)\chi(\tau^\prime_B)e^{-i\Delta \tau_B \Delta E_B}G_w(x_B,x_B^\prime) 
					\nonumber 
					\\ 
					&& \equiv  C_B^2\gamma \gamma^*  P^{ \prime \prime}_B+C_A C_B\gamma \alpha^*  P_{AB}^{\prime} +C_A C_B\gamma^* \alpha  P^{{ \prime *}}_{AB}+C_A^2\alpha \alpha^* P_A~.
					\label{B3}
				\end{eqnarray}
				To obtain \eqref{b0}-\eqref{B3} from \eqref{R1}, we change the dummy variables $\tau_A^\prime \rightarrow \tau_A$ and $\tau_B \rightarrow \tau_B^\prime$, in some expressions to manintain consistancy of notation.

				Next, we calculate $R^{(2)}_{n_A,n_B,\bar n_A,\bar n_B}$. This is given by
				\begin{eqnarray}
					&&R^{(2)}_{n_A,n_B,\bar n_A,\bar n_B}= -Tr_{\phi}[\bra{ E_{n_A}^A E_{n_B}^B}\frac{T(S_{int}S_{int}^\prime)}{2}|(\alpha\ket{ E_0^A E_0^B}+\gamma\ket{ E_1^A E_1^B})\ket{0}\bra{0}(\alpha^\star\bra{ E_0^A E_0^B}+\gamma^\star\bra{ E_1^A E_1^B}){\mathrm{I}}\ket{ E_{\bar n_A}^A E_{\bar n_B}^B}] 
					\nonumber 
					\\
					&&		=-(\alpha^*\delta^0_{\bar n_A}\delta^0_{\bar n_B}+\gamma^*\delta^1_{\bar n_A}\delta^1_{\bar n_B})* \bra{ E_{n_A}^A E_{n_B}^B}\bra{0}\frac{T}{2}(\int C_A \chi_A(\tau_A)m_A(\tau_A)\phi_A(x_A)d\tau_A + \int C_B \chi_B(\tau_B)m_B(\tau_B)\phi_B(x_B)d\tau_B) 
					\nonumber 
					\\
					&& \times (\int C_A \chi_A(\tau_A^\prime)m_A(\tau_A^\prime)\phi_A(x^\prime_A)d\tau_A^\prime + \int C_B \chi_B(\tau_B^\prime)m_B(\tau_B^\prime)\phi_B(x^\prime_B)d\tau_B^\prime)\ket{0}(\ket{\alpha E_0^A E_0^B + \gamma  E_1^A E_1^B}) 
					\nonumber 
					\\
					&&		=-(\alpha^*\delta^0_{\bar n_A}\delta^0_{\bar n_B}+\gamma^*\delta^1_{\bar n_A}\delta^1_{\bar n_B})*
					\bra{ E_{n_A}^A E_{n_B}^B}\bra{0}\frac{1}{2}(\int \int d\tau_A d\tau^\prime_A  C_A^2)\chi(\tau_A)\chi(\tau^\prime_A) T[ m_A(\tau_A)m_A(\tau^\prime_A)\phi(x_A)\phi(x^\prime_A)]
					\nonumber 
					\\
					&& +\int \int d\tau_A d\tau^\prime_B  C_A C_B\chi(\tau_A)\chi(\tau^\prime_B) T[m_A(\tau_A)m_B(\tau^\prime_B)\phi(x_A)\phi(x^\prime_B)] 
					\nonumber
					\\
					&&+ \int \int d\tau_B d\tau^\prime_A  C_A C_B\chi(\tau_B)\chi(\tau^\prime_A) T[ m_B(\tau_B)m_A(\tau^\prime_A)\phi(x_B)\phi(x^\prime_A)]
					\nonumber 
					\\
					&& + \int \int d\tau_B d\tau^\prime_B   C_B^2\chi(\tau_B)\chi(\tau^\prime_B) T[m_B(\tau_B)m_B(\tau^\prime_B)\phi(x_B)\phi(x^\prime_B)])(\ket{\alpha E_0^A E_0^B + \gamma  E_1^A E_1^B})
					\nonumber 
					\\
					&& =-(\alpha^*\delta^0_{\bar n_A}\delta^0_{\bar n_B}+\gamma^*\delta^1_{\bar n_A}\delta^1_{\bar n_B})*
					\bra{ E_{n_A}^A E_{n_B}^B}\bra{0}\frac{1}{2}(\int \int d\tau_A d\tau^\prime_A  C_A^2)\chi(\tau_A)\chi(\tau^\prime_A) T[ m_A(\tau_A)m_A(\tau^\prime_A)]T[\phi(x_A)\phi(x^\prime_A)]
					\nonumber 
					\\
					&&+ \int \int d\tau_A d\tau^\prime_B  C_A C_B\chi(\tau_A)\chi(\tau^\prime_B) T[m_A(\tau_A)m_B(\tau^\prime_B)]T[\phi(x_A)\phi(x^\prime_B)] 
					\nonumber
					\\
					&&+\int\int d\tau_B d\tau^\prime_A  C_A C_B\chi(\tau_B)\chi(\tau^\prime_A) T[ m_B(\tau_B)m_A(\tau^\prime_A)]T[\phi(x_B)\phi(x^\prime_A)]
					\nonumber 
					\\
					&& + \int \int d\tau_B d\tau^\prime_B   C_B^2\chi(\tau_B)\chi(\tau^\prime_B) T[m_B(\tau_B)m_B(\tau^\prime_B)]T[\phi(x_B)\phi(x^\prime_B)])(\ket{\alpha E_0^A E_0^B + \gamma  E_1^A E_1^B})
					\nonumber
					\\
                  	&&		=-(\alpha^*\delta^0_{\bar n_A}\delta^0_{\bar n_B}+\gamma^*\delta^1_{\bar n_A}\delta^1_{\bar n_B})*
					\bra{ E_{n_A}^A E_{n_B}^B}\bra{0}(\int \int d\tau_A d\tau^\prime_A  C_A^2)\chi(\tau_A)\chi(\tau^\prime_A) \Theta(\tau_A-\tau_A^\prime) m_A(\tau_A)m_A(\tau^\prime_A)T[\phi(x_A)\phi(x^\prime_A)]
					\nonumber 
					\\
					&&+ \int \int d\tau_A d\tau^\prime_B  C_A C_B\chi(\tau_A)\chi(\tau^\prime_B) \Theta(\tau_A-\tau_B^\prime)m_A(\tau_A)m_B(\tau^\prime_B)T[\phi(x_A)\phi(x^\prime_B)] 
					\nonumber
					\\
					&&+\int\int d\tau_B d\tau^\prime_A  C_A C_B\chi(\tau_B)\chi(\tau^\prime_A)  \Theta(\tau_B-\tau_A^\prime)m_B(\tau_B)m_A(\tau^\prime_A)T[\phi(x_B)\phi(x^\prime_A)]
					\nonumber 
					\\
					&& + \int \int d\tau_B d\tau^\prime_B   C_B^2\chi(\tau_B)\chi(\tau^\prime_B)\Theta(\tau_B-\tau_B^\prime) m_B(\tau_B)m_B(\tau^\prime_B)T[\phi(x_B)\phi(x^\prime_B)])(\ket{\alpha E_0^A E_0^B + \gamma  E_1^A E_1^B})
					\nonumber
					\\
					&&		=-(\alpha^*\delta^0_{\bar n_A}\delta^0_{\bar n_B}+\gamma^*\delta^1_{\bar n_A}\delta^1_{\bar n_B})*
					\{\alpha \delta^0_{n_B}\Sigma_k\bra{E_{n_A}^A}m_A(0)\ket{E^A_k}\bra{E_k^A}m_A(0)\ket{E^A_0}
					\nonumber
					\\
					&& \times \int \int d\tau_A d\tau^\prime_A   C_A^2 \chi(\tau_A)\chi(\tau^\prime_A) e^{i\tau_A(E_{n_A}^A-E_{k}^A)} e^{-i\tau^\prime_A(E_{0}^A-E_{k}^A)}\Theta(\tau_A-\tau^\prime_A)\langle{0}|T\phi(x_A)\phi(x^\prime_A)|0\rangle 
					\nonumber
					\\
					&&		+\alpha\bra{E_{n_A}^A}m_A(0)\ket{E^A_0}\bra{E_{n_B}^B}m_B(0)\ket{E^B_0}C_A C_B 
					\nonumber
					\\
					&& \times \int \int d\tau_A d\tau^\prime_B    \chi(\tau_A)\chi(\tau^\prime_B) e^{i\tau_A(E_{n_A}^A-E_{0}^A)} e^{-i\tau^\prime_B(E_{0}^B-E_{n_B}^B)}\Theta(\tau_A-\tau^\prime_B)\langle{0}|T\phi(x_A)\phi(x^\prime_B)|0\rangle
					\nonumber
					\\
					&&		+\alpha\bra{E_{n_B}^B}m_B(0)\ket{E^B_0}\bra{E_{n_A}^A}m_A(0)\ket{E^A_0}C_A C_B
					\nonumber
					\\
					&& \times \int \int d\tau_B d\tau^\prime_A    \chi(\tau_B)\chi(\tau^\prime_A) e^{i\tau_B(E_{n_B}^B-E_{0}^B)} e^{-i\tau^\prime_A(E_{0}^A-E_{n_A}^A)}\Theta(\tau_B-\tau^\prime_A)\langle{0}|T\phi(x_B)\phi(x^\prime_A)|0\rangle
					\nonumber
					\\
					&& +\alpha \delta^0_{n_A}\Sigma_k\bra{E_{n_B}^B}m_B(0)\ket{E^B_k}\bra{E_k^B}m_B(0)\ket{E^B_0}
					\nonumber
					\\
					&& \times \int \int d\tau_B d\tau^\prime_B   C_B^2 \chi(\tau_B)\chi(\tau^\prime_B) e^{i\tau_B(E_{n_B}^B-E_{k}^B)} e^{-i\tau^\prime_B(E_{0}^B-E_{k}^B)}\Theta(\tau_B-\tau^\prime_B)\langle{0}|T\phi(x_A)\phi(x^\prime_A)|0\rangle 
					\nonumber
					\\
					&& +\gamma \delta^1_{n_B}\Sigma_k\bra{E_{n_A}^A}m_A(0)\ket{E^A_k}\bra{E_k^A}m_A(0)\ket{E^A_1}
					\nonumber
					\\
					&& \times \int \int d\tau_A d\tau^\prime_A   C_A^2 \chi(\tau_A)\chi(\tau^\prime_A) e^{i\tau_A(E_{n_A}^A-E_{k}^A)} e^{-i\tau^\prime_A(E_{1}^A-E_{k}^A)}\Theta(\tau_A-\tau^\prime_A)\langle{0}|T\phi(x_A)\phi(x^\prime_A)|0\rangle 
					\nonumber
					\\
					&&		+\gamma\bra{E_{n_A}^A}m_A(0)\ket{E^A_1}\bra{E_{n_B}^B}m_B(0)\ket{E^B_1}C_A C_B
					\nonumber
					\\
					&& \times \int \int d\tau_A d\tau^\prime_B    \chi(\tau_A)\chi(\tau^\prime_B) e^{i\tau_A(E_{n_A}^A-E_{1}^A)} e^{-i\tau^\prime_B(E_{1}^B-E_{n_B}^B)}\Theta(\tau_A-\tau^\prime_B)\langle{0}|T\phi(x_A)\phi(x^\prime_B)|0\rangle
					\nonumber
					\end{eqnarray}
					\begin{eqnarray}
					&&		+\gamma\bra{E_{n_B}^B}m_B(0)\ket{E^B_1}\bra{E_{n_A}^A}m_A(0)\ket{E^A_1}C_A C_B
					\nonumber
					\\
					&& \times \int \int d\tau_B d\tau^\prime_A    \chi(\tau_B)\chi(\tau^\prime_A) e^{i\tau_B(E_{n_B}^B-E_{1}^B)} e^{-i\tau^\prime_A(E_{1}^A-E_{n_A}^A)}\Theta(\tau_B-\tau^\prime_A)\langle{0}|T\phi(x_B)\phi(x^\prime_A)|0\rangle
					\nonumber
					\\
					&& +\gamma \delta^1_{n_A}\Sigma_k\bra{E_{n_B}^B}m_B(0)\ket{E^B_k}\bra{E_k^B}m_B(0)\ket{E^B_0}
					\nonumber
					\\
					&& \times \int \int d\tau_B d\tau^\prime_B   C_B^2 \chi(\tau_B)\chi(\tau^\prime_B) e^{i\tau_B(E_{n_B}^B-E_{k}^B)} e^{-i\tau^\prime_B(E_{0}^B-E_{k}^B)}\Theta(\tau_B-\tau^\prime_B)\langle{0}|T\phi(x_A)\phi(x^\prime_A)|0\rangle\}~. \nonumber\\
					\label{RR-gen}
				\end{eqnarray}
				Now, in the third term of \eqref{RR-gen}, we replace $\tau^\prime_A \rightarrow \tau_A $,   $\tau_B \rightarrow \tau^\prime_B$ and use 
				$\Theta(  \tau_A-\tau^\prime_B)+\Theta(\tau^\prime_B-\tau_A)=1$
				to add this with the second term. 
			We, follow the same steps for the sixth and seventh terms as well.
				Also, since the Feynman propagator follows the property, 
				\begin{equation}
					G_F(x,x^\prime)= G_F(x^\prime,x)=
					\langle{0}|T\phi(x_B)\phi(x^\prime_A)|0\rangle=\langle{0}|T\phi(x^\prime_A)\phi(x_B)|0\rangle~,
				\end{equation}
				 Eq. \eqref{RR-gen} simplifies to
				\begin{eqnarray}
					&&R^{(2)}_{n_A,n_B,\bar n_A,\bar n_B}=  -(\alpha^*\delta^0_{\bar n_A}\delta^0_{\bar n_B}+\gamma^*\delta^1_{\bar n_A}\delta^1_{\bar n_B})* \Big[\alpha \delta_{n_B}^0\sum \bra{E_{n_A}^A}m_A\ket{E_{k}^A}\bra{E_{k}^A}m_A\ket{E_{0}^A}
					\nonumber
					\\
					&& \times \int \int d\tau_A d\tau_A^\prime e^{i\tau_A (E_{n_A}^A-E_{k}^A)}e^{-i\tau_A^\prime (E_{0}^A-E_{k}^A)}\Theta(\tau_A-\tau_A^\prime ) \chi(\tau_A)\chi(\tau_A^\prime)\bra{0}T\phi(x_A)\phi(x_A^\prime)\ket{0} \nonumber
					\\ 
					&& + \alpha \bra{E_{n_A}^A}m_A\ket{E_{0}^A}\bra{E_{n_B}^B}m_B\ket{E_{0}^B}
					\nonumber
					\\
					&& \times \int \int d\tau_A d\tau_B^\prime e^{i\tau_A (E_{n_A}^A-E_{0}^A)}e^{-i\tau_B^\prime (E_{0}^B-E_{n_B}^B)} \chi(\tau_A)\chi(\tau_B^\prime)\bra{0}T\phi(x_A)\phi(x_B^\prime)\ket{0} 
					\nonumber
					\\ 
					&&+\alpha \delta_{n_A}^0\sum \bra{E_{n_B}^B}m_B\ket{E_{k}^B}\bra{E_{k}^B}m_B\ket{E_{0}^B}
					\nonumber
					\\
					&& \times \int \int d\tau_B d\tau_B^\prime e^{i\tau_B (E_{n_B}^B-E_{k}^B)}e^{-i\tau_B^\prime (E_{0}^B-E_{k}^B)}
					\Theta(\tau_B-\tau_B^\prime ) \chi(\tau_B)\chi(\tau_B^\prime)\bra{0}T\phi(x_B)\phi(x_B^\prime)\ket{0} \nonumber
					\\ 
					&& +\gamma \delta_{n_B}^1\sum \bra{E_{n_A}^A}m_A\ket{E_{k}^A}\bra{E_{k}^A}m_A\ket{E_{1}^A}
					\nonumber
					\\
					&& \times \int \int d\tau_A d\tau_A^\prime e^{i\tau_A (E_{n_A}^A-E_{k}^A)}e^{-i\tau_A^\prime (E_{1}^A-E_{k}^A)}
					\Theta(\tau_A-\tau_A^\prime ) \chi(\tau_A)\chi(\tau_A^\prime)\bra{0}T\phi(x_A)\phi(x_A^\prime)\ket{0} \nonumber
					\\ 
					&& +\gamma  \bra{E_{n_A}^A}m_A\ket{E_{1}^A}\bra{E_{n_B}^B}m_B\ket{E_{1}^B}
					\nonumber
					\\
					&& \times \int \int d\tau_A d\tau_B^\prime e^{i\tau_A (E_{n_A}^A-E_{1}^A)} e^{-i\tau_B^\prime (E_{1}^B-E_{n_A}^B)}
					\chi(\tau_A)\chi(\tau_B^\prime)\bra{0}T\phi(x_A)\phi(x_B^\prime)\ket{0} 
					\nonumber
					\\ 
					&& +\gamma \delta_{n_A}^1\sum \bra{E_{n_B}^B}m_B\ket{E_{k}^B}\bra{E_{k}^B}m_B\ket{E_{1}^B}
					\nonumber
					\\
					&& \times \int \int d\tau_B d\tau_B^\prime e^{i\tau_B (E_{n_B}^B-E_{k}^B)}e^{-i\tau_B^\prime (E_{1}^B-E_{k}^B)}
					\Theta(\tau_B-\tau_B^\prime ) \chi(\tau_B)\chi(\tau_B^\prime)\bra{0}T\phi(x_B)\phi(x_B^\prime)\ket{0}\Big]~.
					\label{R2}
				\end{eqnarray}
				Like earlier, we have denoted $m_j(0)=m_j$ as well.

				From the above, one finds $	R^{2}_{0001}=0, R^{2}_{0010}=0, R^{2}_{0100}=0, R^{2}_{0101}=0, R^{2}_{0110}=0, R^{2}_{0111}=0, R^{2}_{1000}=0, R^{2}_{1001}=0, R^{2}_{1010}=0, R^{2}_{1011}=0, R^{2}_{1101}=0, $ and $R^{2}_{1110}=0$, while the non-vanishing components are as follows.
				\begin{eqnarray}
					&&R^2_{0000}=-\alpha \alpha^* C_A^2|\bra{E_0^A}m_A\ket{E_1^A}|^2\int \int d\tau_A d\tau^\prime_A \chi(\tau_A)\chi(\tau^\prime_A)e^{i\Delta \tau_A \Delta E_A}\Theta(\tau_A-\tau^\prime_A)G_w(x_A,x^\prime_A) \nonumber 
					\\
					&& -\alpha \alpha^* C_B^2|\bra{E_0^B}m_B\ket{E_1^B}|^2\int \int d\tau_B d\tau^\prime_B \chi(\tau_B)\chi(\tau^\prime_B)e^{i\Delta \tau_B \Delta E_B}\Theta(\tau_B-\tau^\prime_B)G_w(x_B,x^\prime_B) 
					\nonumber 
					\\
					&& - \alpha^*\gamma C_A C_B \bra{E_0^A}m_A\ket{E_1^A}\bra{E_0^B}m_B\ket{E_1^B}\int \int d\tau_A d\tau^\prime_B \chi(\tau_A)\chi(\tau^\prime_B)e^{-i \tau_A \Delta E_A}e^{-i \tau^\prime_B \Delta E_B}\{iG_F(x_A,x^\prime_B)\}~.
					\label{R2-0000}
				\end{eqnarray}
			
				\begin{eqnarray}
					&&R^{2}_{0011}= -\gamma^*\alpha C_A^2\bra{E_0^A}m_A\ket{E_1^A}\bra{E_1^A}m_A\ket{E_0^A}\int \int d\tau_A d\tau^\prime_A \chi(\tau_A)\chi(\tau^\prime_A)e^{-i \tau_A \Delta E_A}e^{i \tau^\prime_A \Delta E_A}\Theta(\tau_A-\tau^\prime_A)G_W(x_A,x^\prime_A)	
					\nonumber 
					\\
					&& -\gamma\alpha^*  C_B^2 \bra{E_0^B}m_B\ket{E_1^B}\bra{E_1^B}m_B\ket{E_0^B}\int \int d\tau_B d\tau^\prime_B \chi(\tau_B)\chi(\tau^\prime_B)e^{-i \tau_B \Delta E_B}e^{i \tau^\prime_B \Delta E_B}\Theta(\tau_B-\tau^\prime_B)G_W(x_B,x^\prime_B)	
					\nonumber 
					\\
					&& - \gamma^*\gamma C_B C_A \bra{E_0^A}m_A\ket{E_1^A}\bra{E_0^B}m_B\ket{E_1^B}\int \int d\tau_A d\tau^\prime_B \chi(\tau_A)\chi(\tau^\prime_B)e^{-i \tau_A \Delta E_A}e^{-i \tau^\prime_B \Delta E_B}\{iG_F(x_A,x^\prime_B)\}~,
					\\
					&&R^{2}_{1100}=- \alpha^*\alpha C_B C_A \bra{E_1^A}m_A\ket{E_0^A}\bra{E_1^B}m_B\ket{E_0^B}\int \int d\tau_A d\tau^\prime_B \chi(\tau_A)\chi(\tau^\prime_B)e^{i \tau_A \Delta E_A}e^{i \tau^\prime_B \Delta E_B}\{iG_F(x_A,x^\prime_B)\}	
					\nonumber 
					\\
					&&		-\gamma\alpha^*C_A^2
					|\bra{E_0^A}m_A\ket{E_1^A}|^2\int \int d\tau_A d\tau^\prime_A \chi(\tau_A)\chi(\tau^\prime_A)e^{i \tau_A \Delta E_A}e^{-i \tau^\prime_A \Delta E_A}\Theta(\tau_A-\tau^\prime_A)G_W(x_A,x^\prime_A)	
					\nonumber 
					\\
					&&		-\gamma\alpha^*  C_B^2 |\bra{E_0^B}m_B\ket{E_1^B}|^2\int \int d\tau_B d\tau^\prime_B \chi(\tau_B)\chi(\tau^\prime_B)e^{i \tau_B \Delta E_B}e^{-i \tau^\prime_B \Delta E_B}\Theta(\tau_B-\tau^\prime_B)G_W(x_B,x^\prime_B)~,	
					\\
					&&R^{2}_{1111}= - \gamma^*\alpha C_B C_A \bra{E_1^A}m_A\ket{E_0^A}\bra{E_1^B}m_B\ket{E_0^B}\int \int d\tau_A d\tau^\prime_B \chi(\tau_A)\chi(\tau^\prime_B)e^{i \tau_A \Delta E_A}e^{i \tau^\prime_B \Delta E_B}\{iG_F(x_A,x^\prime_B)\}	
					\nonumber 
					\\
					&& -\gamma\gamma^*C_A^2
					|\bra{E_0^A}m_A\ket{E_1^A}|^2\int \int d\tau_A d\tau^\prime_A \chi(\tau_A)\chi(\tau^\prime_A)e^{i \tau_A \Delta E_A}e^{-i \tau^\prime_A \Delta E_A}\Theta(\tau_A-\tau^\prime_A) G_W(x_A,x^\prime_A)	
					\nonumber 
					\\
					&& -\gamma\gamma^*  C_B^2 |\bra{E_0^B}m_B\ket{E_1^B}|^2\int \int d\tau_B d\tau^\prime_B \chi(\tau_B)\chi(\tau^\prime_B)e^{i \tau_B \Delta E_B}e^{-i \tau^\prime_B \Delta E_B}\Theta(\tau_B-\tau^\prime_B) G_W(x_B,x^\prime_B)~.
				\end{eqnarray}
				Next, interchanging $\tau_A \leftrightarrow  \tau^\prime_A$ and $\tau^\prime_B \leftrightarrow \tau_B$ in the first and second terms of \eqref{R2-0000}, we obtain
				\begin{eqnarray}
					&&R^2_{0000}=-\alpha \alpha^* C_A^2|\bra{E_0^A}m_A\ket{E_1^A}|^2\int \int d\tau_A d\tau^\prime_A \chi(\tau_A)\chi(\tau^\prime_A)e^{-i\Delta \tau_A \Delta E_A}\Theta(\tau^\prime_A-\tau_A)G_w(x^\prime_A,x_A) \nonumber 
					\\
					&& -\alpha \alpha^* C_B^2|\bra{E_0^B}m_B\ket{E_1^B}|^2\int \int d\tau_B d\tau^\prime_B \chi(\tau_B)\chi(\tau^\prime_B)e^{-i\Delta \tau_B \Delta E_B}\Theta(\tau^\prime_B-\tau_B)G_w(x^\prime_B,x_B) 
					\nonumber 
					\\
					&& - \alpha^*\gamma C_A C_B \bra{E_0^A}m_A\ket{E_1^A}\bra{E_0^B}m_B\ket{E_1^B}\int \int d\tau_A d\tau^\prime_B \chi(\tau_A)\chi(\tau^\prime_B)e^{-i \tau_A \Delta E_A}e^{-i \tau^\prime_B \Delta E_A} {iG_F(x_A,x^\prime_B)}~.
					\label{R2-0000-interchanged}
				\end{eqnarray}

				Taking the complex conjugate of \eqref{R2-0000}, we get
				\begin{eqnarray}	
					&&{R^{2*}_{0000}}=-\alpha \alpha^* C_A^2|\bra{E_0^A}m_A\ket{E_1^A}|^2\int \int d\tau_A d\tau^\prime_A \chi(\tau_A)\chi(\tau^\prime_A)e^{-i\Delta \tau_A \Delta E_A}\Theta(\tau_A-\tau^\prime_A)G_w(x^\prime_A,x_A) 
					\nonumber 
					\\
					&& -\alpha \alpha^* C_B^2|\bra{E_0^B}m_B\ket{E_1^B}|^2\int \int d\tau_B d\tau^\prime_B \chi(\tau_B)\chi(\tau^\prime_B)e^{-i\Delta \tau_B \Delta E_B}\Theta(\tau_B-\tau^\prime_B)G_w(x^\prime_B,x_B) 
					\nonumber 
					\\
					&& - \alpha\gamma^* C_A C_B \bra{E_1^A}m_A\ket{E_0^A}\bra{E_1^B}m_B\ket{E_0^B}\int \int d\tau_A d\tau^\prime_B \chi(\tau_A)\chi(\tau^\prime_B)e^{i \tau_A \Delta E_A}e^{i \tau^\prime_B \Delta E_A} \{iG_F(x_A,x^\prime_B)\}^*~.
					\label{R2-0000*}
				\end{eqnarray}
				Therefore, adding \eqref{R2-0000-interchanged} and \eqref{R2-0000*}  and  using $\Theta(\tau_A-\tau^\prime_A)+\Theta(\tau^\prime_A-\tau_A)=1$ we have :
				\begin{eqnarray}
					&& R^{2}_{0000}+R^{2*}_{0000}=-\alpha \alpha^* C_A^2|\bra{E_0^A}m_A\ket{E_1^A}|^2\int \int d\tau_A d\tau^\prime_A \chi(\tau_A)\chi(\tau^\prime_A)e^{-i\Delta \tau_A \Delta E_A}G_w(x^\prime_A,x_A) 
					\nonumber 
					\\
					&& -\alpha \alpha^* C_B^2|\bra{E_0^B}m_B\ket{E_1^B}|^2\int \int d\tau_B d\tau^\prime_B \chi(\tau_B)\chi(\tau^\prime_B)e^{-i\Delta \tau_B \Delta E_B}G_w(x^\prime_B,x_B) 
					\nonumber 
					\\
					&& -\alpha^*\gamma C_A C_B \bra{E_0^A}m_A\ket{E_1^A}\bra{E_0^B}m_B\ket{E_1^B}\int \int d\tau_A d\tau^\prime_B \chi(\tau_A)\chi(\tau^\prime_B)e^{-i \tau_A \Delta E_A}e^{-i \tau^\prime_B \Delta E_B}\times \{iG_F(x_A,x^\prime_B)\}
					\nonumber 
					\\
					&& - \alpha\gamma^* C_A C_B \bra{E_1^A}m_A\ket{E_0^A}\bra{E_1^B}m_B\ket{E_0^B}\int \int d\tau_A d\tau^\prime_B \chi(\tau_A)\chi(\tau^\prime_B)e^{i \tau_A \Delta E_A}e^{i \tau^\prime_B \Delta E_A}\times \{i G_F(x^\prime_B,x_A)\}^*
					\nonumber 
					\\ 
					&& \equiv -C_A ^2\alpha^*\alpha P_A- C_B^2\alpha^*\alpha P_B-C_A C_B\alpha^*\gamma Y_{AB}-C_A C_B\alpha\gamma^* Y_{AB}^*~.
				\end{eqnarray}
				Similarly, we can calculate, 
				\begin{eqnarray}
					&&R^{(2)}_{1111}+R^{(2)*}_{1111}=- C_A^2\gamma \gamma^* P_A^{\prime \prime} - C_B^2\gamma \gamma^* P_B^{\prime \prime} - C_A C_B\alpha^*\gamma \zeta^*_{AB}-C_A C_B\alpha\gamma^* \zeta_{AB}~.\\
					\nonumber
				\end{eqnarray}
				Further, we have, 
				\begin{eqnarray}
					&&R^{(2)}_{0011}+R^{(2)*}_{1100}=-\gamma^*\alpha C_A^2
							|\bra{E_0^A}m_A\ket{E_1^A}|^2
					\nonumber
					\\
					&&\times \int \int d\tau_A d\tau^\prime_A \chi(\tau_A)\chi(\tau^\prime_A)e^{-i \tau_A \Delta E_A}e^{i \tau^\prime_A \Delta E_A}\Theta(\tau_A-\tau^\prime_A)(G_W(x_A,x^\prime_A)+G_W(x^\prime_A,x_A))	
					\nonumber 
					\\
					&& - \alpha\gamma^* C_B^2 |\bra{E_0^B}m_B\ket{E_1^B}|^2
					\int \int d\tau_B d\tau^\prime_B \chi(\tau_B)\chi(\tau^\prime_B)e^{-i \tau_B \Delta E_B}e^{i \tau^\prime_B \Delta E_B}\Theta(\tau_B-\tau^\prime_B)(G_W(x_B,x^\prime_B)+G_W(x^\prime_B,x_B))	
					\nonumber 
					\\
					&& - \gamma^*\gamma C_B C_A \bra{E_0^A}m_A\ket{E_1^A}\bra{E_0^B}m_B\ket{E_1^B}\int \int d\tau_A d\tau^\prime_B \chi(\tau_A)\chi(\tau^\prime_B)e^{-i \tau_A \Delta E_A}e^{-i \tau^\prime_B \Delta E_B}iG_F(x_A,x^\prime_B)	
					\nonumber 
					\\
					&& - \alpha^*\alpha C_B C_A \bra{E_0^A}m_A\ket{E_1^A}\bra{E_0^B}m_B\ket{E_1^B}\int \int d\tau_A d\tau^\prime_B \chi(\tau_A)\chi(\tau^\prime_B)e^{-i \tau_A \Delta E_A}e^{-i \tau^\prime_B \Delta E_B}(iG_F(x_A,x^\prime_B))^*	
					\nonumber 
					\\
					&& \equiv -C_A^2\gamma^*\alpha M_A^*-C_B^2\gamma^*\alpha  M_B^* -C_A C_B\gamma\gamma^* Y_{AB}-C_A C_B\alpha \alpha^*\zeta_{AB}^*~.
				\end{eqnarray}
				
				Similarly, one finds
				\begin{eqnarray}
					&&R^{(2)}_{1100}+R^{(2)*}_{0011}= -C_A C_B \alpha \alpha^* \zeta_{AB}-C_A C_B\gamma \gamma^* Y^*_{AB} - C_A^2\alpha ^* \gamma M_A - C_B^2\gamma\alpha^* { M_B}~.
					\nonumber 
					\\
				\end{eqnarray}

				Now, using the above results, we finally calculate the elements of density matrix
				\begin{eqnarray}
					&&\rho_{0001}=0~, \quad \rho_{0010}=0~, \quad \rho_{0100}=0~, \quad \rho_{0111}=0~, \quad \rho_{1011}=0~, \quad \rho_{1000}=0~, \quad \rho_{1101}=0~, \quad \rho_{1110}=0~,
					\nonumber 
					\\
					&&\rho_{0000}= \alpha \alpha^*-C_A^2\alpha^*\alpha P_A-C_B^2\alpha^*\alpha P_B-C_A C_B\alpha^*\gamma Y_{AB}-C_A C_B\alpha\gamma^* Y_{AB}^*~,
					\nonumber 
					\\
					&&\rho_{0011}=\alpha \gamma^*-C_A^2\gamma^*\alpha M_A^*-C_B^2\gamma^*\alpha  M_B^* -C_A C_B\gamma\gamma^* Y_{AB}-C_A C_B\alpha \alpha^*\zeta_{AB}^*~,
					\nonumber 
					\\
					&&\rho_{0101}= C_A^2\gamma \gamma^* P_A^{\prime \prime}+C_A C_B\alpha^* \gamma \bar P^{\prime}_{AB} +C_A C_B\alpha \gamma^*\bar P^{\prime *}_{AB}+ C_B^2\alpha^*  \alpha{P_B}~,
					\nonumber 
					\\
					&& \rho_{0110}=C_A^2\alpha^* \gamma \bar P_A +  C_B^2\alpha \gamma^* \bar P^{\prime }_{B} +C_A C_B\gamma \gamma^* X_{AB}+C_A C_B\alpha \alpha^* P^*_{AB}
					\nonumber  
					\\
					&& \rho_{1001}=C_A ^2\alpha \gamma^* \bar P_A^\prime + C_A C_B\alpha^* \alpha P_{AB}+C_A C_B\gamma \gamma^* X^*_{AB}+ C_B^2\gamma \alpha^*\bar P_B~,
					\nonumber 
					\\
					&& \rho_{1010}=C_A^2\alpha \alpha^*  P_A + C_B^2\gamma \gamma^*  P_B^{\prime \prime} + C_A C_B\alpha^*\gamma  P^{\prime}_{AB}+C_A C_B\alpha\gamma^*  P^{\prime *}_{AB}~,
					\nonumber
					\\
					&& \rho_{1100}=\gamma \alpha^* -C_A C_B \alpha \alpha^* \zeta_{AB}-C_A C_B\gamma \gamma^* Y^*_{AB} - C_A^2\alpha ^* \gamma M_A - C_B^2\gamma\alpha^* { M_B}~,
					\nonumber 
					\\
					&& \rho_{1111}= \gamma \gamma^* - C_A^2\gamma \gamma^* P_A^{\prime \prime} - C_B^2\gamma \gamma^* P_B^{\prime \prime} - C_A C_B\alpha^*\gamma \zeta^*_{AB}-C_A C_B\alpha\gamma^* \zeta_{AB}~.
				\end{eqnarray}
				Then in matrix form we obtain Eq. \eqref{gen-mat}.
	
\section{\label{AppJHEP1} EIGENVALUES (\ref{L1}) -- (\ref{L4}) ARE REAL}
It is easily seen from Eq.(9) of our manuscript that $P_j = P_j^* $ and $P_j^{\prime \prime}=P_j^{\prime \prime *} $, i.e they are real (can be shown by change in variable $\tau \leftrightarrow \tau^ \prime$ in of the expressions ).
Now, note that 
\vskip 3mm
\begin{eqnarray} 
  a_1 &=& \gamma \gamma^* - C_A^2\gamma \gamma^* P_A^{\prime \prime} -  C_B^2\gamma \gamma^* P_B^{\prime \prime} 
  - C_A C_B(\underbrace{\alpha\gamma^* \zeta_{AB}+ \{\alpha\gamma^* \zeta_{AB}\}^* }_{2\mathfrak{R}(\alpha \gamma^* \zeta_{AB})}) ,\nonumber \\
  b_1&=& C_A^2\alpha \alpha^*  P_A + C_B^2\gamma \gamma^*  P_B^{\prime \prime} + C_A C_B (\underbrace{\alpha^*\gamma  P^{\prime}_{AB}+ \{\alpha^*\gamma  P^{\prime}_{AB}\}^\star}_{2\mathfrak{R}(\alpha^* \gamma P_{AB}^\prime)}) , 
  \nonumber \\
  c_2 &=&  C_A^2\gamma^*\gamma P_A^{\prime \prime} + C_B^2\alpha \alpha^* P_B +C_A C_B(\underbrace{\gamma \alpha^*\bar P^{\prime}_{AB}+ \{\gamma \alpha^*\bar P^{\prime}_{AB}\}^\star}_{2\mathfrak{R}(\alpha^* \gamma \bar P_{AB}^\prime)})
   ,
  \nonumber  \\
  d_2&=&\alpha \alpha^*-C_A^2\alpha \alpha^*P_A- C_B^2\alpha \alpha^*P_B- C_A C_B(\underbrace{\alpha^*\gamma Y_{AB}
  + \{\alpha^*\gamma Y_{AB}\}^\star}_{2\mathfrak{R}(\alpha^* \gamma Y_{AB})})  ,
\end{eqnarray}
and thus, $a_1, b_1, c_2, d_2$ in Eq. (\ref{eqn_dyna_elements}) are real. Next using the following result
\begin{eqnarray}
  &&\bar P_j^\star = \bra{E_1^j}m_j(0)\ket{E_0^j}\bra{E_1^j}m_j(0)\ket{E_0^j}\int \int d\tau_j d\tau^\prime_j \chi(\tau_j)\chi(\tau^\prime_j)e^{i \bar \tau_j \Delta E_j}G_w(x_j,x^\prime_j) \nonumber 
  \\&& \textrm{(interchanging the variables}\hspace{.1cm}\tau_j\leftrightarrow \tau^\prime_j\hspace{.1cm} \textrm{and} \hspace{.1cm}x_j\leftrightarrow x^\prime_j)\nonumber 
  \\
  &&               =\bar P_j^\prime  
\end{eqnarray}
one finds $c_1 = b_2^\star$ and $d_1 = a_2^\star$. Use of these in (\ref{L1}) -- (\ref{L4}) shows that these eigenvalues are real.

Having shown that the general expressions for $\lambda_{1,2,3,4}$ will be real, we further show that the expressions for $\lambda_{1,2,3,4}$ relevant for our further computation up to $\mathcal{O}(C^2)$ will also be real.
Let us start with first two eigenvalues:
\begin{align}
	\lambda_{1,2} &= \frac{1}{2}\left\{b_1+c_2\pm\sqrt{(b_1-c_2)^2+4a_2d_1}\right\} \nonumber \\	
	& = \frac{C^2}{2} \big[\alpha^2 ( P_A + P_B) + \gamma^2 (P_B^{\prime \prime} + P_A^{\prime \prime})  + \alpha \gamma  \{(P^{\prime}_{AB} + \bar P^{\prime}_{AB}) + (P^{\prime}_{AB} + \bar P^{\prime}_{AB})^* \}\big]
	\nonumber \\
				  &\pm\frac{1}{2}\sqrt{ \begin{aligned}4\{ (\gamma \alpha)^2 - C^2(\gamma \alpha)\big[\alpha^2( \zeta_{AB} + \zeta^*_{AB}) + \gamma^2 (Y^*_{AB}+Y_{AB}) + \alpha \gamma (M_A+ M^*_A ) 
					+  \alpha \gamma (M_B+ { M^*_B})]\} + \mathcal{O}(C^4)\\   
   \end{aligned}
   }  \nonumber \\
   & \approx \frac{C^2}{2} \big[\alpha^2 ( P_A + P_B) + \gamma^2 (P_B^{\prime \prime} + P_A^{\prime \prime})  + \alpha \gamma  \{(P^{\prime}_{AB} + \bar P^{\prime}_{AB}) + (P^{\prime}_{AB} + \bar P^{\prime}_{AB})^* \}\big]
	\nonumber \\
				  &\pm \begin{aligned}\{ (\gamma \alpha) - \frac{C^2}{2}\big[(\alpha^2 \zeta_{AB}+\gamma^2 Y_{AB} + \alpha \gamma M_A +  \alpha \gamma M_B) + (\alpha^2 \zeta^*_{AB} + \gamma^2 Y^*_{AB}+ \alpha \gamma M^*_A  + \alpha \gamma M^*_B)]\} \\   
   \end{aligned}
     \nonumber \\
	 &\approx {\bar{A}}_r + A_c + A_c^* ~,
\label{gen-L12-R-derive}
  \end{align}
where, 
\begin{eqnarray}
	\bar{A}_{r} &=& \Big[\frac{C^2}{2} \big\{\alpha^2 ( P_A + P_B) + \gamma^2 (P_B^{\prime \prime} + P_A^{\prime \prime})\big\} \pm  (\gamma \alpha)^2\Big]~; 
	\nonumber \\
	A_c &= & \frac{C^2}{2}\Big[\alpha \gamma (P^{\prime}_{AB} + \bar P^{\prime}_{AB}) \mp (\alpha^2 \zeta_{AB}+\gamma^2 Y_{AB} + \alpha \gamma M_A +  \alpha \gamma M_B) \Big]~.
\end{eqnarray}
As explained earlier $\bar{A}_{r}$ is real. Therefore
from \eqref{gen-L12-R-derive} it is now clear that $\lambda_{1,2}$ is real as the imaginary contributions from $A_c$ cancels with that from $A_c^*$.
Thus we have 
\begin{align}
	\lambda_{1,2} 
	 & \approx \mathfrak{R}\Big[\frac{C^2}{2} \{\sum_{j}(\alpha^2  P_j + \gamma^2 P_j^{\prime \prime})  + 2\alpha \gamma  ({ {P^{\prime}_{AB}}} + {{\bar P^{\prime}_{AB}}})\}
	 \nonumber \\
				   &\pm [(\gamma \alpha) - C^2\{\alpha^2 \zeta_{AB}  + \gamma^2  Y_{AB} + \alpha \gamma   z \}]\Big]~.
  \end{align}

Next, the remaining eigenvalues can also be shown to be real in a similar manner:
  \begin{align}
	\lambda_{3,4} & = \frac{1}{2}\left\{a_1+d_2+\sqrt{(a_1-d_2)^2+4b_2c_1}\right\} \nonumber \\
	& = \frac{1}{2}\Big[\gamma^2 + \alpha^2 - C^2\big[\gamma^2( P_A^{\prime \prime} + P_B^{\prime \prime}) + \alpha^2 (P_A + P_B) + \alpha\gamma  (\zeta_{AB}
	 + Y_{AB}) + \alpha\gamma  (\zeta_{AB}
	 + Y_{AB})^* \big]
	\nonumber \\
				  &\pm\sqrt{ \begin{aligned}\Big\{\gamma^2 - \alpha^2 - C^2\big[\gamma^2( P_A^{\prime \prime} + P_B^{\prime \prime}) - \alpha^2 (P_A + P_B) + \alpha\gamma  (\zeta_{AB}
					- Y_{AB}) + \alpha\gamma  (\zeta_{AB}
					- Y_{AB})^*\big]\Big\}^2 + \mathcal{O}(C^4)\\   
   \end{aligned}
   }\hspace{.3cm}\Big]  \nonumber \\
   & \approx \frac{1}{2}\Big[\gamma^2 + \alpha^2 - C^2\big[\gamma^2( P_A^{\prime \prime} + P_B^{\prime \prime}) + \alpha^2 (P_A + P_B) + \alpha\gamma  (\zeta_{AB}
	 + Y_{AB}) + \alpha\gamma  (\zeta_{AB}
	 + Y_{AB})^* \big]
	\nonumber \\
				  &\pm \begin{aligned}\Big\{(\gamma^2 - \alpha^2) - C^2\big[\gamma^2( P_A^{\prime \prime} + P_B^{\prime \prime}) - \alpha^2 (P_A + P_B) + \alpha\gamma  (\zeta_{AB}
					- Y_{AB}) + \alpha\gamma  (\zeta_{AB}
					- Y_{AB})^*\big]\Big\} \\   
   \end{aligned}
   \hspace{.3cm}\Big]  \nonumber \\
   & \approx B_r + B_c + B_c^* 
   \label{gen-L34-R-derive}
  \end{align}	
where, 
\begin{eqnarray}
	\bar{B}_{r} &=&  \frac{1}{2}\Big[\Big\{(\gamma^2 + \alpha^2) - C^2\{\gamma^2( P_A^{\prime \prime} + P_B^{\prime \prime}) + \alpha^2 (P_A + P_B)\}\Big\} \pm \Big\{(\gamma^2 - \alpha^2) - C^2\{\gamma^2( P_A^{\prime \prime} + P_B^{\prime \prime}) - \alpha^2 (P_A + P_B)\} \Big\}\Big]\nonumber \\ 
	B_c &= & -\frac{C^2}{2}\Big[\alpha\gamma  (\zeta_{AB}
	+ Y_{AB}) \pm \alpha\gamma  (\zeta_{AB}
	- Y_{AB})\Big]\nonumber \\
\end{eqnarray}
Since $\bar{B}_{r}$ is real, then following earlier argument one finds again that these are real. 
Thus, we  obtain
\begin{eqnarray}
	\lambda_{3} &=& \gamma^2 + \mathcal{O}(C^2)~;\nonumber \\
	\lambda_{4} &=& \alpha^2 + \mathcal{O}(C^2)~.
\end{eqnarray}

Thus it is clear, although $M_j$, $\zeta_{AB}$ and $Y_{AB}$ are complex, only their real parts contribute to the eigenvalues to quadratic order in the couplings.

			\section{CALCULATION OF INTEGRALS}\label{integrals-all}
					
				\subsection{PARALLELLY ACCELERATING DETECTORS}
					\label{app-grn-same}
									
					
					\subsubsection{(1+1)-dimensions}	\label{app-2d-integral}
					Here, we calculate the integrals in Eq. \eqref{eqn_dyna_integrals} those are required  for the analysis of the Eq. \eqref{negativity}. First, we find
						\begin{eqnarray}
							&&P^\beta_{j_R} = \left|\bra{E_0^j}m_j\ket{E_1^j}\right|^2 \int \int d\tau_j d\tau^\prime_j \chi(\tau_j)\chi(\tau^\prime_j)e^{-i\Delta \tau_j \Delta E_j}G^{\beta +}_{R}(x^\prime_j,x_j) \nonumber \\
							&&=\left|\bra{E_0^j}m_j\ket{E_1^j}\right|^2 \int_0^{\infty}dw\int \int d\tau_j d\tau^\prime_j e^{-i\Delta \tau_j \Delta E_j}\frac{1}{2\pi w} \times \nonumber \\
							&&[\frac{1}{1-e^{-\beta w}}\{e^{-iw \Delta \tau}\frac{1}{1-e^{\frac{-2 \pi w}{a_j}}}+e^{iw \Delta \tau}\frac{1}{e^{\frac{2 \pi w}{a_j}}-1}\} +
							\frac{1}{e^{\beta w}-1}\{e^{iw \Delta \tau}\frac{1}{1-e^{\frac{-2 \pi w}{a_j}}}+e^{-iw \Delta \tau}\frac{1}{e^{\frac{2 \pi w}{a_j}}-1}\}	]~.
						\end{eqnarray}
Now, changing  variables $\tau^\prime_A-\tau_A \rightarrow u $ and $\tau^\prime_A+\tau_A \rightarrow v $, which incorporates the Jacobian as $\frac{1}{2}$, one finds
					\begin{eqnarray}\label{integral_P}
							&&P^\beta_{j_R} =\left|\bra{E_0^j}m_j\ket{E_1^j}\right|^2 \int_0^{\infty}dw\int \frac{1}{2}\int dvdu e^{-i u \Delta E_j}\frac{1}{2\pi w} \times \nonumber \\
							&&[\frac{1}{1-e^{-\beta w}}\{e^{-iw  u}\frac{1}{1-e^{\frac{-2 \pi w}{a_j}}}+e^{iw u}\frac{1}{e^{\frac{2 \pi w}{a_j}}-1}\} +
							\frac{1}{e^{\beta w}-1}\{e^{iw u}\frac{1}{1-e^{\frac{-2 \pi w}{a_A}}}+e^{-iw u}\frac{1}{e^{\frac{2 \pi w}{a_j}}-1}\}]\nonumber \\
						\end{eqnarray}
						After performing the  $v$ integral, we obtain,
							\begin{eqnarray}
							&&P^\beta_{j_R} =\left|\bra{E_0^j}m_j\ket{E_1^j}\right|^2 \int_0^{\infty}dw  \delta(0) \pi \int du \frac{1}{2\pi w} \times \nonumber \\
							&&[\frac{1}{1-e^{-\beta w}}\{e^{-i(w+ \Delta E_j) u}\frac{1}{1-e^{\frac{-2 \pi w}{a_j}}}+e^{i(w-\Delta E_j) u}\frac{1}{e^{\frac{2 \pi w}{a_j}}-1}\} +
							\frac{1}{e^{\beta w}-1}\{e^{i(w-\Delta E_j) u}\frac{1}{1-e^{\frac{-2 \pi w}{a_j}}}+e^{-i(w+\Delta E_j) u}\frac{1}{e^{\frac{2 \pi w}{a_j}}-1}\}] \nonumber\\
						\end{eqnarray}
						After performing the  $u$ integral, we obtain,
							\begin{eqnarray}
							&&P^\beta_{j_R}=\left|\bra{E_0^j}m_j\ket{E_1^j}\right|^2 \int_0^{\infty}dw  \delta(0) \pi  \frac{1}{w} \times \nonumber 
							\\
							&&[\frac{1}{1-e^{-\beta w}}\{\delta(w+ \Delta E_j)\frac{1}{1-e^{\frac{-2 \pi w}{a_j}}}+\delta(w- \Delta E_j)\frac{1}{e^{\frac{2 \pi w}{a_j}}-1}\} +
							\frac{1}{e^{\beta w}-1}\{\delta(w-\Delta E_j)\frac{1}{1-e^{\frac{-2 \pi w}{a_j}}}+\delta(w-\Delta E_j)\frac{1}{e^{\frac{2 \pi w}{a_j}}-1}\}] \nonumber\\
							&&=\left|\bra{E_0^j}m_j\ket{E_1^j}\right|^2  \frac{\pi}{\Delta E_j}\delta(0)[\frac{1}{1-e^{-\beta \Delta E_j}}\frac{1}{e^{\frac{2\pi \Delta E_j}{a_j}}-1}+\frac{1}{e^{\beta \Delta E_j}-1}\frac{1}{1-e^{-\frac{2\pi \Delta E_j}{a_j}}}] 
						\end{eqnarray}
The  quantity $P^{\prime \prime \beta}_{j_R}$ can be calculated exactly in the same way .

						Next we concentrate on the following term,
                        \begin{eqnarray}
						M_j&=& \left|\bra{E_0^j}m_j\ket{E_1^j}\right|^2\int \int d\tau_j d\tau_j^\prime e^{-i \Delta \tau_j\Delta E_j}\Theta(-\Delta \tau_j )\Big[{G^{+ \beta}_R(x^\prime_A,x_A)+G^{+ \beta}_R(x_A,x^\prime_A)}\Big]	
						\nonumber
						\\		
						 &\equiv& M_j^1+M_j^2~.
						\end{eqnarray}
						The first part is evaluated to be 
						\begin{eqnarray}
							&&M_j^1=\left|\bra{E_0^j}m_j\ket{E_1^j}\right|^2\int \int d\tau_j d\tau_j^\prime e^{-i \Delta \tau_j \Delta E_j }\Theta(-\Delta \tau_j )		\int_0^\infty dw\frac{1}{2 \pi w}\Big[\frac{1}{1-e^{-\beta w}}\{e^{-iw \Delta \tau_j}\frac{1}{1-e^{\frac{-2 \pi w}{a_A}}}+e^{iw \Delta \tau_j}\frac{1}{e^{\frac{2 \pi w}{a_A}}-1}\}\Big] \nonumber\\ 
                            && \hspace{5 cm }+	\Big[\frac{1}{e^{\beta w}-1}\{e^{iw \Delta \tau_j}\frac{1}{1-e^{\frac{-2 \pi w}{a_A}}}+e^{-iw \Delta \tau_j}\frac{1}{e^{\frac{2 \pi w}{a_A}}-1}\}	\Big]\nonumber \\
							&&=\frac{1}{2}\left|\bra{E_0^j}m_j\ket{E_1^j}\right|^2 	\int_0^\infty dw\frac{1}{2 \pi w} \int_{-\infty}^{\infty} \int_{-\infty}^{\infty}  dv du  e^{-i  u \Delta E_j}\Theta(-u )\Big[\frac{1}{1-e^{-\beta w}}\{e^{-iw u}\frac{1}{1-e^{\frac{-2 \pi w}{a_A}}}+e^{iw u}\frac{1}{e^{\frac{2 \pi w}{a_A}}-1}\}\Big] \nonumber\\ 
                            && \hspace{5 cm }+	\Big[\frac{1}{e^{\beta w}-1}\{e^{iw u}\frac{1}{1-e^{\frac{-2 \pi w}{a_A}}}+e^{-iw u}\frac{1}{e^{\frac{2 \pi w}{a_A}}-1}\}	\Big]\nonumber \\
						\end{eqnarray}
						After performing the $v$ integral, we obtain,
							\begin{eqnarray}
							&&M_j^1=\frac{1}{2}\left|\bra{E_0^j}m_j\ket{E_1^j}\right|^2 	\int_0^\infty dw\frac{1}{2 \pi w}2\pi \delta(0) \int_{-\infty}^{0} du\Big[\frac{1}{1-e^{-\beta w}}\{e^{-i(w+\Delta E_j) u}\frac{1}{1-e^{\frac{-2 \pi w}{a_A}}}+e^{i(w-\Delta E_j) u}\frac{1}{e^{\frac{2 \pi w}{a_A}}-1}\}\Big] \nonumber\\ 
                            && \hspace{5 cm }+	\Big[\frac{1}{e^{\beta w}-1}\{e^{i(w-\Delta E_j) u}\frac{1}{1-e^{\frac{-2 \pi w}{a_A}}}+e^{-i(w+\Delta E_j) u}\frac{1}{e^{\frac{2 \pi w}{a_A}}-1}\}	\Big]~.
						\end{eqnarray}
						Now, using 
                         \begin{eqnarray}
							 \lim_{\epsilon \rightarrow \infty}\int_{-\infty}^{0} du \hspace{.1cm} e^{-iu(a+i\epsilon)}=\pi \delta(a) - \frac{P}{i}\Big(\frac{1}{a}\Big)~,
							 \label{BB5}
						 \end{eqnarray}
the $du$ integration can be performed. Here $P$ denotes the principal value. When it contributes at $\mathcal{O}(C^4)$
in the couplings, it makes a divergent contribution to $M_j$ and so will require special treatment if our results are extended to higher order. However, in our case (valid till order $C^2$), in Appendix \ref{AppJHEP1} and Sec \ref{about_late_eigen}, we noticed that although some terms like $M_j, \zeta_{AB}, Y_{AB}$ can be complex, ultimately we will only require the real parts corresponding to these terms to analyze our $\mathcal{ \delta N}$. In fact the term corresponding to principal value will not cotribute to our required quantities. Thus, we focus only on the real part of these expressions in the rest of our manuscript by dropping the principal value contribution and for compactification of our notation we omit explicit mention of $\mathfrak{R}$. Finally, performing the $dw$ integral and concentrating on the real part, we obtain the required expresssion:
\begin{eqnarray}
							&&M_j^1=\frac{1}{2}\left|\bra{E_0^j}m_j\ket{E_1^j}\right|^2 \frac{\pi\delta(0)}{\Delta E_j}[\frac{1}{1-e^{-\beta \Delta E_j}}\frac{\pi}{e^{\frac{2\pi \Delta E_j}{a_j}}-1}+\frac{1}{e^{\beta \Delta E_j}-1}\frac{\pi}{1-e^{\frac{-2\pi \Delta E_j}{a_j}}}]~.	
						\end{eqnarray}
Similarly, $M_j^2$ can be calculated and $M_j$ is hereby obtained as
\begin{eqnarray}
							&&M_j^2=\frac{1}{2}\left|\bra{E_0^j}m_j\ket{E_1^j}\right|^2 \frac{\pi\delta(0)}{\Delta E_j}[\frac{1}{1-e^{-\beta \Delta E_j}}\frac{\pi}{1-e^{\frac{-2\pi \Delta E_j}{a_j}}}+\frac{1}{e^{\beta \Delta E_j}-1}\frac{\pi}{e^{\frac{2\pi \Delta E_j}{a_j}}-1}]~.		
						\end{eqnarray}
Therefore, we find
						\begin{eqnarray}
							M_j=\left|\bra{E_0^j}m_j\ket{E_1^j}\right|^2 \frac{\pi}{2\Delta E_j}\delta(0)[\frac{1+e^{\frac{2\pi \Delta E_j}{a_j}}}{e^{\frac{2\pi\Delta E_j}{a_j}}-1}\times \frac{e^{\beta \Delta E_j}+1}{e^{\beta \Delta E_j}-1}]~.
						\end{eqnarray}
						Now as $z^\beta = \sum_j M_j$, one obtains
						\begin{eqnarray}
							&&	z^{\beta}=\sum_{j}\left|\bra{E_0^j}m_j\ket{E_1^j}\right|^2 \frac{\pi}{2\Delta E_j}\delta(0)[\frac{1+e^{\frac{2\pi \Delta E_j}{a_j}}}{e^{\frac{2\pi\Delta E_j}{a_j}}-1}\times \frac{e^{\beta \Delta E_j}+1}{e^{\beta \Delta E_j}-1}]~.
					        \nonumber \\
						\end{eqnarray}
						
$\bar P^{\beta}_{AB_{same}}$, $ Y^{\beta}_{AB_{same}}$, $ P^{\beta}_{AB_{same}}$, $\zeta^{\beta}_{AB_{same}}$ can be calculated from an earlier paper (see Eqs. (42), (43) and (46)  in \cite{Barman:2021bbw}).

\subsubsection{(1+3)-dimensions}\label{AppB2}
						
Using 
						\begin{eqnarray}\label{3d-expansion}
							&&\int_{-\infty}^{\infty} \frac{d^2k_p}{{(2 \pi)}^{4}}\mathcal{K}\left[\frac{iw}{a_A},\frac{\left|k_p\right|e^{a_A\zeta_A}}{a_A}\right] \mathcal{K}\left[\frac{iw}{a_B},\frac{\left|k_p\right|e^{a_B\zeta_B}}{a_B}\right]  \nonumber \\
							&& = \int_{0}^{\infty} \frac{  2\pi k_p dk_p}{{(2 \pi)}^{4}}\mathcal{K}\left[\frac{iw}{a_A},\frac{\left|k_p\right|e^{a_A\zeta_A}}{a_A}\right] \mathcal{K}\left[\frac{iw}{a_B},\frac{\left|k_p\right|e^{a_B\zeta_B}}{a_B}\right] \nonumber \\
							&& = \frac{1}{(2\pi)^3}\Upsilon(w,a_B,a_A)~,
			\end{eqnarray}
			in (\ref{BRMP1}), we obtain
			\begin{eqnarray}
							G^{+\beta}_{{3D_R}}(x_B^\prime,x_A)&=&\int_{0}^{\infty} dw \frac{2}{\sqrt{a_A a_B}}\Big[\frac{e^{-iw\Delta \eta }e^{\frac{\pi w}{2}\left(\frac{1}{a_A}+\frac{1}{a_B}\right)}+e^{iw\Delta \eta }e^{\frac{-\pi w}{2}\left(\frac{1}{a_A}+\frac{1}{a_B}\right)}}{1-e^{-\beta w}}
							\nonumber
							\\
							&&+	\frac{e^{iw\Delta \eta }e^{\frac{\pi w}{2}\left(\frac{1}{a_A}+\frac{1}{a_B}\right)}+e^{-iw\Delta \eta }e^{\frac{-\pi w}{2}\left(\frac{1}{a_A}+\frac{1}{a_B}\right)}}{1-e^{-\beta w}}\Big] \frac{1}{(2\pi)^3}\Upsilon(w,a_B,a_A)~.
							\label{BRMP2}
						\end{eqnarray}
			Now, to compute $P^\beta_{j_R}$, we use Eq. \eqref{BRMP2} and follow the same procedure that was used to obtain \eqref{integral_P}. This yields
							\begin{eqnarray}
								&&P^\beta_{j_R} = \left|\bra{E_0^j}m_j\ket{E_1^j}\right|^2 \int \int d\tau_j d\tau^\prime_j \chi(\tau_j)\chi(\tau^\prime_j)e^{-i\Delta \tau_j \Delta E_j}G^{\beta +}_{3D_{R}}(x^\prime_j,x_j) \nonumber \\
								&&=\left|\bra{E_0^j}m_j\ket{E_1^j}\right|^2 \delta(0)\frac{\Upsilon(\Delta E_j,a_j,a_j)}{\pi{a_j}  }\left[\frac{e^{-\frac{\pi \Delta E_j}{a_j}}}{1-e^{-\beta \Delta E_j}}+\frac{e^{\frac{\pi \Delta E_j}{a_j}}}{e^{\beta \Delta E_j}-1}\right]~.\nonumber \\
							\end{eqnarray}
							All the remaining expressions can be derived using Eq. \eqref{3d-expansion} and the steps are followed as in the earlier calculations for (1+1)D case.
											
			\subsection{ANTI-PARALLELLY ACCELERATING DETECTORS} 
				\label{app-opp}
				\subsubsection{(1+1)-dimensions}
				
				The expressions for $P^{\beta}_{j_J}$, $P^{\prime \prime \beta}_{j_J}$ , where $j = A(B)$ and $J = R(L)$ are derived similar to Eq. \eqref{integral_P}. The following term is calculated using $iG^{\beta}_{ F}(x_A,x^\prime_B)=G^{\beta}_{W}(x_A,x^\prime_B)+iG^{\beta}_R(x_B^{\prime},x_A) =G^{\beta}_{W}(x_A,x^\prime_B)+\Theta(\tau_B^\prime-\tau_A)\Big[G^{\beta}_{W}(x^\prime_B,x_A)-G^{\beta}_{W}(x_A,x^\prime_B)\Big] $, as follows.
				\begin{eqnarray}
					&&\zeta^{\beta}_{AB_{opp}} = \bra{E_1^B}m_B\ket{E_0^B}\bra{E_1^A}m_A\ket{E_0^A}\int \int d\tau_A d\tau^\prime_B \chi(\tau_A)\chi(\tau^\prime_B)e^{i \tau_A \Delta E_A}  e^{i \tau^\prime_B \Delta E_B}(iG^{\beta}_{ F}(x_A,x^\prime_B)) \nonumber \\
					&&=\bra{E_1^B}m_B\ket{E_0^B}\bra{E_1^A}m_A\ket{E_0^A}\int \int d\tau_A d\tau^\prime_B \chi(\tau_A)\chi(\tau^\prime_B)e^{i \tau_A \Delta E_A}  e^{i \tau^\prime_B \Delta E_B}(G^{\beta}_{W}(x_A,x^\prime_B)+iG^{\beta}_R(x_B^{\prime},x_A)) \nonumber \\
					&&\equiv I_E^\beta + I_R^\beta~.
				\end{eqnarray}
				Now change in variables $\tau^\prime_B-\tau_A \rightarrow u $ and $\tau^\prime_B+\tau_A \rightarrow v $ yields
				\begin{eqnarray}
					&&I^{\beta}_E= \bra{E_1^B}m_B\ket{E_0^B}\bra{E_1^A}m_A\ket{E_0^A}\int \int d\tau_A d\tau^\prime_B e^{i \tau_A \Delta E_A}  e^{i \tau^\prime_B \Delta E_B}G^{\beta}_{W}(x_A,x^\prime_B)\nonumber \\
					&&= \bra{E_1^B}m_B\ket{E_0^B}\bra{E_1^A}m_A\ket{E_0^A}\frac{1}{2}\int \int dv du e^{i \Delta E u}  e^{i \bar{E}v} \times \nonumber \\
					&&\int_0^\infty dw\frac{1}{4 \pi w}\frac{1}{\sqrt{\sinh \frac{\pi w}{a_B} \sinh \frac{\pi w}{a_A}}}[\frac{1}{1-e^{-\beta w}}\{e^{\frac{\pi}{2}w(\frac{1}{a_A}-\frac{1}{a_B})}e^{-iw v}+e^{-{\frac{\pi}{2}w(\frac{1}{a_A}-\frac{1}{a_B})}}e^{iw v}\} \nonumber \\
					&&+\frac{1}{e^{\beta w}-1}\{e^{\frac{\pi}{2}w(\frac{1}{a_A}-\frac{1}{a_B})}e^{iw v}+e^{-{\frac{\pi}{2}w(\frac{1}{a_A}-\frac{1}{a_B})}}e^{-iw v }\}]	
					\nonumber
					\\
					&&=	\bra{E_1^B}m_B\ket{E_0^B}\bra{E_1^A}m_A\ket{E_0^A}\delta({\Delta E_A - \Delta E_B})\frac{\pi}{\bar E }\big[\frac{e^{\frac{\pi}{2} \bar E(\frac{1}{a_A}-\frac{1}{a_B})}}{1-e^{-\beta \bar E}}+\frac{e^{\frac{-\pi}{2} \bar E(\frac{1}{a_A}-\frac{1}{a_B})}}{e^{\beta \bar E}-1}\big]\frac{1}{\sqrt{\sinh\frac{\pi \bar E}{a_A}\sinh\frac{\pi \bar E}{a_B}}}~,
					\label{I_E}
				\end{eqnarray}
and
				\begin{eqnarray}
					&&I^{\beta}_R=\bra{E_1^B}m_B\ket{E_0^B}\bra{E_1^A}m_A\ket{E_0^A}\int \int d\tau_A d\tau^\prime_B e^{i \tau_A \Delta E_A}  e^{i \tau^\prime_B \Delta E_B}iG^{\beta}_{R}(x^\prime_B,x_A)\nonumber \\
					&&= \bra{E_1^B}m_B\ket{E_0^B}\bra{E_1^A}m_A\ket{E_0^A}\frac{1}{2}\int \int dv du e^{i \Delta E u}  e^{i \bar{E}v} \{\Theta(u)\Big[G^{\beta}_{W}(x^\prime_B,x_A)-G^{\beta}_{W}(x_A,x^\prime_B)\Big]\}
					\nonumber \\
				\end{eqnarray}
					Integrating over $u$ and $ v$ by using 
                         \begin{eqnarray}
							 \lim_{\epsilon \rightarrow \infty}\int^{\infty}_{0} du \hspace{.1cm}e^{iu(a+i\epsilon)}=\pi \delta(a) - \frac{P}{i}\Big(\frac{1}{a}\Big)~,
							 \label{BB18}
						 \end{eqnarray}
						 and then finally integrating over $ w$, we obtain
				    \begin{eqnarray}
					&&I^{\beta}_R=-\bra{E_1^B}m_B\ket{E_0^B}\bra{E_1^A}m_A\ket{E_0^A}\delta({\Delta E_A - \Delta E_B})\frac{\pi}{\bar E }\sinh{\frac{\pi\bar E(\frac{1}{a_A}-\frac{1}{a_B})}{2}}\}\frac{1}{\sqrt{\sinh\frac{\pi \bar E}{a_A}\sinh\frac{\pi \bar E}{a_B}}}~.
				\end{eqnarray}
				In the above the principal value, which accounts an imaginary part, has been dropped. This is because, as discussed in Appendix \ref{AppB2} the imaginary part will ultimately cancel out when we analyze the $\mathcal{\delta N}$, and therefore here as well we focus only on the real parts of the necessary expressions.
				Thus one finds (which is only the real part)
				\begin{eqnarray}
					&&\zeta^{\beta}_{AB_{opp}}=\bra{E_1^B}m_B\ket{E_0^B}\bra{E_1^A}m_A\ket{E_0^A}\delta({\Delta E_A - \Delta E_B})\frac{\pi}{\bar E }\{\frac{e^{\frac{\pi}{2} \bar E(\frac{1}{a_A}-\frac{1}{a_B})}}{1-e^{-\beta \bar E}}+\frac{e^{\frac{-\pi}{2} \bar E(\frac{1}{a_A}-\frac{1}{a_B})}}{e^{\beta \bar E}-1}\nonumber \\
					&&-\sinh{\frac{\pi\bar E(\frac{1}{a_A}-\frac{1}{a_B})}{2}}\}\frac{1}{\sqrt{\sinh\frac{\pi \bar E}{a_A}\sinh\frac{\pi \bar E}{a_B}}}~.
				\end{eqnarray}
				$Y^{\beta}_{AB_{opp}}$ is calculated exactly using the same method which is complex in general; but ultimately only the real part will contribute to change in negativity. Again, $\bar  P^{\prime \beta}_{AB_{opp}}$ and $ P^{\prime \beta}_{AB_{opp}}$ are calculated as \eqref{I_E} and they are real in general.
				
				For $(1+3)$ dimensions, all expressions are derived using the similar procedure. So, we skip the details here. The final forms are given in the main text.
	
\section{\label{KMS}Proof of KMS relations}
From (\ref{V21}), in the proper frame of a detector (accelerating with constant value $a$) we have
\begin{eqnarray} 
  G(\Delta \eta)&=&\int_0^\infty \frac{dw}{4 \pi w}\frac{1}{\sinh \left( \frac{\pi w}{a}\right)}
  \Big[\frac{e^{\frac{\pi w}{a}} e^{-iw\Delta\eta}+ e^{\frac{-\pi w}{a}}e^{ iw\Delta\eta}}{1-e^{-\beta w}} + \frac{e^{\frac{\pi w}{a}} e^{iw\Delta\eta}+ e^{\frac{-\pi w}{a}}e^{ -iw\Delta\eta}}{e^{\beta w}-1}\Big] \nonumber \\
  &=& \int_0^\infty \frac{dw}{4 \pi w}\frac{1}{\sinh \left( \frac{\pi w}{a}\right)}
  \Big[e^{\frac{\pi w}{a}}(\frac{e^{-iw\Delta\eta}}{1-e^{-\beta w}}+ \frac{e^{iw\Delta\eta}}{e^{\beta w}-1})+ e^{\frac{-\pi w}{a}}(\frac{e^{iw\Delta\eta}}{1-e^{-\beta w}}+ \frac{e^{-iw\Delta\eta}}{e^{\beta w}-1})\Big] \nonumber \\
  &=& \int_0^\infty \frac{dw}{4 \pi w}\frac{1}{\sinh \left( \frac{\pi w}{a}\right)}\Big[e^{\frac{\pi w}{a}} K(\Delta \eta) + e^{\frac{-\pi w}{a}} \{K(\Delta \eta)\}^\star  \Big]~,
\label{JHEP1}
\end{eqnarray} 
where the following notation has been used:
\begin{eqnarray}
  K(\Delta \eta) &=& \frac{e^{-iw\Delta\eta}}{1-e^{-\beta w}}+ \frac{e^{iw\Delta\eta}}{e^{\beta w}-1}~.
\end{eqnarray}
Therefore one also finds
\begin{eqnarray}
  K(\Delta \eta- i \beta) &=& \frac{e^{-iw\Delta\eta}e^{-\beta w}}{1-e^{-\beta w}}+ \frac{e^{iw\Delta\eta}e^{\beta w}}{e^{\beta w}-1} \nonumber \\
  &=& \frac{e^{-iw\Delta\eta}}{e^{\beta w}-1}+ \frac{e^{iw\Delta\eta}}{1- e^{-\beta w}} \nonumber \\
   &=&  K(- \Delta \eta)~.
\end{eqnarray}
and then consequently we have
\begin{eqnarray}
  \{K(\Delta \eta- i \beta)\}^\star &=&  \{K(- \Delta \eta)\}^\star~.
\end{eqnarray}
Then (\ref{JHEP1}) yields
\begin{eqnarray}
  G(\Delta \eta - i \beta)&=& \int_0^\infty \frac{dw}{4 \pi w}\frac{1}{\sinh \left( \frac{\pi w}{a}\right)}\Big[e^{\frac{\pi w}{a}} K(\Delta \eta - i \beta) + e^{\frac{-\pi w}{a}} \{K(\Delta \eta - i \beta)\}^\star  \Big] \nonumber \\
  &=& \int_0^\infty \frac{dw}{4 \pi w}\frac{1}{\sinh \left( \frac{\pi w}{a}\right)}\Big[e^{\frac{\pi w}{a}} K(- \Delta \eta ) + e^{\frac{-\pi w}{a}} \{K(-\Delta \eta )\}^\star  \Big] \nonumber\\
  &=& G(-\Delta \eta )
\end{eqnarray}
This is indeed the KMS condition with respect to the inverse temperature $\beta$.

Now we will move for the same with respect to inverse Unruh temperature $(2\pi/a)$. For that we re-expresee (\ref{JHEP1}) in the following form:
\begin{eqnarray}
  G(\Delta \eta)&=& \int_0^\infty \frac{dw}{4 \pi w}\frac{1}{\sinh \left( \frac{\pi w}{a}\right)}
  \Big[\frac{1}{1-e^{-\beta w}}(e^{\frac{\pi w}{a}}e^{-iw\Delta \eta}+ e^{\frac{-\pi w}{a}}e^{ iw\Delta\eta}) + \frac{1}{e^{\beta w}-1}(e^{\frac{\pi w}{a}e^{iw\Delta \eta}} + e^{\frac{-\pi w}{a}}e^{- iw\Delta\eta})\Big] \nonumber \\
  &=& \int_0^\infty \frac{dw}{4 \pi w}\frac{1}{\sinh \left( \frac{\pi w}{a}\right)}
  \Big[\frac{1}{1-e^{-\beta w}} L(\Delta \eta)  + \frac{1}{e^{\beta w}-1}  \{L( \Delta \eta)\}^\star \Big]~, 
  \label{JHEP2}
\end{eqnarray}
where
\begin{eqnarray}
  L(\Delta \eta)&=& e^{\frac{\pi w}{a}}e^{-iw\Delta \eta} + e^{\frac{-\pi w}{a}}e^{ iw\Delta \eta}~.
\end{eqnarray}
Therefore one finds
\begin{eqnarray}
  L(\Delta \eta - i \frac{2 \pi}{a})&=& e^{\frac{\pi w}{a}}e^{-iw\Delta \eta}e^{-\frac{2\pi w}{a}} + e^{\frac{-\pi w}{a}}e^{ iw\Delta \eta}e^{\frac{2\pi w}{a}} \nonumber \\
  &=& e^{\frac{-\pi w}{a}}e^{-iw\Delta \eta} + e^{\frac{\pi w}{a}}e^{ iw\Delta \eta}\nonumber \\ 
  &=& L(-\Delta \eta)~,
\end{eqnarray}
and consequently we have
\begin{eqnarray}
  L(\Delta \eta - i \frac{2 \pi}{a}) &=& L(-\Delta \eta)~.
\end{eqnarray}
Hence from (\ref{JHEP2}) we find
\begin{eqnarray}
  G(\Delta \eta - i \frac{2 \pi}{a})&=& \int_0^\infty \frac{dw}{4 \pi w}\frac{1}{\sinh \left( \frac{\pi w}{a}\right)}
  \Big[\frac{1}{1-e^{-\beta w}} L(\Delta \eta - i \frac{2 \pi}{a})  + \frac{1}{e^{\beta w}-1}  \{L( \Delta \eta - i \frac{2 \pi}{a})\}^\star \Big] \nonumber \\
  &=& \int_0^\infty \frac{dw}{4 \pi w}\frac{1}{\sinh \left( \frac{\pi w}{a}\right)}
  \Big[\frac{1}{1-e^{-\beta w}} L(-\Delta \eta )  + \frac{1}{e^{\beta w}-1}  \{L( - \Delta \eta)\}^\star \Big] \nonumber \\
  &= & G(-\Delta \eta)~.
\end{eqnarray}
This is the KMS condition with respect to the inverse Unruh temperature.	

Exactly identical setps can be followed for $(1+3)$ dimensions as well and one will then lead to identical results.

\section{\label{RefJHEP1}$P_j$ is IR cut-off parameter free}			
The integration part of the $P_j$ for $\chi=1$ is given by 
\begin{equation}
P_j\sim\int_{-\infty}^{+\infty}\int_{-\infty}^{+\infty}	d\tau_j d\tau'_j e^{-i\Delta\tau_j\Delta E_j} G_w(x'_j,x_j)~,
\label{V27}
\end{equation}	
where the two dimensional positive frequency Wightman function in position space representation is \cite{Tjoa:2020eqh}
\begin{equation}
G_w(x'_j,x_j) = -\frac{1}{4\pi} \ln\Big((\Delta U_j - i\epsilon)(\Delta V_j - i\epsilon)\Big) - \frac{\ln(-\Lambda^2)}{4\pi}~.
\label{V28}
\end{equation} 	
In the above $U_j=x_j-t_j$, $V_j = x_j+t_j$ are the Minkowski null-null coordinates, while $\Lambda^2$ is the IR cut-off parameter. In an uniformly accelerated frame the first term of (\ref{V28}) yields the known value of $P_j$ whose rate (i.e. per unit time) is Bose distribution in nature (a detailed analysis can be followed from Section $4.4$ of \cite{Book1}, particularly see the analysis around Eq. $(4.54)$ and onwards of this reference). Whereas the IR part of (\ref{V28}) leads to
\begin{equation}
P_j^{\Lambda} \sim \ln(-\Lambda^2)\int_{-\infty}^{+\infty}\int_{-\infty}^{+\infty}	dT_j dT'_j e^{-iT_j\Delta E_j}~,
\label{V29}
\end{equation}
where $T_j = \Delta\tau_j = \tau_j-\tau'_j$ and $T_j = \tau_j+\tau'_j$. Now it may be noted that the integration over $T_j$ yields a Dirac-delta function of the form $\delta(\Delta E_j)$. Since in this case we have $\Delta E_j>0$, this terms vanishes and hence in this case $P_j$ is IR-safe.
						
		\end{appendix}

   	\end{widetext}

	\bibliographystyle{elsarticle-num}
	\bibliography{V3}
\end{document}